\documentclass[a4paper,10pt]{article}
\usepackage{amssymb}

\usepackage{amsmath}
\usepackage{bm}
\usepackage{boxedminipage}
\usepackage[pdftex]{graphicx}
\usepackage{graphicx}
\usepackage{enumerate}

\makeatletter \@addtoreset{equation}{section}

\makeatletter\renewcommand\section{\@startsection {section}{1}{\z@}                                   {-3.5ex \@plus -1ex \@minus -.2ex}                                   {2.3ex \@plus.2ex}                                   {\normalfont\large\bfseries}}
\renewcommand\subsection{\@startsection{subsection}{2}{\z@}                                     {-3.25ex\@plus -1ex \@minus -.2ex}                                     {1.5ex \@plus .2ex}                                     {\normalfont\bfseries}}
\def\Label#1{\label{#1}  \smash{\hbox to0pt{\raise1ex\hbox{\tiny[#1]}\hss}}}
\def\noLabels{\let\Label=\label}
\def\nobbibitem{\let\bbibitem=\bibitem}

\begin{document}

\begin{titlepage}

    \thispagestyle{empty}
    \begin{flushright}
        \hfill{CERN-PH-TH/2008-158}\\
        \hfill{UCB-PTH-08/73}\\
        \hfill{SU-ITP-08/28}\\
    \end{flushright}

    \vspace{10pt}
    \begin{center}
        { \Huge{\textbf{\textbf{$stu$} Black Holes Unveiled}}}

        \vspace{18pt}

        {\large{{\bf Stefano Bellucci$^{\spadesuit}$, \ Sergio Ferrara$^{\diamondsuit,\spadesuit,\flat}$, \ Alessio Marrani$^{\heartsuit,\sharp,\spadesuit}$\\and \ Armen Yeranyan$^{\spadesuit,\clubsuit}$}}}

        \vspace{20pt}

        {$\spadesuit$ \it INFN - Laboratori Nazionali di Frascati, \\
        Via Enrico Fermi 40,00044 Frascati, Italy\\
        \texttt{bellucci,marrani,ayeran@lnf.infn.it}}

        \vspace{10pt}

        {$\diamondsuit$ \it Physics Department,Theory Unit, CERN, \\
        CH 1211, Geneva 23, Switzerland\\
        \texttt{sergio.ferrara@cern.ch}}

        \vspace{10pt}

        {$\flat$ \it Miller Institute for Basic Research in Science \\
        University of California, Berkeley, CA94720, USA}

        \vspace{10pt}

        {$\heartsuit$ \it Museo Storico della Fisica e\\
        Centro Studi e Ricerche ``Enrico Fermi"\\
        Via Panisperna 89A, I-00184 Roma, Italy}

        \vspace{10pt}

        {$\sharp$ \it Stanford Institute for Theoretical Physics\\
        Department of Physics, 382 Via Pueblo Mall, Varian Lab,\\
        Stanford University, Stanford CA, 94305-4060, USA}

        \vspace{10pt}

        {$\clubsuit$ \it Department of Physics, Yerevan State University, \\Alex  Manoogian St., 1, Yerevan,
        375025, Armenia}

        \vspace{5pt}

        \vspace{40pt}

        {ABSTRACT}

    \end{center}


 The general solutions of the radial attractor flow equations for extremal black holes, both for non-BPS with non-vanishing
central charge $Z$ and for $Z=0$, are obtained for the\ so-called
$stu$ model, the minimal \textit{rank-}$3$ $\mathcal{N}=2$ symmetric
supergravity in $d=4$ space-time dimensions.

Comparisons with previous results, as well as the \textit{fake
supergravity (first order) formalism} and an analysis of the
\textit{BPS bound} all along the non-BPS attractor flows and of the
\textit{marginal stability} of corresponding $D$-brane
configurations, are given.


\end{titlepage}
\newpage\tableofcontents

\section{\label{Introduction}Introduction}

The physics of black holes (BHs) \cite{witt}--\nocite
{moore,duff1,hawking1,penrose}\cite{gibbons1} has received much attention in
the last years. This is also due to the issue of the \textit{Attractor
Mechanism}\cite{ferrara1}--\nocite{ferrara2,strominger2}\cite{FGK}, a
general phenomenon which occurs in \textit{extremal} BHs coupled to Maxwell
and scalar fields, as it is the case in supersymmetric theories of gravity
\cite{Sen-old1}--\nocite{GIJT,Sen-old2,K1,TT,G,GJMT,Ebra1,K2,Ira1,Tom,
BFM,AoB-book,FKlast,Ebra2,bellucci1,rotating-attr,K2-bis,Misra1,Lust2,Morales,BFMY,Astefa,CdWMa, DFT07-1,BFM-SIGRAV06,Cer-Dal-1,ADFT-2,Saraikin-Vafa-1,Ferrara-Marrani-1,TT2, ADOT-1,ferrara4,CCDOP,Misra2,Astefanesei,Anber,Myung1,Ceresole,BMOS-1,Hotta, Gao, PASCOS07,Sen-review,Belhaj1,AFMT1,Gaiotto1,BFMS1,GLS-1,ANYY1,bellucci2,Cai-Pang,Vaula, Li,BFMY2,Saidi2,Saidi3,Saidi4,Valencia-RTN-07,FHM,Unattractor,Vaula2,Trigiante}
\cite{Gnecchi-1} (for further developments, see also \textit{e.g.} \cite{OSV}%
--\nocite{OVV,ANV}\cite{GSV}).

Supergravity \cite{VN-Supergravity} can be obtained as the low-energy (small
curvature expansion) limit of \textit{superstrings} \cite{maldacena}--\nocite
{schwarz1,schwarz2} \cite{gasperini} or \textit{M-theory } \cite
{witten,schwarz3,schwarz3-bis}; in such a framework, a certain number of $%
U\left( 1\right) $ gauge fields and moduli fields are coupled to the
Einstein-Hilbert action. This is especially the case for theories in
$d=4$ space-time dimensions, and having $\mathcal{N}\geqslant 2$
\textit{supercharges}, where $4\mathcal{N}$ is the number of
supersymmetries. A popular example is the compactification of Type
II superstring theory on a Calabi-Yau threefold ($\mathcal{N}=2$) or
on a six-torus ($\mathcal{N}=8$).
The fermionic sector of these theories contains a certain number of spin $%
1/2 $ fermions and $\mathcal{N}$ spin $3/2$ Rarita-Schwinger fields, named
\textit{gravitinos} (the gauge fields of local supersymmetry). The vanishing
of the supersymmetric variation of the gravitinos determines whether or not
a certain number of supersymmetries (\textit{BPS property}) is preserved by
the BH background.

In this situation, asymptotically flat charged BH solutions, within a static
and spherically symmetric \textit{Ansatz}, can be regarded as a
generalization of the famous Schwarzschild BH. However, the presence of
additional quantum numbers (such as charges and scalar hair) make their
properties change drastically, and new phenomena appear. A novel important
feature of electrically (and/or magnetically) charged BHs \cite{nordstrom}
as well as rotating ones \cite{smarr} is a somewhat unconventional
thermodynamical property named \textit{extremality} \cite
{gibbons1,kallosh1,SUSY-censor}. \textit{Extremal} BHs are possibly stable
gravitational objects with finite entropy but vanishing temperature, in
which case the contribution to the gravitational energy entirely comes from
the electromagnetic (charges) and rotational \cite{rotating-attr} (angular
momentum/spin) attributes. \textit{Extremality} also means that the inner
(Cauchy) and outer (event) horizons do coincide, thus implying vanishing
\textit{surface gravity} (for a recent review see \textit{e.g.} \cite{FHM},
and Refs. therein).

In the regime of \textit{extremality} a particular relation among entropy,
charges and spin holds, yielding that the Arnowitt-Deser-Misner (ADM)
gravitational mass \cite{arnowitt,deser,bondi} is \textit{not} an
independent quantity. Stationary and spherically symmetric BHs in $d=4$
space-time dimensions and in an environment of scalar fields (typically
described by a non-linear sigma model) have scalar hair (\textit{scalar
charges}), corresponding to the values of the scalars at (asymptotically
flat) spatial infinity. These values may continuously vary, being an
arbitrary point in the moduli space of the theory or, in a more geometrical
language, a point in the target manifold of the scalar non-linear Lagrangian
\cite{ferrara1,gibbons3}. Nevertheless, the BH entropy, as given by the
Bekenstein-Hawking entropy-area formula \cite{hawking2}, is also in this
case independent on the scalar charges (\textit{``no scalar hair''}) and it
only depends on the asymptotic (generally \textit{dyonic}) BH charges.

This apparent puzzle can be resolved thanks to the aforementioned \textit{%
Attractor Mechanism}, a fascinating phenomenon that combines extremal BHs,
dynamical systems, algebraic geometry and number theory \cite{moore}. It was
firstly discovered in the context of supergravity; in a few words, in
constructing extremal dyonic BHs of $\mathcal{N}=2$, $d=4$ \textit{ungauged}
supergravity coupled to vector and hypermultiplets (with no $d=4$ scalar
potential), two phenomena occur: the hyperscalars can take arbitrary
constant values, while the radial evolution of the vector multiplets'
scalars is described by a dynamical system \cite{ferrara2,strominger2}.
Under some mild assumptions, the scalar trajectory flows to a \textit{%
``fixed point''}, located at the BH event horizon, in the target (moduli)
space. The \textit{``fixed point''} (\textit{i.e.} a point of vanishing
\textit{phase velocity}) represents the system in equilibrium, and it is the
analogue of an \textit{attractor} in the dynamical flow of dissipative
systems. In approaching such an \textit{attractor}, the orbits lose
practically all memory of initial conditions (\textit{i.e.} of the ``\textit{%
scalar hair''}), even though the dynamics is fully deterministic. The
scalars at the BH horizon turn out to depend only on the dyonic (asymptotic)
BH charges.

All extremal static, spherically symmetric and asymptotically flat BHs in $%
d=4$ have a \textit{Bertotti-Robinson} \cite{bertotti} $AdS_{2}\times S^{2}$
near-horizon geometry, with vanishing scalar curvature and \textit{%
conformally flat}; in particular, the radius of $AdS_{2}$ coincides with the
radius of $S^{2}$, and it is proportional to the (square root of the) BH
entropy (in turn proportional, through the Bekenstein-Hawking formula \cite
{hawking2}, to the area of the event horizon). Non-BPS (\textit{i.e.}
non-supersymmetric) (see \textit{e.g.} \cite
{FGK,bellucci1,TT2,ferrara4,Gaiotto1,GLS-1,ANYY1}) extremal BHs exist as
well, and they also exhibit an attractor behavior.\bigskip

A particularly remarkable $\mathcal{N}=2$, $d=4$ \textit{ungauged}
supergravity is the so-called $stu$ model, which exhibits the noteworthy
\textit{triality symmetry} \cite
{Magnific-7,Duff-stu,BKRSW,Shmakova,TT,Saraikin-Vafa-1,TT2,BMOS-1}. It has
been recently shown to be relevant for the analogy between pure states of
multipartite entanglement of qubits in quantum information theory and
extremal stringy BHs \cite{duff1}.

The $3$ complex scalars coming from the $3$ Abelian vector multiplets
coupled to the supergravity one span the \textit{rank-}$3$, completely
factorized special K\"{a}hler manifold $\frac{G}{H}=\left( \frac{SU\left(
1,1\right) }{U\left( 1\right) }\right) ^{3}$, with $dim_{\mathbb{C}}=3$,
\begin{equation}
G=\left( SU\left( 1,1\right) \right) ^{3}\sim \left( SO\left( 2,1\right)
\right) ^{3}\sim \left( SL\left( 2,\mathbb{R}\right) \right) ^{3}\sim \left(
Sp\left( 2,\mathbb{R}\right) \right) ^{3}  \label{GG}
\end{equation}
being the $d=4$ $U$-duality group\footnote{%
With a slight abuse of language, we refer to $U$-duality group as to the
\textit{continuous} version, valid for large values of charges, of the
string duality group introduced by Hull and Townsend \cite{Hull}.}, and $%
H=\left( U\left( 1\right) \right) ^{3}\sim \left( SO\left( 2\right) \right)
^{3}$ its \textit{maximal compact subgroup}. Such a space is nothing but the
element $n=2$ of the sequence of reducible homogeneous symmetric special
K\"{a}hler manifolds $\frac{SU(1,1)}{U(1)}\otimes \frac{SO(2,n)}{%
SO(2)\otimes SO(n)}$ (see \textit{e.g.}~\cite{bellucci1} and Refs. therein).

The $stu$ model has $2$ non-BPS $Z\neq 0$ \textit{flat} directions, spanning
the moduli space $SO\left( 1,1\right) \times SO\left( 1,1\right) $ (\textit{%
i.e.} the scalar manifold of the $stu$ model in $d=5$), but \textit{no}
non-BPS $Z=0$ \textit{massless Hessian modes} at all \cite{ferrara4} (see
also \cite{TT2} and \cite{Ferrara-Marrani-1}). In other words, the $6\times
6 $ Hessian matrix of the effective BH potential at its non-BPS $Z\neq 0$
critical points has $4$ strictly positive and $2$ vanishing eigenvalues
(these latter correspond to \textit{massless Hessian modes}), whereas at its
non-BPS $Z=0$ critical points all the eigenvalue are strictly positive.
After \cite{FGK}, $\frac{1}{2}$-BPS critical points of $V_{BH}$ in $\mathcal{%
N}=2$, $d=4$ supergravity are \textit{all} stable, and thus they determine
attractors in a strict sense. It is here worth pointing out that the $d=6$
uplift of the $stu$ model is $\left( 1,0\right) $ supergravity coupled to $%
n_{T}=1$ tensor multiplet. The BH charge orbits supporting the various
classes of non-degenerate attractors have been studied in \cite{bellucci1}
(see also \cite{BMOS-1}).

Concerning its stringy origins, the $stu$ model can be interpreted \textit{%
e.g.} as the low-energy limit of Type $IIA$ superstrings compactified on a
six-torus $T^{6}$ factorized as $T^{2}\times T^{2}\times T^{2}$. The $%
D0-D2-D4-D6$ branes wrapping the various $T^{2}$s determine the $4$ magnetic
and $\mathit{4}$ electric BH charges.

Remarkably, the $stu$ model is a sector of \textit{all} $\mathcal{N}>2$, $%
d=4 $ supergravities, as well as of \textit{all} $\mathcal{N}=2$, $d=4$
supergravities based on homogeneous (both \textit{symmetric} \cite
{GST1,Magnific-7,CFG,CVP} and \textit{non-symmetric} - see \textit{e.g.}
\cite{dWVVP,dWVP} -) scalar manifolds based on cubic geometries. Thus, $stu$
model captures the essential features of extremal BHs in all such theories
(see \textit{e.g.} the $stu$ interpretation of $\mathcal{N}=8$, $d=4$
attractors \cite{FKlast}, and the observations in \cite{GLS-1}).\bigskip

Recently, the $stu$ model has been object of detailed investigation
concerning the integration of the equations of motion of the scalars in the
background of a given BH charge configuration, in particular of those ones
which support non-supersymmetric flows:

\begin{itemize}
\item  In \cite{K2-bis} the non-BPS $Z\neq 0$ attractor flow was
investigated for the first time. The $\frac{1}{2}$-BPS attractor flow
solution was previously studied in \cite{Cvetic-Youm}-\nocite
{Tseytlin,Gauntlett,Bala,BPS-flow-1,BPS-flow-2,BPS-flow-3,Bates-Denef}\cite
{bala-foam}, and its most general form is known to be obtained simply by
replacing the BH charges with the corresponding, symplectic covariant
harmonic functions in the the most general horizon, critical solution,
obtained in \cite{BKRSW,Shmakova}. In \cite{K2-bis} such a feature has been
shown to hold also for non-BPS $Z\neq 0$ attractor flow, for the $D2-D6$ (%
\textit{electric}) and $D0-D2-D4-D6$ configurations, for particular cases
\textit{without }$B$\textit{-fields}.

\item  In \cite{Hotta} the $D0-D4$ (\textit{magnetic}) system was
considered, and new exact non-BPS $Z\neq 0$ attractor flow solutions were
derived, with non-vanishing dynamical axion. Furthermore, the result of \cite
{K2-bis} was proved to be non-general. Indeed, within the $D0-D4$ (\textit{%
magnetic}) configuration (\textit{dual} to the \textit{electric} one,
studied in \cite{K2-bis}), it was pointed out that non-BPS $Z\neq 0$
attractor flow differ from the $\frac{1}{2}$-BPS one, because the most
general non-BPS $Z\neq 0$ solution \textit{cannot} be obtained by replacing
the BH charges with the corresponding harmonic functions in the most general
horizon critical solution, as instead it holds for the supersymmetric case
\cite{Cvetic-Youm}-\nocite
{Tseytlin,Gauntlett,Bala,BPS-flow-1,BPS-flow-2,BPS-flow-3,Bates-Denef}\cite
{bala-foam}, and actually also for the non-BPS $Z=0$ case (see Sect. \ref
{Non-BPS-Z=0-Flow}). This is actually due to the presence of non-trivial
so-called $B$\textit{-fields} in the non-BPS $Z\neq 0$ attractor flow, as
well as to the presence of \textit{flat} directions (spanning a related
\textit{moduli space}) \textit{all along} such a flow.

\item  In \cite{GLS-1} the $stu$ model was further studied in the $D0-D4$ (%
\textit{magnetic}) as well as in the $D0-D6$ configurations, by fully
exploiting the contribution of the $B$\textit{-fields}, and (as also done in
\cite{K2-bis}) performing the relevant $U$-duality transformations in order
to relate different BH charge configurations supporting the same attractor
flow (\textit{i.e.} belonging to the same BH charge orbit of $U$-duality
\cite{bellucci1}). The \textit{ADM mass} $M_{ADM}$ of the extremal BH was
computed in the $\frac{1}{2}$-BPS and in the non-BPS $Z\neq 0$ cases, and
the \textit{marginal stability} \cite{Marginal-Refs} of the corresponding
physical states was studied, founding that the \textit{marginal bound} \cite
{Marginal-Refs} was saturated in the non-supersymmetric case, contrarily to
the BPS case. The difference between the squared non-BPS $Z\neq 0$ ADM mass $%
M_{ADM,non-BPS,Z\neq 0}^{2}$ and $\left| Z\right| ^{2}$ was computed at the
radial infinity along the non-BPS $Z\neq 0$ attractor flow, showing that the
\textit{BPS bound} \cite{gibbons2} actually holds also at the infinity, even
if dependent on the asymptotical values of the scalars (see Eq. (4.8) of
\cite{GLS-1}). Moreover, in such a paper the two non-BPS $Z\neq 0$ \textit{%
flat} directions of the $stu$ model \cite{TT2,ferrara4} were shown to hold
also along the whole corresponding attractor flow, as mentioned above.

\item  The analysis of \cite{GLS-1} was further developed in \cite{Cai-Pang}%
, in which the non-BPS $Z\neq 0$ equations of motion of the scalars were
solved for the $D2-D6$ (\textit{electric}) and $D0-D2-D4$ supporting BH
charge configurations.\bigskip
\end{itemize}

The present paper is devoted to a detailed, complete study of the attractor
flow equations of the $stu$ model, whose fundamental facts are summarized in
Sect. \ref{stu}. All the classes of \textit{non-degenerate} (\textit{i.e.}
with non-vanishing classical Bekenstein-Hawking \cite{hawking2} BH entropy)
attractor flow solutions are determined, in their most general form (with
all $B$\textit{-fields} switched on). The main results of our investigation
are listed below:

\begin{itemize}
\item  As mentioned above, the $\frac{1}{2}$-BPS attractor flow solution is
known since \cite{Cvetic-Youm}-\nocite
{Tseytlin,Gauntlett,Bala,BPS-flow-1,BPS-flow-2,BPS-flow-3,Bates-Denef}\cite
{bala-foam}, and it is reviewed in Sect. \ref{BPS-Flow}. In Sect. \ref
{Non-BPS-Z=0-Flow} the non-BPS $Z=0$ attractor flow solution, untreated so
far, is determined for the most general supporting BH charge configuration,
and its relation to the supersymmetric flow, both at and away from the event
horizon radius $r_{H}$, is established, consistently with the results of
\cite{BMOS-1}.

\item  Sect. \ref{Non-BPS-Z<>0-Flow} is devoted to the study of the non-BPS $%
Z\neq 0$ attractor flow solution in full generality. By using suitable $U$%
-duality transformations (Subsect. \ref{U-Duality-Transf}), and starting
from the $D0-D6$ configuration (Subsect. \ref{D0-D6}), the non-BPS $Z\neq 0$
attractor flow supported by the most general $D0-D2-D4-D6$ configuration
(with \textit{all} charges switched on) is explicitly derived in Subsect.
\ref{D0-D2-D4-D6}. This completes and generalizes the analyses and the
results of \cite{K2-bis}, \cite{Hotta}, \cite{GLS-1} and \cite{Cai-Pang}.
The above mentioned finding of \cite{GLS-1} is confirmed in such a general
framework: the moduli space $\left( SO\left( 1,1\right) \right) ^{2}$, known
to exist at the non-BPS $Z\neq 0$ critical points of $V_{BH}$ \cite
{ferrara4,TT2}, is found to be present \textit{all along} the non-BPS
attractor flow, \textit{i.e.} for every $r\geqslant r_{H}$.

\item  In Sect. \ref{Analysis-Particular} a detailed analysis of particular
configurations, namely $D0-D4$ (\textit{magnetic}, Subsect. \ref{D0-D4}),
its \textit{dual} $D2-D6$ (\textit{electric}, Subsect. \ref{D2-D6}), and $%
D0-D2-D4$ (Subsect. \ref{D0-D2-D4}), is performed.

\item  The so-called \textit{first order (fake supergravity) formalism},
introduced in \cite{Fake-Refs}, has been recently developed in \cite
{Cer-Dal-1} and \cite{ADOT-1} in order to describe $d=4$ \textit{extremal}
BHs; in general, it is based on a suitably defined real, scalar-dependent,
\textit{fake superpotential}\footnote{%
It is worth pointing out that the \textit{first order formalism}, as
(re)formulated in \cite{Cer-Dal-1} and \cite{ADOT-1} for $d=4$ \textit{%
extremal} BHs, automatically selects the solutions which do \textit{not}
blow up at the BH event horizon. In other words, the \textit{(covariant)
scalar charges} $\Sigma _{i}$ built in terms of the \textit{fake
superpotential} $\mathcal{W}$ (see Eq. (\ref{d-2}) further below) satisfy by
construction all the conditions in order for the \textit{Attractor Mechanism}
to hold.
\par
It should be here recalled that for \textit{extremal} BHs the solution
converging at the BH event horizon ($r\rightarrow r_{H}^{+}$) does \textit{%
not} depend on the initial, asymptotical values of the scalar fields. See%
\textit{\ e.g.} discussions in \cite{Astefa} and \cite{Astefanesei}.}\textit{%
\ }$\mathcal{W}$. In the framework of $stu$ model, we explicitly build up $%
\mathcal{W}$ in the non-trivial cases represented by the non-BPS attractor
flows. For the non-BPS $Z=0$ attractor flow (Sect. \ref{Non-BPS-Z=0-Flow}) a
manifestly $\mathit{H}$\textit{-invariant} $\mathcal{W}$ is determined, in
three different (but equivalent) \textit{``polarizations''}. Furthermore,
the difference between the squared non-BPS $Z=0$ \textit{fake superpotential}
$\mathcal{W}_{non-BPS,Z=0}^{2}$ and $\left| Z\right| ^{2}$ is computed along
the non-BPS $Z=0$ attractor flow, and the \textit{BPS bound} \cite{gibbons2}
is found to hold \textit{all along the attractor flow} (\textit{i.e.} not
only on the BH event horizon $r=r_{H}$, but $\forall r\geqslant r_{H}$). On
the other hand, for non-BPS $Z\neq 0$ attractor flow (Subsects. \ref{D0-D6}
and \ref{D0-D2-D4-D6}) the fake superpotential is manifestly \textit{not }$%
\mathit{H}$\textit{-invariant}.

As it will be commented in Sect. \ref{Conclusion}, this is not inconsistent
with the treatment of \cite{Cer-Dal-1} and \cite{ADOT-1}. Indeed, the
\textit{fake superpotential} is \textit{not} unique within the same
attractor flow, the various equivalent superpotentials being related through
a (possibly scalar-dependent) $R$-matrix satisfying the conditions (2.28)
and (2.29) of \cite{Cer-Dal-1} (see in general Subsect. 2.2 of \cite
{Cer-Dal-1}). Moreover, in Sect. \ref{Analysis-Particular} the difference
between the squared non-BPS $Z\neq 0$ \textit{fake superpotential} $\mathcal{%
W}_{non-BPS,Z\neq 0}^{2}$ and $\left| Z\right| ^{2}$ is computed
along the non-BPS $Z\neq 0$ attractor flow in the \textit{magnetic
(}Subsect. \ref {D0-D4}), \textit{electric} (Subsect. \ref{D2-D6})
and $D0-D2-D4$ (Subsect. \ref{D0-D2-D4}) charge configurations.
Analogously to the non-BPS $Z=0$ case, the \textit{BPS bound
}\cite{gibbons2}\textit{\ }is found to hold \textit{all along the
attractor flow} (\textit{i.e.} not only on the BH event horizon
$r=r_{H}$, but $\forall r\geqslant r_{H}$). In particular, for the
\textit{magnetic} charge configuration at radial infinity the result
given by Eq. (4.8) of \cite{GLS-1} is recovered\footnote{We are
grateful to E. G. Gimon for a clarifying discussion about the
definition of ``gap" above the \textit{BPS bound} as given in
\cite{GLS-1}.}.

\item  Within the \textit{first order (fake supergravity) formalism}, for
all attractor flows we compute the \textit{covariant scalar charges }as well
as the \textit{ADM mass}, studying the issue of \textit{marginal stability}
\cite{Marginal-Refs}. We thus complete the analysis and the results of \cite
{Hotta}, \cite{GLS-1} and \cite{Cai-Pang}. As expected due to the strict
similarity to the $\frac{1}{2}$-BPS attractor flow (Sect. \ref{BPS-Flow}),
also in the non-BPS $Z=0$ case the \textit{marginal bound} is \textit{not}
saturated (Sect. \ref{Non-BPS-Z=0-Flow}), as instead we confirm to hold in
general in the (non-BPS $Z\neq 0$-supporting branch of the) $D0-D2-D4-D6$
configuration (Subsect. \ref{D0-D2-D4-D6}).

\item  Final remarks, comments, and outlook for further developments are
given in the concluding Sect. \ref{Conclusion}.
\end{itemize}

\section{\label{stu}Basics of the $stu$ Model}

We here recall some basic facts of the above mentioned $stu$ model \cite
{Magnific-7,Duff-stu,BKRSW,Shmakova,TT,K2-bis,Saraikin-Vafa-1,Ferrara-Marrani-1,TT2,ferrara4,BMOS-1}%
, fixing our notations and conventions. The three complex moduli~of the
model are defined as
\begin{equation}
z^{1}\equiv x^{1}-iy^{1}\equiv s,~~z^{2}\equiv x^{2}-iy^{2}\equiv
~t,~~z^{3}\equiv x^{3}-iy^{3}\equiv ~u,  \label{boss-3-bis}
\end{equation}
with $y^{i}\in \mathbb{R}_{0}^{+}$~\cite{Gilmore} (see also the treatment of
\cite{Ceresole}). In \textit{special coordinates} (see e.g. \cite{4} and
Refs. therein) the prepotential determining the relevant special K\"{a}hler
geometry simply reads
\begin{equation}
f=stu.  \label{prepotential-stu}
\end{equation}
Due to the aforementioned \textit{triality symmetry} among $s$, $t$ and $u$,
the expressions of all the relevant geometric quantities acquire quite
elegant form. Some formul\ae\ can be found \textit{e.g.} in the treatments
of \cite{BKRSW,Shmakova,K2-bis,BMOS-1,GLS-1}; for completeness, below we
list the expressions of the K\"{a}hler potential, contravariant metric
tensor, non-vanishing components of the Christoffel symbols of the second
kind and of the $C$-tensor, holomorphic central charge (also named
superpotential) and BH effective potential ($i=1,2,3$ throughout):
\begin{equation}
\begin{array}{l}
K=-ln\left[ -i(s-\overline{s})(t-\overline{t})(u-\overline{u})\right]
\Rightarrow exp\left( -K\right) =8y^{1}y^{2}y^{3}; \\[0.5em]
\\
g^{i\bar{j}}=-diag((s-\overline{s})^{2},(t-\overline{t})^{2},(u-\overline{u}%
)^{2}); \\
\\
\Gamma _{~11}^{1}=-2(s-\overline{s})^{-1},~\Gamma _{~22}^{2}=-2(t-\overline{t%
})^{-1},~\Gamma _{~33}^{3}=-2(u-\overline{u})^{-1}; \\
\\
C_{stu}=\frac{i}{(s-\bar{s})(t-\bar{t})(u-\bar{u})}=exp\left( K\right) ; \\
\\
W(s,t,u)=q_{0}+q_{1}s+q_{2}t+q_{3}u+p^{0}stu-p^{1}tu-p^{2}su-p^{3}st\,; \\
\\
V_{BH}=exp\left( K\right) \cdot \left[ \,|W(s,t,u)|^{2}+|W(\bar{s}%
,t,u)|^{2}+|W(s,\bar{t},u)|^{2}+|W(s,t,\bar{u})|^{2}\right] .
\end{array}
\label{stu-geometry}
\end{equation}
Thus, the (covariantly holomorphic) \textit{central charge function} for the
$stu$ model reads (see \textit{e.g.} \cite{4} and Refs. therein)
\begin{eqnarray}
Z\left( s,t,u,\overline{s},\overline{t},\overline{u}\right) &\equiv
&e^{K/2}W\left( s,t,u\right) =  \notag \\
&=&\frac{1}{\sqrt{-i(s-\overline{s})(t-\overline{t})(u-\overline{u})}}\left(
q_{0}+q_{1}s+q_{2}t+q_{3}u+p^{0}stu-p^{1}tu-p^{2}su-p^{3}st\,\right) .
\notag \\
&&  \label{g-1}
\end{eqnarray}

The definition of the BH charges $p^{\Lambda }$ (magnetic) and $q_{\Lambda }$
(electric) ($\Lambda =0,1,2,3$ throughout), the effective $1$-dim.
(quasi-)geodesic Lagrangian of the $stu$ model, and the corresponding Eqs.
of motion for the scalars can be found in Subsects. 2.2 and 2.3, as well as
in appendix A (treating the case $D0-D4$ in detail), of \cite{GLS-1}.

Through the Bekenstein-Hawking entropy-area formula \cite{hawking2}, the
entropy of an extremal BH in the $stu$ model in the Einsteinian
approximation reads as follows:
\begin{equation}
S_{BH}=\frac{A_{H}}{4}=\pi \left. V_{BH}\right| _{\partial V_{BH}=0}=\pi
\sqrt{\left| \mathcal{I}_{4}\left( \Gamma \right) \right| },
\end{equation}
where the $\left( 2n_{V}+2\right) \times 1$ vector of BH charges
\begin{equation}
\Gamma \equiv \left( p^{\Lambda },q_{\Lambda }\right) ,  \label{Gamma}
\end{equation}
was introduced, $n_{V}$ denoting the number of Abelian vector multiplets
coupled to the supergravity one (in the case under consideration $n_{V}=3$).
Furthermore, $\mathcal{I}_{4}\left( \Gamma \right) $ denotes the unique
invariant of the \textit{tri-fundamental representation} $\left( \mathbf{2},%
\mathbf{2},\mathbf{2}\right) $ of the $U$-duality group $G$, reading as
follows (see \textit{e.g.} Eq. (4.10) and Sect. 5 of \cite{BMOS-1}, and
Refs. therein):
\begin{equation}
\mathcal{I}_{4}\left( \Gamma \right) =-\left( p^{\Lambda }q_{\Lambda
}\right)
^{2}+4%
\sum_{i<j}p^{i}q_{i}p^{j}q_{j}-4p^{0}q_{1}q_{2}q_{3}+4q_{0}p^{1}p^{2}p^{3}=-Det\left( \Psi \right) ,
\label{I4}
\end{equation}
where $Det\left( \Psi \right) $ is the so-called Cayley's hyperdeterminant
\cite{duff1}.\smallskip

In the next three Sections we will discuss the explicit solutions of the
equations of motion of the scalars $s$, $t$ and $u$ in the dyonic background
of an extremal BH of the $stu$ model, also named \textit{Attractor Flow
Equations}. We will consider only \textit{non-degenerate} attractor flows,
\textit{i.e.} those flows determining a regular, non-vanishing area of the
horizon in the Einsteinian approximation.

As mentioned above, $3$ classes of non-degenerate attractor flows exist in
the $stu$ model:

\begin{itemize}
\item  $\frac{1}{2}$-BPS (Sect. \ref{BPS-Flow});

\item  non-BPS $Z=0$ (Sect. \ref{Non-BPS-Z=0-Flow});

\item  non-BPS $Z\neq 0$ (Sects. \ref{Non-BPS-Z<>0-Flow}) and (\ref
{Analysis-Particular}).
\end{itemize}

\section{\label{BPS-Flow}The Most General $\frac{1}{2}$-BPS Attractor Flow}

The explicit expression of the attractor flow solution supported by the most
general $\frac{1}{2}$-BPS BH charge configuration in $\mathcal{N}=2$, $d=4$
\textit{ungauged} supergravity coupled to $n_{V}$ Abelian vector multiplets
(and exhibiting a unique $U$-invariant $\mathcal{I}_{4}$) is known after
\cite{Cvetic-Youm}-\nocite
{Tseytlin,Gauntlett,Bala,BPS-flow-1,BPS-flow-2,BPS-flow-3,Bates-Denef}\cite
{bala-foam} (as well as the third of Refs. \cite{Marginal-Refs}):
\begin{eqnarray}
exp\left[ -4U_{\frac{1}{2}-BPS}\left( \tau \right) \right] &=&\mathcal{I}%
_{4}\left( \mathcal{H}\left( \tau \right) \right) ;  \notag \\
z_{\frac{1}{2}-BPS}^{i}\left( \tau \right) &=&\frac{H^{i}\left( \tau \right)
+i\partial _{H_{i}}\mathcal{I}_{4}^{1/2}\left( \mathcal{H}\left( \tau
\right) \right) }{H^{0}\left( \tau \right) +i\partial _{H_{0}}\mathcal{I}%
_{4}^{1/2}\left( \mathcal{H}\left( \tau \right) \right) },  \label{BPS-sol}
\end{eqnarray}
where $\partial _{H_{i}}\equiv \frac{\partial }{\partial H_{i}}$, and the $%
\left( 2n_{V}+2\right) \times 1$($=8\times 1$ in the model under
consideration) symplectic vector
\begin{equation}
\mathcal{H}\left( \tau \right) \equiv \left( H^{\Lambda }\left( \tau \right)
,H_{\Lambda }\left( \tau \right) \right) ,  \label{H}
\end{equation}
was introduced, where $H^{\Lambda }\left( \tau \right) $ and $H_{\Lambda
}\left( \tau \right) $ are \textit{harmonic functions} defined as follows ($%
\tau \equiv \left( r_{H}-r\right) ^{-1}\in \mathbb{R}^{-}$):
\begin{equation}
\begin{array}{l}
H^{\Lambda }\left( \tau \right) \equiv p_{\infty }^{\Lambda }+p^{\Lambda
}\tau ; \\
\\
H_{\Lambda }\left( \tau \right) =q_{\Lambda ,\infty }+q_{\Lambda }\tau ,
\end{array}
\label{HH}
\end{equation}
such that $\mathcal{H}\left( \tau \right) $ can be formally rewritten as
\begin{equation}
\mathcal{H}\left( \tau \right) =\Gamma _{\infty }+\Gamma \tau .  \label{HHH}
\end{equation}
The asymptotical constants $\Gamma _{\infty }$ must satisfy the following
\textit{integrability conditions}:
\begin{equation}
\mathcal{I}_{4}\left( \Gamma _{\infty }\right) =1,~~\left\langle \Gamma
,\Gamma _{\infty }\right\rangle =0,  \label{constraints}
\end{equation}
where $\left\langle \cdot ,\cdot \right\rangle $ is the scalar product
defined by the $\left( 2n_{V}+2\right) \times \left( 2n_{V}+2\right) $
symplectic metric. Under such conditions, the flow (\ref{BPS-sol}) is the
most general solution of the so-called $\frac{1}{2}$\textit{-BPS
stabilization Eqs. }(see \textit{e.g.} the recent treatment of \cite{K2-bis}%
):
\begin{equation}
\mathcal{H}^{T}\left( \tau \right) =2e^{K\left( z\left( \tau \right) ,%
\overline{z}\left( \tau \right) \right) }Im\left[ W\left( z\left( \tau
\right) ,\mathcal{H}\left( \tau \right) \right) \left(
\begin{array}{c}
\overline{X}^{\Lambda }\left( \overline{z}\left( \tau \right) \right) \\
\\
\overline{F}_{\Lambda }\left( \overline{z}\left( \tau \right) \right)
\end{array}
\right) \right] ,  \label{BPS-stabilization-Eqs.}
\end{equation}
obtained from the $\frac{1}{2}$\textit{-BPS Attractor Eqs.} (see \textit{e.g.%
} the treatment in \cite{AoB-book}, and Refs. therein)
\begin{equation}
\Gamma ^{T}=2e^{K\left( z,\overline{z}\right) }Im\left[ W\left( z,\Gamma
\right) \left(
\begin{array}{c}
\overline{X}^{\Lambda }\left( \overline{z}\right) \\
\\
\overline{F}_{\Lambda }\left( \overline{z}\right)
\end{array}
\right) \right]  \label{BPS-AEs}
\end{equation}
by simply replacing $\Gamma $ with $\mathcal{H}\left( \tau \right) $ (see
\textit{e.g.} \cite{BPS-flow-1} and Refs. therein). Consistently, Eq. (\ref
{BPS-AEs}) is the \textit{near-horizon} ($\tau \rightarrow -\infty $)
\textit{limit }of Eq. (\ref{BPS-stabilization-Eqs.}).

Moreover, the BH charge configurations supporting the $\frac{1}{2}$-BPS
attractors at the BH event horizon satisfy the following constraints,
defining the $\frac{1}{2}$-BPS orbit (see Appendix II of \cite{bellucci1})
\begin{equation}
\mathcal{O}_{\frac{1}{2}-BPS}=\frac{\left( SU\left( 1,1\right) \right) ^{3}}{%
\left( U\left( 1\right) \right) ^{2}}  \label{Wed-5}
\end{equation}
of the \textit{tri-fundamental} representation $\left( \mathbf{2},\mathbf{2},%
\mathbf{2}\right) $ of the $U$-duality group $\left( SU\left( 1,1\right)
\right) ^{3}$ \cite{bellucci1,BMOS-1}:
\begin{equation}
\begin{array}{l}
\mathcal{I}_{4}\left( \Gamma \right) >0; \\
\\
p^{2}p^{3}-p^{0}q_{1}\gtrless 0; \\
\\
p^{1}p^{3}-p^{0}q_{2}\gtrless 0; \\
\\
p^{1}p^{2}-p^{0}q_{3}\gtrless 0.
\end{array}
\label{Wed-1}
\end{equation}
Correspondingly, $\mathcal{H}\left( \tau \right) $ is constrained as follows
along the $\frac{1}{2}$-BPS attractor flow ($\forall \tau \in \mathbb{R}^{-}$%
):
\begin{equation}
\begin{array}{l}
\mathcal{I}_{4}\left( \mathcal{H}\left( \tau \right) \right) >0; \\
\\
H^{2}\left( \tau \right) H^{3}\left( \tau \right) -H^{0}\left( \tau \right)
H_{1}\left( \tau \right) \gtrless 0; \\
\\
H^{1}\left( \tau \right) H^{3}\left( \tau \right) -H^{0}\left( \tau \right)
H_{2}\left( \tau \right) \gtrless 0; \\
\\
H^{1}\left( \tau \right) H^{2}\left( \tau \right) -H^{0}\left( \tau \right)
H_{3}\left( \tau \right) \gtrless 0.
\end{array}
\label{Wed-2}
\end{equation}

In the \textit{near-horizon limit} $\tau \rightarrow -\infty $, Eq. (\ref
{BPS-sol}) yields the \textit{purely charge-dependent}, \textit{critical}
expressions of the scalars at the BH event horizon, \textit{e.g.} given by
Eq. (3.1) of \cite{GLS-1}. In the same limit, the constraints (\ref{Wed-2})
consistently yield the constraints (\ref{Wed-1}).

Consistently with the analysis of \cite{Ceresole}, the general $\frac{1}{2}$%
-BPS attractor flow solution (\ref{BPS-sol}) of the $stu$ model can be
\textit{axion-free} only for the configurations $D0-D6$, $D0-D4$ (\textit{%
magnetic}) and $D2-D6$ (\textit{electric}).

As found in \cite{Ferrara-Gimon} and observed also in \cite{GLS-1}, an
immediate consequence of Eq. (\ref{BPS-sol}) is that $\Gamma _{\infty }$
satisfies the $\frac{1}{2}$\textit{-BPS Attractor Eqs. }\cite{BPS-flow-1}.
This determines a sort of \textit{``Attractor Mechanism at spatial
infinity'',} mapping the $6$ \textit{real} moduli $\left(
x^{1},x^{2},x^{3},y^{1},y^{2},y^{3}\right) $ into the $8$ \textit{real}
constants $\left( p_{\infty }^{1},p_{\infty }^{2},p_{\infty
}^{3},q_{1,\infty },q_{2,\infty },q_{3,\infty }\right) $, arranged as $%
\Gamma _{\infty }$ and constrained by the $2$ \textit{real} conditions (\ref
{constraints}).

As noticed in \cite{GLS-1}, the absence of \textit{flat} directions in the $%
\frac{1}{2}$-BPS attractor flow (which is a general feature of $\mathcal{N}%
=2 $, $d=4$ \textit{ungauged} supergravity coupled to Abelian vector
multiplets, \textit{at least} as far as the metric of the scalar manifold is
strictly positive definite $\forall \tau \in \mathbb{R}^{-}$ \cite{FGK}) is
crucial for the validity of the expression (\ref{BPS-sol}).\medskip

Now, by exploiting the \textit{first order formalism} \cite{Fake-Refs} for $%
d=4$ extremal BHs \cite{Cer-Dal-1,ADOT-1} (see also \cite{FHM} and \cite
{Gnecchi-1}), one can compute the relevant BH parameters of the $\frac{1}{2}$%
-BPS attractor flow of the $stu$ model starting from the expression of the $%
\frac{1}{2}$-BPS \textit{fake superpotential} $\mathcal{W}_{\frac{1}{2}-BPS}$%
. For instance, the \textit{ADM mass} and \textit{covariant scalar charges}
respectively read (see \textit{e.g.} the treatments in \cite{FHM} and \cite
{Gnecchi-1}):
\begin{eqnarray}
M_{ADM}\left( z_{\infty },\overline{z}_{\infty },\Gamma \right) &=&\mathcal{W%
}\left( z_{\infty },\overline{z}_{\infty },\Gamma \right) \equiv lim_{\tau
\rightarrow 0^{-}}\mathcal{W}\left( z\left( \tau \right) ,\overline{z}\left(
\tau \right) ,\Gamma \right) ;  \label{d-1} \\
&&  \notag \\
\Sigma _{i}\left( z_{\infty },\overline{z}_{\infty },\Gamma \right)
&=&\left( \partial _{i}\mathcal{W}\right) \left( z_{\infty },\overline{z}%
_{\infty },\Gamma \right) \equiv lim_{\tau \rightarrow 0^{-}}\left( \partial
_{i}\mathcal{W}\right) \left( z\left( \tau \right) ,\overline{z}\left( \tau
\right) ,\Gamma \right) ,  \label{d-2}
\end{eqnarray}
where the subscript \textit{``}$\infty $'' denotes the evaluation at the
moduli at spatial infinity ($r\rightarrow \infty \Leftrightarrow \tau
\rightarrow 0^{-}$). Notice that Eq. (\ref{d-1}) provides, within the
considered \textit{first order formalism}, an alternative (eventually
simpler) formula for the computation of $M_{ADM}$, with respect to the
general definition in terms of the \textit{warp factor} $U$ (see \textit{e.g.%
} \cite{FGK}):
\begin{equation}
M_{ADM}=lim_{\tau \rightarrow 0^{-}}\frac{dU\left( \tau \right) }{d\tau }.
\end{equation}

Recalling that for all $\mathcal{N}=2$, $d=4$ \textit{ungauged}
supergravities it holds that $\mathcal{W}_{\frac{1}{2}-BPS}=\left| Z\right| $%
, Eqs. (\ref{g-1}) and (\ref{d-1}) yield the following expressions of the
\textit{ADM mass} of the $\frac{1}{2}$-BPS attractor flow of the $stu$
model:
\begin{eqnarray}
M_{ADM,\frac{1}{2}-BPS}\left( z_{\infty },\overline{z}_{\infty },\Gamma
\right) &\equiv &lim_{\tau \rightarrow 0^{-}}\left| Z\right| \left( z\left(
\tau \right) ,\overline{z}\left( \tau \right) ,\Gamma \right) =  \notag \\
&=&\frac{\left| q_{0}+q_{1}s_{\infty }+q_{2}t_{\infty }+q_{3}u_{\infty
}+p^{0}s_{\infty }t_{\infty }u_{\infty }-p^{1}t_{\infty }u_{\infty
}-p^{2}s_{\infty }u_{\infty }-p^{3}s_{\infty }t_{\infty }\right| }{\sqrt{%
-i(s_{\infty }-\overline{s}_{\infty })(t_{\infty }-\overline{t}_{\infty
})(u_{\infty }-\overline{u}_{\infty })}}\,.  \notag \\
&&  \label{M-BPS-gen}
\end{eqnarray}
Eq. (\ref{M-BPS-gen}) yields that the \textit{marginal bound} \cite
{Marginal-Refs} is not \textit{saturated} by $\frac{1}{2}$-BPS states,
because $M_{ADM,\frac{1}{2}-BPS}$ is \textit{not} equal to the sum of the
\textit{ADM masses} of four $D6$-branes with appropriate fluxes (for further
detail, see the discussion in \cite{GLS-1}).

Concerning the \textit{(covariant) scalar charges} of the $\frac{1}{2}$-BPS
attractor flow of the $stu$ model, they can be straightforwardly computed by
using Eqs. (\ref{g-1}) and (\ref{d-2}):
\begin{gather}  \label{Sigma-BPS-s}
\Sigma _{s,\frac{1}{2}-BPS}\left( z_{\infty },\overline{z}_{\infty },\Gamma
\right) \equiv lim_{\tau \rightarrow 0^{-}}\left( \partial _{s}\left|
Z\right| \right) \left( z\left( \tau \right) ,\overline{z}\left( \tau
\right) ,\Gamma \right) =lim_{\tau \rightarrow 0^{-}}\frac{\left[ \left(
\partial _{s}Z\right) \overline{Z}+Z\partial _{s}\overline{Z}\right] }{%
2\left| Z\right| }\left( z\left( \tau \right) ,\overline{z}\left( \tau
\right) ,\Gamma \right) =  \notag \\
=lim_{\tau \rightarrow 0^{-}}\frac{e^{K/2}}{2}\left[\left( \partial
_{s}K\right) \left| W\right| +\left( \partial _{s}W\right) \sqrt{\frac{%
\overline{W}}{W}}\right] \left( z\left( \tau \right) ,\overline{z}\left(
\tau \right) ,\Gamma \right) =  \notag \\
=\frac{1}{2\sqrt{-i(s_{\infty }-\overline{s}_{\infty })(t_{\infty }-%
\overline{t}_{\infty })(u_{\infty }-\overline{u}_{\infty })}}\cdot  \notag \\
\cdot \left[ \frac{\left| q_{0}+q_{1}s_{\infty }+q_{2}t_{\infty
}+q_{3}u_{\infty }+p^{0}s_{\infty }t_{\infty }u_{\infty }-p^{1}t_{\infty
}u_{\infty }-p^{2}s_{\infty }u_{\infty }-p^{3}s_{\infty }t_{\infty }\right|
}{(s_{\infty }-\overline{s}_{\infty })}\right. +  \notag \\
\notag \\
+\left( q_{1}+p^{0}t_{\infty }u_{\infty }-p^{2}u_{\infty }-p^{3}t\,_{\infty
}\right) \cdot  \notag \\
\left. \cdot \sqrt{\frac{q_{0}+q_{1}\overline{s}_{\infty }+q_{2}\overline{t}%
_{\infty }+q_{3}\overline{u}_{\infty }+p^{0}\overline{s}_{\infty }\overline{t%
}_{\infty }\overline{u}_{\infty }-p^{1}\overline{t}_{\infty }\overline{u}%
_{\infty }-p^{2}\overline{s}_{\infty }\overline{u}_{\infty }-p^{3}\overline{s%
}_{\infty }\overline{t}_{\infty }\,}{q_{0}+q_{1}s_{\infty }+q_{2}t_{\infty
}+q_{3}u_{\infty }+p^{0}s_{\infty }t_{\infty }u_{\infty }-p^{1}t_{\infty
}u_{\infty }-p^{2}s_{\infty }u_{\infty }-p^{3}s_{\infty }t_{\infty }}}\right]
.  \notag \\
\end{gather}
\begin{gather}  \label{Sigma-BPS-t}
\Sigma _{t,\frac{1}{2}-BPS}\left( z_{\infty },\overline{z}_{\infty },\Gamma
\right) \equiv lim_{\tau \rightarrow 0^{-}}\left( \partial _{t}\left|
Z\right| \right) \left( z\left( \tau \right) ,\overline{z}\left( \tau
\right) ,\Gamma \right) =lim_{\tau \rightarrow 0^{-}}\frac{\left[ \left(
\partial _{t}Z\right) \overline{Z}+Z\partial _{t}\overline{Z}\right] }{%
2\left| Z\right| }\left( z\left( \tau \right) ,\overline{z}\left( \tau
\right) ,\Gamma \right) =  \notag \\
=lim_{\tau \rightarrow 0^{-}}\frac{e^{K/2}}{2}\left[\left( \partial
_{t}K\right) \left| W\right| +\left( \partial _{t}W\right) \sqrt{\frac{%
\overline{W}}{W}}\right] \left( z\left( \tau \right) ,\overline{z}\left(
\tau \right) ,\Gamma \right) =  \notag \\
=\frac{1}{2\sqrt{-i(s_{\infty }-\overline{s}_{\infty })(t_{\infty }-%
\overline{t}_{\infty })(u_{\infty }-\overline{u}_{\infty })}}\cdot  \notag \\
\cdot \left[ \frac{\left| q_{0}+q_{1}s_{\infty }+q_{2}t_{\infty
}+q_{3}u_{\infty }+p^{0}s_{\infty }t_{\infty }u_{\infty }-p^{1}t_{\infty
}u_{\infty }-p^{2}s_{\infty }u_{\infty }-p^{3}s_{\infty }t_{\infty }\right|
}{(t_{\infty }-\overline{t}_{\infty })}\right. +  \notag \\
\notag \\
+\left( q_{2}+p^{0}s_{\infty }u_{\infty }-p^{1}u_{\infty }-p^{3}s\,_{\infty
}\right) \cdot  \notag \\
\left. \cdot \sqrt{\frac{q_{0}+q_{1}\overline{s}_{\infty }+q_{2}\overline{t}%
_{\infty }+q_{3}\overline{u}_{\infty }+p^{0}\overline{s}_{\infty }\overline{t%
}_{\infty }\overline{u}_{\infty }-p^{1}\overline{t}_{\infty }\overline{u}%
_{\infty }-p^{2}\overline{s}_{\infty }\overline{u}_{\infty }-p^{3}\overline{s%
}_{\infty }\overline{t}_{\infty }\,}{q_{0}+q_{1}s_{\infty }+q_{2}t_{\infty
}+q_{3}u_{\infty }+p^{0}s_{\infty }t_{\infty }u_{\infty }-p^{1}t_{\infty
}u_{\infty }-p^{2}s_{\infty }u_{\infty }-p^{3}s_{\infty }t_{\infty }}}\right]
.  \notag \\
\end{gather}
\begin{gather}  \label{Sigma-BPS-u}
\Sigma _{u,\frac{1}{2}-BPS}\left( z_{\infty },\overline{z}_{\infty },\Gamma
\right) \equiv lim_{\tau \rightarrow 0^{-}}\left( \partial _{u}\left|
Z\right| \right) \left( z\left( \tau \right) ,\overline{z}\left( \tau
\right) ,\Gamma \right) =lim_{\tau \rightarrow 0^{-}}\frac{\left[ \left(
\partial _{u}Z\right) \overline{Z}+Z\partial _{u}\overline{Z}\right] }{%
2\left| Z\right| }\left( z\left( \tau \right) ,\overline{z}\left( \tau
\right) ,\Gamma \right) =  \notag \\
=lim_{\tau \rightarrow 0^{-}}\frac{e^{K/2}}{2}\left[ \left( \partial
_{u}K\right) \left| W\right| +\left( \partial _{u}W\right) \sqrt{\frac{%
\overline{W}}{W}}\right] \left( z\left( \tau \right) ,\overline{z}\left(
\tau \right) ,\Gamma \right) =  \notag \\
=\frac{1}{2\sqrt{-i(s_{\infty }-\overline{s}_{\infty })(t_{\infty }-%
\overline{t}_{\infty })(u_{\infty }-\overline{u}_{\infty })}}\cdot  \notag \\
\cdot \left[ \frac{\left| q_{0}+q_{1}s_{\infty }+q_{2}t_{\infty
}+q_{3}u_{\infty }+p^{0}s_{\infty }t_{\infty }u_{\infty }-p^{1}t_{\infty
}u_{\infty }-p^{2}s_{\infty }u_{\infty }-p^{3}s_{\infty }t_{\infty }\right|
}{(u_{\infty }-\overline{u}_{\infty })}\right. +  \notag \\
\notag \\
+\left( q_{3}+p^{0}s_{\infty }t_{\infty }-p^{1}t_{\infty }-p^{2}s_{\infty
}\right) \cdot  \notag \\
\left. \cdot \sqrt{\frac{q_{0}+q_{1}\overline{s}_{\infty }+q_{2}\overline{t}%
_{\infty }+q_{3}\overline{u}_{\infty }+p^{0}\overline{s}_{\infty }\overline{t%
}_{\infty }\overline{u}_{\infty }-p^{1}\overline{t}_{\infty }\overline{u}%
_{\infty }-p^{2}\overline{s}_{\infty }\overline{u}_{\infty }-p^{3}\overline{s%
}_{\infty }\overline{t}_{\infty }\,}{q_{0}+q_{1}s_{\infty }+q_{2}t_{\infty
}+q_{3}u_{\infty }+p^{0}s_{\infty }t_{\infty }u_{\infty }-p^{1}t_{\infty
}u_{\infty }-p^{2}s_{\infty }u_{\infty }-p^{3}s_{\infty }t_{\infty }}}\right]
.  \notag \\
\end{gather}

\section{\label{Non-BPS-Z=0-Flow}The Most General Non-BPS $Z=0$ Attractor
Flow}

Let us now investigate the non-BPS $Z=0$ case.

As shortly noticed in \cite{GLS-1}, in spite of the fact that this attractor
flow is non-supersymmetric, it has many common features with the
supersymmetric ($\frac{1}{2}$-BPS) case.

As yielded by the analysis of \cite{BMOS-1}, the non-BPS $Z=0$ horizon
attractor solutions can be obtained from $\frac{1}{2}$-BPS ones simply by
changing the signs of any two imaginary parts of the moduli (\textit{dilatons%
}) and consistently imposing specific constraints on BH charges. For
example, one can choose (without any loss of generality, due to \textit{%
triality symmetry}) to flip the dilatons as follows:
\begin{equation}
y^{1}\rightarrow y^{1},\;y^{2}\rightarrow -y^{2},\;y^{3}\rightarrow -y^{3}.
\label{Wed-6}
\end{equation}
This yields the following constraints on the BH charge configurations
supporting the non-BPS $Z=0$ attractors at the BH event horizon ($\tau
\rightarrow -\infty $) \cite{BMOS-1}:
\begin{equation}
\begin{array}{l}
\mathcal{I}_{4}\left( \Gamma \right) >0; \\
\\
p^{2}p^{3}-p^{0}q_{1}\lessgtr 0; \\
\\
p^{1}p^{3}-p^{0}q_{2}\gtrless 0; \\
\\
p^{1}p^{2}-p^{0}q_{3}\gtrless 0.
\end{array}
\label{Wed-3}
\end{equation}
The constraints (\ref{Wed-3}) defines the non-BPS $Z=0$ orbit of the \textit{%
tri-fundamental} representation $\left( \mathbf{2},\mathbf{2},\mathbf{2}%
\right) $ of the $U$-duality group $\left( SU\left( 1,1\right) \right) ^{3}$
(see Appendix II of \cite{bellucci1})
\begin{equation}
\mathcal{O}_{non-BPS,Z=0}=\frac{\left( SU\left( 1,1\right) \right) ^{3}}{%
\left( U\left( 1\right) \right) ^{2}}.  \label{Wed-4}
\end{equation}
Notice that such an orbit shares the same coset expression of $\mathcal{O}_{%
\frac{1}{2}-BPS}$ given by Eq. (\ref{Wed-5}). However, they do \textit{not}
coincide, but instead they are two \textit{separated} branches of a \textit{%
disconnected} manifold, classified by the local value of the function $%
sgn\left( \left| Z\right| ^{2}-\left| D_{s}Z\right| ^{2}\right) $ (see
Appendix II of \cite{bellucci1}; $D_{s}Z$ is defined below Eq. (\ref{Wt-geom}%
)).

The same holds \textit{all along the attractor flow}, \textit{i.e.} $\forall
\tau \in \mathbb{R}^{-}$. Indeed, the most general non-BPS $Z=0$ attractor
flow can be obtained by taking the most general $\frac{1}{2}$-BPS attractor
flow, and flipping any two out of the three dilatons. Thus, by taking Eq. (%
\ref{BPS-sol}) and flipping the dilatons as given by Eq. (\ref{Wed-6}), one
achieves the following result:
\begin{eqnarray}
exp\left[ -4U_{non-BPS,Z=0}\left( \tau \right) \right] &=&\mathcal{I}%
_{4}\left( \mathcal{H}\left( \tau \right) \right) ;  \notag \\
&&  \notag \\
z_{non-BPS,Z=0}^{1}(\tau ) &=&\frac{H^{\Lambda }(\tau )H_{\Lambda }(\tau
)-2H^{1}(\tau )H_{1}(\tau )-i\mathcal{I}_{4}^{1/2}(\mathcal{H}(\tau ))}{2%
\left[ H^{2}(\tau )H^{3}(\tau )-H^{0}(\tau )H_{1}(\tau )\right] }=z_{\frac{1%
}{2}-BPS}^{1}(\tau );  \notag \\
&&  \notag \\
z_{non-BPS,Z=0}^{2}(\tau ) &=&\frac{H^{\Lambda }(\tau )H_{\Lambda }(\tau
)-2H^{2}(\tau )H_{2}(\tau )+i\mathcal{I}_{4}^{1/2}(\mathcal{H}(\tau ))}{2%
\left[ H^{1}(\tau )H^{3}(\tau )-H^{0}(\tau )H_{2}(\tau )\right] }=\overline{%
z_{\frac{1}{2}-BPS}^{2}}(\tau );  \notag \\
&&  \notag \\
z_{non-BPS,Z=0}^{3}(\tau ) &=&\frac{H^{\Lambda }(\tau )H_{\Lambda }(\tau
)-2H^{3}(\tau )H_{3}(\tau )+i\mathcal{I}_{4}^{1/2}(\mathcal{H}(\tau ))}{2%
\left[ H^{2}(\tau )H^{1}(\tau )-H^{0}(\tau )H_{3}(\tau )\right] }=\overline{%
z_{\frac{1}{2}-BPS}^{3}}(\tau ).  \label{non-BPS-Z=0-flow}
\end{eqnarray}
This is the most general expression of the non-BPS $Z=0$ attractor flow, in
the \textit{``polarization'' }given by Eq. (\ref{Wed-6}), which, due to the
underlying \textit{triality symmetry} of the $stu$ model, does \textit{not}
imply any loss of generality.

Consistently with the constraints (\ref{Wed-3}), $\mathcal{H}\left( \tau
\right) $ is constrained as follows along the non-BPS $Z=0$ attractor flow ($%
\forall \tau \in \mathbb{R}^{-}$):
\begin{equation}
\begin{array}{l}
\mathcal{I}_{4}\left( \mathcal{H}\left( \tau \right) \right) >0; \\
\\
H^{2}\left( \tau \right) H^{3}\left( \tau \right) -H^{0}\left( \tau \right)
H_{1}\left( \tau \right) \lessgtr 0; \\
\\
H^{1}\left( \tau \right) H^{3}\left( \tau \right) -H^{0}\left( \tau \right)
H_{2}\left( \tau \right) \gtrless 0; \\
\\
H^{1}\left( \tau \right) H^{2}\left( \tau \right) -H^{0}\left( \tau \right)
H_{3}\left( \tau \right) \gtrless 0.
\end{array}
\label{Wed-7}
\end{equation}

In the \textit{near-horizon limit} $\tau \rightarrow -\infty $, Eq. (\ref
{non-BPS-Z=0-flow}) yields the \textit{purely charge-dependent}, \textit{%
critical} expressions of the scalars at the BH event horizon, given by Eq.
(4.9) of \cite{BMOS-1}. In the same limit, the constraints (\ref{Wed-7})
consistently yield the contraints (\ref{Wed-3}). The \textit{integrability
conditions} (\ref{constraints}) clearly hold also in this case.

Consistently with the analysis of \cite{Ceresole}, the general non-BPS $Z=0$
attractor flow solution (\ref{non-BPS-Z=0-flow}) of the $stu$ model can be
\textit{axion-free} only for the configurations $D0-D6$, $D0-D4$ (\textit{%
magnetic}) and $D2-D6$ (\textit{electric}).

A consequence of Eq. (\ref{non-BPS-Z=0-flow}) is that $\Gamma _{\infty }$
satisfies the \textit{non-BPS }$\mathit{Z=0}$ \textit{Attractor Eqs. }(see
\textit{e.g.} \cite{AoB-book} and \cite{BFMY}). Analogously to what happens
for the $\frac{1}{2}$-BPS attractor flow, this determines a sort of \textit{%
``Attractor Mechanism at spatial infinity''}.

Analogously to what happens in the $\frac{1}{2}$-BPS case, the absence of
\textit{flat} directions in the non-BPS $Z=0$ attractor flow (which is
\textit{not} a general feature of $\mathcal{N}=2$, $d=4$ \textit{ungauged}
supergravity coupled to Abelian vector multiplets, but however holds for the
$stu$ model \cite{Ferrara-Marrani-1,ferrara4}) is crucial for the validity
of the expression (\ref{non-BPS-Z=0-flow}).

By exploiting the strict relation with the $\frac{1}{2}$-BPS attractor flow,
one can also determine the explicit expression of the \textit{fake
superpotential }$\mathcal{W}_{non-BPS,Z=0}$ for the non-BPS $Z=0$ attractor
flow. Considering the absolute value of the $\mathcal{N}=2$, $d=4$ \textit{%
central charge function} $Z$ given by Eq. (\ref{g-1}) and flipping two
dilatons out of three in the \textit{``polarization'' }given by Eq. (\ref
{Wed-6}), one obtains the following non-BPS $Z=0$ \textit{fake superpotential%
} (notice that $K$, as given by the first Eq. of (\ref{stu-geometry}), is
invariant under such a flipping):
\begin{eqnarray}
\mathcal{W}_{non-BPS,Z=0,s} &=&e^{K/2}\left| q_{0}+q_{1}s+q_{2}\overline{t}%
+q_{3}\overline{u}+p^{0}s\overline{t}\overline{u}-p^{1}\overline{t}\overline{%
u}-p^{2}s\overline{u}-p^{3}s\overline{t}\right| =  \notag \\
&=&\left| Z\left( s,\overline{t},\overline{u}\right) \right| =\mathcal{W}_{%
\frac{1}{2}-BPS}\left( s,\overline{t},\overline{u}\right) ,  \label{Ws}
\end{eqnarray}
where the subscript \textit{``}$s$\textit{'' }denotes the modulus untouched
by the considered flipping of dilatons; in the last step we used that
aforementioned fact that for all $\mathcal{N}=2$, $d=4$ \textit{ungauged}
supergravities it holds that $\mathcal{W}_{\frac{1}{2}-BPS}=\left| Z\right| $%
.

Clearly, the flipping (\ref{Wed-6}) is not the only possible one. By
triality symmetry, two other equivalent flippings exist, namely
\begin{eqnarray}
y^{1} &\rightarrow &-y^{1},\;y^{2}\rightarrow y^{2},\;y^{3}\rightarrow
-y^{3};  \label{Wed-6-bis} \\
&&  \notag \\
y^{1} &\rightarrow &-y^{1},\;y^{2}\rightarrow -y^{2},\;y^{3}\rightarrow
y^{3},  \label{Wed-6-tris}
\end{eqnarray}
obtained by cyclic permutations from (\ref{Wed-6}). Such equivalent
flippings respectively determine the following (respectively \textit{``}$t$%
\textit{-polarized''} and \textit{``}$u$\textit{-polarized''}) non-BPS $Z=0$
\textit{fake superpotentials}, completely equivalent to the \textit{``}$s$%
\textit{-polarized''} one given by Eq. (\ref{Ws}):
\begin{eqnarray}
\mathcal{W}_{non-BPS,Z=0,t} &=&e^{K/2}\left| q_{0}+q_{1}\overline{s}%
+q_{2}t+q_{3}\overline{u}+p^{0}\overline{s}t\overline{u}-p^{1}t\overline{u}%
-p^{2}\overline{s}\overline{u}-p^{3}\overline{s}t\right| =  \notag \\
&=&\left| Z\left( \overline{s},t,\overline{u}\right) \right| =\mathcal{W}_{%
\frac{1}{2}-BPS}\left( \overline{s},t,\overline{u}\right) ;  \label{Wt} \\
&&  \notag \\
\mathcal{W}_{non-BPS,Z=0,u} &=&e^{K/2}\left| q_{0}+q_{1}\overline{s}+q_{2}%
\overline{t}+q_{3}u+p^{0}\overline{s}\overline{t}u-p^{1}\overline{t}u-p^{2}%
\overline{s}u-p^{3}\overline{s}\overline{t}\right| =  \notag \\
&=&\left| Z\left( \overline{s},\overline{t},u\right) \right| =\mathcal{W}_{%
\frac{1}{2}-BPS}\left( \overline{s},\overline{t},u\right) .  \label{Wu}
\end{eqnarray}
It can be shown by straightforward computations that the \textit{real},
moduli- and charge- dependent functions given by Eqs. (\ref{Ws}), (\ref{Wt})
and (\ref{Wu}) do satisfy all the conditions defining a \textit{first order
fake superpotential} (see the treatment in \cite{Cer-Dal-1,ADOT-1}, recently
reviewed in \cite{FHM,Gnecchi-1}), and thus they respectively are an \textit{%
``}$s$\textit{-polarized''}, \textit{``}$t$\textit{-polarized''} and \textit{%
``}$u$\textit{-polarized''} non-BPS $Z=0$ \textit{fake superpotential}.

Eqs. (\ref{Ws}), (\ref{Wt}) and (\ref{Wu}) can also be rewritten
respectively as follows:
\begin{eqnarray}
\mathcal{W}_{non-BPS,Z=0,s} &=&\sqrt{g^{1\overline{1}}\left( D_{s}Z\right)
\overline{D}_{\overline{s}}\overline{Z}};  \label{Ws-geom} \\
\mathcal{W}_{non-BPS,Z=0,t} &=&\sqrt{g^{2\overline{2}}\left( D_{t}Z\right)
\overline{D}_{\overline{t}}\overline{Z}};  \label{Wu-geom} \\
\mathcal{W}_{non-BPS,Z=0,u} &=&\sqrt{g^{3\overline{3}}\left( D_{u}Z\right)
\overline{D}_{\overline{u}}\overline{Z}},  \label{Wt-geom}
\end{eqnarray}
where $D_{s}Z\equiv \left[ \partial _{s}+\frac{1}{2}\left( \partial
_{s}K\right) \right] Z$ is the covariant derivarive of $Z$ along the
direction $s$ (and analogously for the moduli $t$ and $u$).

Due to the complete factorization of the manifold $\left( \frac{SU(1,1)}{%
U\left( 1\right) }\right) ^{3}$ (determining the diagonality of the metric $%
g^{i\overline{j}}$, given by the second Eq. of (\ref{stu-geometry})), Eqs. (%
\ref{Ws-geom}), (\ref{Wt-geom}) and (\ref{Wu-geom}) are \textit{manifestly} $%
H$\textit{-invariant}.\textbf{\ }This result is consistent with the $H$%
-invariance imposed by the relation\footnote{%
Actually, in the treatment given in the present paper, Eq. (\ref{cc-1}) has
been crucial in order to guess (and thus check) the analytical form of the
\textit{fake superpotential }$\mathcal{W}\left( z\left( \tau \right) ,%
\overline{z}\left( \tau \right) ,\Gamma \right) $, by knowing the
analyitical solution for the \textit{warp factor} $U\left( \tau ,\Gamma
\right) $ relevant to the considered charge configuration.} (given by Eq.
(2.21) of \cite{Cer-Dal-1}, as well as by Eq. (13) of \cite{ADOT-1}, and
reported here for completeness' sake)
\begin{equation}
\frac{dU\left( \tau ,\Gamma \right) }{d\tau }=e^{U\left( \tau \right)
}W\left( z\left( \tau \right) ,\overline{z}\left( \tau \right) ,\Gamma
\right)  \label{cc-1}
\end{equation}
between $\mathcal{W}$ and the \textit{warp factor} $U\left( \tau \right) $
appearing in the \textit{Ansatz} for the static, spherically symmetric,
asymptotically flat, \textit{extremal} dyonic BH metric:
\begin{equation}
ds^{2}=-e^{2U\left( \tau \right) }dt^{2}+e^{-2U\left( \tau \right) }d%
\overrightarrow{x}^{2}.  \label{c-1}
\end{equation}

Under the \textit{integrability conditions} (\ref{constraints}), the flow (%
\ref{non-BPS-Z=0-flow}) is the most general solution (within the \textit{``}$%
s$\textit{-polarization''} defined by Eq. (\ref{Wed-6})) of the ``non-BPS $%
Z=0$ analogue'' of the $\frac{1}{2}$\textit{-BPS stabilization Eqs. }(\ref
{BPS-stabilization-Eqs.}), namely of:
\begin{eqnarray}
\mathcal{H}^{T}\left( \tau \right) &=&2e^{K\left( z\left( \tau \right) ,%
\overline{z}\left( \tau \right) \right) }Im\left[ g^{i\overline{j}}\left(
z\left( \tau \right) ,\overline{z}\left( \tau \right) \right) \left(
\begin{array}{c}
\left( D_{i}X^{\Lambda }\right) \left( z\left( \tau \right) ,\overline{z}%
\left( \tau \right) \right) \\
\\
\left( D_{i}F_{\Lambda }\right) \left( z\left( \tau \right) ,\overline{z}%
\left( \tau \right) \right)
\end{array}
\right) \left( \overline{\partial }_{\overline{j}}\overline{W}\right) \left(
\overline{z}\left( \tau \right) ,\mathcal{H}\left( \tau \right) \right)
\right] ,  \notag \\
&&  \label{non-BPS-Z=0-stabilization-Eqs.}
\end{eqnarray}
which we can refer to as the \textit{non-BPS }$\mathit{Z=0}$\textit{\
stabilization Equations}. It is easy realized that they can be obtained from
the \textit{non-BPS }$\mathit{Z=0}$\textit{\ Attractor Eqs.} (see the
treatment in \cite{AoB-book})
\begin{equation}
\Gamma ^{T}=2e^{K\left( z,\overline{z}\right) }Im\left[ g^{i\overline{j}%
}\left( z,\overline{z}\right) \left(
\begin{array}{c}
\left( D_{i}X^{\Lambda }\right) \left( z,\overline{z}\right) \\
\\
\left( D_{i}F_{\Lambda }\right) \left( z,\overline{z}\right)
\end{array}
\right) \left( \overline{\partial }_{\overline{j}}\overline{W}\right) \left(
\overline{z},\Gamma \right) \right]  \label{non-BPS-Z=0-AEs}
\end{equation}
by simply replacing $\Gamma $ with $\mathcal{H}\left( \tau \right) $.
Consistently, Eq. (\ref{non-BPS-Z=0-AEs}) is the \textit{near-horizon} ($%
\tau \rightarrow -\infty $) \textit{limit }of Eq. (\ref
{non-BPS-Z=0-stabilization-Eqs.}).

A remarkable consequence of the \textit{first order formalism} for the
non-BPS $Z=0$ attractor flow is that Eqs. (\ref
{non-BPS-Z=0-stabilization-Eqs.}) and (\ref{non-BPS-Z=0-AEs}) can actually
be recast in the following \textit{(}$\frac{1}{2}$\textit{-)BPS-like} forms
in terms of the non-BPS $Z=0$ \textit{fake superpotential}(\textit{s}),
respectively:
\begin{eqnarray}
\mathcal{H}^{T}\left( \tau \right) &=&2e^{K\left( z\left( \tau \right) ,%
\overline{z}\left( \tau \right) \right) }Im\left[ \frak{W}%
_{non-BPS,Z=0}\left( z\left( \tau \right) ,\overline{z}\left( \tau \right) ,%
\mathcal{H}\left( \tau \right) \right) \left(
\begin{array}{c}
\overline{X}_{f}^{\Lambda }\left( z\left( \tau \right) ,\overline{z}\left(
\tau \right) \right) \\
\\
\overline{F}_{\Lambda ,f}\left( z\left( \tau \right) ,\overline{z}\left(
\tau \right) \right)
\end{array}
\right) \right] ;  \notag \\
&&  \label{non-BPS-Z=0-BPS} \\
\Gamma ^{T} &=&2e^{K\left( z,\overline{z}\right) }Im\left[ \frak{W}%
_{non-BPS,Z=0}\left( z,\overline{z},\Gamma \right) \left(
\begin{array}{c}
\overline{X}_{f}^{\Lambda }\left( z,\overline{z}\right) \\
\\
\overline{F}_{\Lambda ,f}\left( z,\overline{z}\right)
\end{array}
\right) \right] ,  \label{non-BPS-Z=0-BPS-2}
\end{eqnarray}
where
\begin{eqnarray}
\frak{W}_{non-BPS,Z=0}\left( z,\overline{z},\Gamma \right) &=&Z\left( s,%
\overline{t},\overline{u},\Gamma \right) ~\text{in the }\mathit{``}s\mathit{%
-polarization"~}\text{(Eq. \ref{Wed-6})};  \label{bossa-1} \\
\frak{W}_{non-BPS,Z=0}\left( z,\overline{z},\Gamma \right) &=&Z\left(
\overline{s},t,\overline{u},\Gamma \right) ~\text{in the }\mathit{``}t%
\mathit{-polarization"~}\text{(Eq. \ref{Wed-6-bis})};  \label{bossa-2} \\
\frak{W}_{non-BPS,Z=0}\left( z,\overline{z},\Gamma \right) &=&Z\left(
\overline{s},\overline{t},u,\Gamma \right) ~\text{in the }\mathit{``}u%
\mathit{-polarization"}\text{~(Eq. \ref{Wed-6-tris})},  \label{bossa-3}
\end{eqnarray}
such that in general
\begin{equation}
\mathcal{W}_{non-BPS,Z=0}=\left| \frak{W}_{non-BPS,Z=0}\right| .
\end{equation}
The subscript \textit{``}$f$\textit{''} in Eqs. (\ref{non-BPS-Z=0-BPS}) and (%
\ref{non-BPS-Z=0-BPS-2}) indicates that a flipping of the dilatons has been
performed ((Eqs. (\ref{Wed-6}), (\ref{Wed-6-bis}) and (\ref{Wed-6-tris}),
respectively for the choices (\ref{bossa-1}), (\ref{bossa-2}) and (\ref
{bossa-3})). Notice that such a flipping destroys the holomorphicity of $%
X^{\Lambda }$ and $F_{\Lambda }$ in the moduli.

Eq. (\ref{non-BPS-Z=0-flow}) is the most general solution of Eq. (\ref
{non-BPS-Z=0-BPS}) with the choice (\ref{bossa-1}), and equivalent
expressions for the most general non-BPS $Z=0$ attractor flow can be
obtained by solving Eq. (\ref{non-BPS-Z=0-BPS}) with the choice (\ref
{bossa-2}) or (\ref{bossa-3}).

Such a \textit{(}$\frac{1}{2}$\textit{-)BPS-like} reformulation of the
\textit{non-BPS }$\mathit{Z=0}$\textit{\ stabilization Equations }(\ref
{non-BPS-Z=0-stabilization-Eqs.}) and of their \textit{near-horizon} ($\tau
\rightarrow -\infty $) \textit{limit} given by the \textit{non-BPS }$\mathit{%
Z=0}$\textit{\ Attractor Eqs. }(\ref{non-BPS-Z=0-AEs}) is possible due to
the strict similarity between the most general $\frac{1}{2}$-BPS and non-BPS
$Z=0$ attractor flows in the considered $stu$ model, which actually are
related through a flipping of two dilatons out of three. This in turn is
related once again to the absence of \textit{flat} directions along such
flows, such that \textit{all} moduli are explicitly determined as functions
of dyonic BH charges (and as functions of $\tau $) \textit{all along the
attractor flow}.\medskip

Now, by exploiting the \textit{first order formalism} \cite{Fake-Refs} for $%
d=4$ extremal BHs \cite{Cer-Dal-1,ADOT-1} (see also \cite{FHM} and \cite
{Gnecchi-1}), one can compute the relevant BH parameters of the non-BPS $Z=0$
attractor flow of the $stu$ model starting from the expression of the
non-BPS $Z=0$ \textit{fake superpotential} $\mathcal{W}_{non-BPS,Z=0}$ given
by Eqs. (\ref{Ws}), (\ref{Wt}) or (\ref{Wu}). The choice of \textit{``}$s$%
\textit{-polarization''}, \textit{``}$t$\textit{-polarization''} or \textit{%
``}$u$\textit{-polarization'' }is immaterial, due to the underlying \textit{%
triality symmetry} of the moduli $s$, $t$ and $u$. Thus, without loss of
generality, we choose to perform computations in the \textit{``}$s$\textit{%
-polarization''} (equivalent results in the other two \textit{%
``polarizations''} can be obtained by cyclic permutations of the moduli).

Eqs. (\ref{Ws}) and (\ref{d-1}) yield the following expressions of the
\textit{ADM mass} of the non-BPS $Z=0$ attractor flow of the $stu$ model:
\begin{eqnarray}
M_{ADM,non-BPS,Z=0}\left( z_{\infty },\overline{z}_{\infty },\Gamma \right)
&\equiv &lim_{\tau \rightarrow 0^{-}}\mathcal{W}_{non-BPS,Z=0,s}\left(
z\left( \tau \right) ,\overline{z}\left( \tau \right) ,\Gamma \right) =
\notag \\
&=&lim_{\tau \rightarrow 0^{-}}\left| Z\left( s\left( \tau \right) ,%
\overline{t}\left( \tau \right) ,\overline{u}\left( \tau \right) \right)
\right| =  \notag \\
&=&\frac{\left| q_{0}+q_{1}s_{\infty }+q_{2}\overline{t}_{\infty }+q_{3}%
\overline{u}_{\infty }+p^{0}s_{\infty }\overline{t}_{\infty }\overline{u}%
_{\infty }-p^{1}\overline{t}_{\infty }\overline{u}_{\infty }-p^{2}s_{\infty }%
\overline{u}_{\infty }-p^{3}s_{\infty }\overline{t}_{\infty }\right| }{\sqrt{%
-i(s_{\infty }-\overline{s}_{\infty })(t_{\infty }-\overline{t}_{\infty
})(u_{\infty }-\overline{u}_{\infty })}}\,.  \notag \\
&&  \label{M-non-BPS-Z=0-gen}
\end{eqnarray}
Eq. (\ref{M-non-BPS-Z=0-gen}) yields that the \textit{marginal bound} \cite
{Marginal-Refs} is not \textit{saturated} by non-BPS $Z=0$ states, because $%
M_{ADM,non-BPS,Z=0}$ is \textit{not} equal to the sum of the \textit{ADM
masses} of four $D6$-branes with appropriate fluxes (for further detail, see
the discussion in \cite{GLS-1}). This is actually expected, due to the
strict similarity, discussed above, between $\frac{1}{2}$-BPS and non -BPS $%
Z=0$ attractor flows in the considered $stu$ model; such a similarity can be
explained by noticing that both such flows can be uplifted to the \textit{%
same} $\frac{1}{8}$-BPS \textit{non-degenerate} attractor flow of $\mathcal{N%
}=8$, $d=4$ supergravity (see \textit{e.g.} the discussion in \cite{BMOS-1}).

Concerning the \textit{(covariant) scalar charges} of the non-BPS $Z=0$
attractor flow of the $stu$ model, they can be straightforwardly computed
(in the \textit{``}$s$\textit{-polarization''}, without loss of generality)
by using Eqs. (\ref{Ws}) and (\ref{d-2}):
\begin{gather}
\Sigma _{s,non-BPS,Z=0}\left( z_{\infty },\overline{z}_{\infty },\Gamma
\right) \equiv lim_{\tau \rightarrow 0^{-}}\left( \partial _{s}\mathcal{W}%
_{non-BPS,Z=0,s}\right) \left( z\left( \tau \right) ,\overline{z}\left( \tau
\right) ,\Gamma \right) =  \notag \\
=lim_{\tau \rightarrow 0^{-}}\partial _{s}\left| Z\left( s\left( \tau
\right) ,\overline{t}\left( \tau \right) ,\overline{u}\left( \tau \right)
\right) \right| =  \notag \\
=lim_{\tau \rightarrow 0^{-}}\frac{e^{K/2}}{2}\left[\left( \partial
_{s}K\right) \left| W\left( s,\overline{t},\overline{u}\right) \right|
+\left( \partial _{s}W\left( s,\overline{t},\overline{u}\right) \right)
\sqrt{\frac{\overline{W}\left( \overline{s},t,u\right) }{W\left( s,\overline{%
t},\overline{u}\right) }}\right] =  \notag \\
=\frac{1}{2\sqrt{-i(s_{\infty }-\overline{s}_{\infty })(t_{\infty }-%
\overline{t}_{\infty })(u_{\infty }-\overline{u}_{\infty })}}\cdot  \notag \\
\cdot \left[ \frac{\left| q_{0}+q_{1}s_{\infty }+q_{2}\overline{t}_{\infty
}+q_{3}\overline{u}_{\infty }+p^{0}s_{\infty }\overline{t}_{\infty }%
\overline{u}_{\infty }-p^{1}\overline{t}_{\infty }\overline{u}_{\infty
}-p^{2}s_{\infty }\overline{u}_{\infty }-p^{3}s_{\infty }\overline{t}%
_{\infty }\right| }{(s_{\infty }-\overline{s}_{\infty })}\right. +  \notag \\
\notag \\
+\left( q_{1}+p^{0}\overline{t}_{\infty }\overline{u}_{\infty }-p^{2}%
\overline{u}_{\infty }-p^{3}\overline{t}\,_{\infty }\right) \cdot  \notag \\
\left. \cdot \sqrt{\frac{q_{0}+q_{1}\overline{s}_{\infty }+q_{2}t_{\infty
}+q_{3}u_{\infty }+p^{0}\overline{s}_{\infty }t_{\infty }u_{\infty
}-p^{1}t_{\infty }u_{\infty }-p^{2}\overline{s}_{\infty }u_{\infty }-p^{3}%
\overline{s}_{\infty }t_{\infty }\,}{q_{0}+q_{1}s_{\infty }+q_{2}\overline{t}%
_{\infty }+q_{3}\overline{u}_{\infty }+p^{0}s_{\infty }\overline{t}_{\infty }%
\overline{u}_{\infty }-p^{1}\overline{t}_{\infty }\overline{u}_{\infty
}-p^{2}s_{\infty }\overline{u}_{\infty }-p^{3}s_{\infty }\overline{t}%
_{\infty }}}\right] =  \notag \\
\notag \\
=\left. \Sigma _{s,\frac{1}{2}-BPS}\right| _{t_{\infty }\rightarrow
\overline{t}_{\infty },u_{\infty }\rightarrow \overline{u}_{\infty }}.
\label{Sigma-non-BPS-Z=0-s}
\end{gather}
\begin{gather}  \label{Sigma-non-BPS-Z=0-t}
\Sigma _{t,non-BPS,Z=0}\left( z_{\infty },\overline{z}_{\infty },\Gamma
\right) \equiv lim_{\tau \rightarrow 0^{-}}\left( \partial _{t}\mathcal{W}%
_{non-BPS,Z=0,s}\right) \left( z\left( \tau \right) ,\overline{z}\left( \tau
\right) ,\Gamma \right) =  \notag \\
=lim_{\tau \rightarrow 0^{-}}\partial _{t}\left| Z\left( s\left( \tau
\right) ,\overline{t}\left( \tau \right) ,\overline{u}\left( \tau \right)
\right) \right| =  \notag \\
=lim_{\tau \rightarrow 0^{-}}\frac{e^{K/2}}{2}\left[ \left( \partial
_{t}K\right) \left| W\left( s,\overline{t},\overline{u}\right) \right|
+\left( \partial _{t}\overline{W}\left( \overline{s},t,u\right) \right)
\sqrt{\frac{W\left( s,\overline{t},\overline{u}\right) }{\overline{W}\left(
\overline{s},t,u\right) }}\right] =  \notag \\
=\frac{1}{2\sqrt{-i(s_{\infty }-\overline{s}_{\infty })(t_{\infty }-%
\overline{t}_{\infty })(u_{\infty }-\overline{u}_{\infty })}}\cdot  \notag \\
\cdot \left[ \frac{\left| q_{0}+q_{1}s_{\infty }+q_{2}\overline{t}_{\infty
}+q_{3}\overline{u}_{\infty }+p^{0}s_{\infty }\overline{t}_{\infty }%
\overline{u}_{\infty }-p^{1}\overline{t}_{\infty }\overline{u}_{\infty
}-p^{2}s_{\infty }\overline{u}_{\infty }-p^{3}s_{\infty }\overline{t}%
_{\infty }\right| }{(t_{\infty }-\overline{t}_{\infty })}\right. +  \notag \\
\notag \\
+\left( q_{2}+p^{0}\overline{s}_{\infty }u_{\infty }-p^{1}u_{\infty }-p^{3}%
\overline{s}_{\infty }\right) \cdot  \notag \\
\left. \cdot \sqrt{\frac{\,q_{0}+q_{1}s_{\infty }+q_{2}\overline{t}_{\infty
}+q_{3}\overline{u}_{\infty }+p^{0}s_{\infty }\overline{t}_{\infty }%
\overline{u}_{\infty }-p^{1}\overline{t}_{\infty }\overline{u}_{\infty
}-p^{2}s_{\infty }\overline{u}_{\infty }-p^{3}s_{\infty }\overline{t}%
_{\infty }}{q_{0}+q_{1}\overline{s}_{\infty }+q_{2}t_{\infty
}+q_{3}u_{\infty }+p^{0}\overline{s}_{\infty }t_{\infty }u_{\infty
}-p^{1}t_{\infty }u_{\infty }-p^{2}\overline{s}_{\infty }u_{\infty }-p^{3}%
\overline{s}_{\infty }t_{\infty }}}\right] .  \notag \\
\end{gather}
\begin{gather}  \label{Sigma-non-BPS-Z=0-u}
\Sigma _{u,non-BPS,Z=0}\left( z_{\infty },\overline{z}_{\infty },\Gamma
\right) \equiv lim_{\tau \rightarrow 0^{-}}\left( \partial _{u}\mathcal{W}%
_{non-BPS,Z=0,s}\right) \left( z\left( \tau \right) ,\overline{z}\left( \tau
\right) ,\Gamma \right) =  \notag \\
=lim_{\tau \rightarrow 0^{-}}\partial _{u}\left| Z\left( s\left( \tau
\right) ,\overline{t}\left( \tau \right) ,\overline{u}\left( \tau \right)
\right) \right| =  \notag \\
=lim_{\tau \rightarrow 0^{-}}\frac{e^{K/2}}{2}\left[ \left( \partial
_{u}K\right) \left| W\left( s,\overline{t},\overline{u}\right) \right|
+\left( \partial _{u}\overline{W}\left( \overline{s},t,u\right) \right)
\sqrt{\frac{W\left( s,\overline{t},\overline{u}\right) }{\overline{W}\left(
\overline{s},t,u\right) }}\right] =  \notag \\
=\frac{1}{2\sqrt{-i(s_{\infty }-\overline{s}_{\infty })(t_{\infty }-%
\overline{t}_{\infty })(u_{\infty }-\overline{u}_{\infty })}}\cdot  \notag \\
\cdot \left[ \frac{\left| q_{0}+q_{1}s_{\infty }+q_{2}\overline{t}_{\infty
}+q_{3}\overline{u}_{\infty }+p^{0}s_{\infty }\overline{t}_{\infty }%
\overline{u}_{\infty }-p^{1}\overline{t}_{\infty }\overline{u}_{\infty
}-p^{2}s_{\infty }\overline{u}_{\infty }-p^{3}s_{\infty }\overline{t}%
_{\infty }\right| }{(u_{\infty }-\overline{u}_{\infty })}\right. +  \notag \\
\notag \\
+\left( q_{3}+p^{0}\overline{s}_{\infty }t_{\infty }-p^{1}t_{\infty }-p^{2}%
\overline{s}_{\infty }\right) \cdot  \notag \\
\left. \cdot \sqrt{\frac{\,q_{0}+q_{1}s_{\infty }+q_{2}\overline{t}_{\infty
}+q_{3}\overline{u}_{\infty }+p^{0}s_{\infty }\overline{t}_{\infty }%
\overline{u}_{\infty }-p^{1}\overline{t}_{\infty }\overline{u}_{\infty
}-p^{2}s_{\infty }\overline{u}_{\infty }-p^{3}s_{\infty }\overline{t}%
_{\infty }}{q_{0}+q_{1}\overline{s}_{\infty }+q_{2}t_{\infty
}+q_{3}u_{\infty }+p^{0}\overline{s}_{\infty }t_{\infty }u_{\infty
}-p^{1}t_{\infty }u_{\infty }-p^{2}\overline{s}_{\infty }u_{\infty }-p^{3}%
\overline{s}_{\infty }t_{\infty }}}\right] .  \notag \\
\end{gather}

Also, it is here worth computing the difference between the squared non-BPS $%
Z=0$ f\textit{ake superpotential} and the squared absolute value of the $%
\mathcal{N}=2$, $d=4$ central charge along the considered non-BPS $Z=0$
attractor flow. This amounts to computing the difference generalizing the
\textit{BPS bound} \cite{gibbons2} to the whole attractor flow (without loss
of generality, we perform calculations in the \textit{``}$s$\textit{%
-polarization''}):
\begin{eqnarray}
&&\Theta \left( \mathcal{X},\mathcal{Y},\Gamma \right) \equiv \mathcal{W}%
_{s,non-BPS,Z=0}^{2}-\left| Z\right| ^{2}=  \notag  \label{gapp} \\
&=&-\frac{1}{8(p^{1}p^{3}-p^{0}q_{2})\mathcal{Y}^{2}}\left[ \left(
p^{\Lambda }q_{\Lambda }-2p^{2}q_{2}-2(p^{1}p^{3}-p^{0}q_{2})\mathcal{X}%
^{2}\right) ^{2}+4(p^{1}p^{3}-p^{0}q_{2})^{2}(\mathcal{Y}^{2})^{2}+\mathcal{I%
}_{4}\right] -  \notag \\
&&-\frac{1}{8(p^{1}p^{2}-p^{0}q_{3})\mathcal{Y}^{2}}\left[ \left( p^{\Lambda
}q_{\Lambda }-2p^{3}q_{3}-2(p^{1}p^{2}-p^{0}q_{3})\mathcal{X}^{3}\right)
^{2}+4(p^{1}p^{2}-p^{0}q_{3})^{2}(\mathcal{Y}^{2})^{2}+\mathcal{I}_{4}\right]
>0.  \notag \\
&&
\end{eqnarray}
Such an expression for the scalar-dependent, strictly positive $\Theta $ was
obtained by using the following relations:
\begin{eqnarray}
(p^{\Lambda }q_{\Lambda }-2p^{2}q_{2})^{2}+\mathcal{I}_{4}
&=&4(p^{1}p^{3}-p^{0}q_{2})(q_{1}q_{3}+p^{2}q_{0});  \notag \\
(p^{\Lambda }q_{\Lambda }-2p^{3}q_{3})^{2}+\mathcal{I}_{4}
&=&4(p^{1}p^{2}-p^{0}q_{3})(q_{1}q_{2}+p^{3}q_{0}).
\end{eqnarray}
Thus, the \textit{BPS bound} \cite{gibbons2} holds not only at the BH event
horizon ($r=r_{H}$), but actually (in a scalar-dependent way) all along the
non-BPS $Z=0$ attractor flow (\textit{i.e.} $\forall r\geqslant r_{H}$).
\medskip

\section{\label{Non-BPS-Z<>0-Flow}The most General Non-BPS $Z\neq 0$
Attractor Flow}

All the features holding for $\frac{1}{2}$-BPS and non-BPS $Z=0$ attractor
flows (respectively treated in Sects. \ref{BPS-Flow} and \ref
{Non-BPS-Z=0-Flow}) do not directly hold for the non-BPS $Z\neq 0$ attractor
flow, which actually turns out to be rather different from (and structurally
much more intricate than) such two attractor flows.

As mentioned in the Introduction, the non-BPS $Z\neq 0$ attractor flow of
the $stu$ model has been already considered in literature in particular
cases, namely for the $D0-D4$ (\textit{magnetic}) \cite{Hotta,GLS-1}, $D0-D6$
\cite{GLS-1}, $D2-D6$ (\textit{electric}) \cite{K2-bis,Cai-Pang} $D0-D2-D4$ (%
\textit{magnetic }with $D2$) \cite{Cai-Pang}, $D0-D2-D4-D6$ (without $B$%
\textit{-fields}) \cite{K2-bis} supporting BH charge
configurations.\smallskip

In the present Section we determine the explicit expression of the non-BPS $%
Z\neq 0$ attractor flow for \textit{the most general} supporting BH charge
configuration, with \textit{all} electric and magnetic charges switched on,
namely for the non-BPS $Z\neq 0$-supporting branch of the $D0-D2-D4-D6$
configuration. Thence, as already done for $\frac{1}{2}$-BPS and non-BPS $%
Z=0 $ attractor flows, by exploiting the \textit{first order (fake
supergravity) formalism} \cite{Fake-Refs, Cer-Dal-1,ADOT-1}, we compute the
\textit{ADM masses} as well as the \textit{covariant scalar charges}, and
study the issue of \textit{marginal stability }\cite{Marginal-Refs},
completing and refining the treatment given in \cite{Hotta,GLS-1,Cai-Pang}.

\subsection{\label{U-Duality-Transf}$U$-Duality Transformations along the
Orbit $\mathcal{O}_{non-BPS,Z\neq 0}$}

In order to derive the explicit expression of the non-BPS $Z\neq 0$
attractor flow when all BH charges are non-vanishing, we exploit a method
already used in \cite{K2-bis}, \cite{GLS-1} and \cite{Cai-Pang}, based on
performing suitable symplectic transformations along the relevant (\textit{%
i.e.} supporting) charge orbit of the $U$-duality group. In Eqs. (\ref{Wed-5}%
) and (\ref{Wed-4}) we recalled the form of the $\frac{1}{2}$-BPS- and
non-BPS $Z=0$- supporting BH charge orbits of the \textit{tri-fundamental}
representation $\left( \mathbf{2},\mathbf{2},\mathbf{2}\right) $ of the $U$%
-duality group $G$ (given by Eq.(\ref{GG})) of the $stu$ model, also
commenting on their separation \cite{bellucci1}. The corresponding non-BPS $%
Z\neq 0$-supporting BH charge orbit reads \cite{bellucci1}
\begin{equation}
\mathcal{O}_{non-BPS,Z\neq 0}=\frac{\left( SU\left( 1,1\right) \right) ^{3}}{%
\left( SO\left( 1,1\right) \right) ^{2}},  \label{a-3}
\end{equation}
defined by the constraint
\begin{equation}
\mathcal{I}_{4}\left( \Gamma \right) <0.  \label{a-2}
\end{equation}

As done in \cite{GLS-1} and \cite{Cai-Pang}, in order to perform a
symplectic transformation along the charge orbit $\mathcal{O}_{non-BPS,Z\neq
0}$ of the $\left( \mathbf{2},\mathbf{2},\mathbf{2}\right) $ representation
of the $U$-duality, we take advantage of the complete factorization of the
special K\"{a}hler manifold $\left( \frac{SU\left( 1,1\right) }{U\left(
1\right) }\right) ^{3}$ (ultimately determining the \textit{triality symmetry%
}), which allows one to deal with the product of three distinct $2\times 2$
matrices of $SL\left( 2,\mathbb{R}\right) $, rather than with a unique $%
8\times 8$ matrix of the $U$-duality group embedded in the relevant
symplectic group $Sp\left( 8,\mathbb{R}\right) $.

The first step is to perform an $Sp\left( 8,\mathbb{R}\right) $%
-transformation from the basis $\left( p^{\Lambda },q_{\Lambda }\right) $ to
a basis $\mathcal{A}_{abc}$ ($a,b,c=0,1$ throughout) of BH charges expicitly
transforming under the $\left( \mathbf{2},\mathbf{2},\mathbf{2}\right) $ of
the $U$-duality. Such a transformation is given by Eq. (5.1) of \cite{GLS-1}
(equivalent to Eq. (3.5) of the second Ref. of \cite{duff1}; see also
Section 5 of \cite{BMOS-1}). The explicit action of a generic symplectic
transformation of the $U$-duality on the BH charges $\mathcal{A}_{abc}$ is
given, up to some change of notation, by Eqs. (5.2) and (5.3) of \cite{GLS-1}%
, which we report below for simplicity's sake:
\begin{eqnarray}
\mathcal{A}_{a^{\prime }b^{\prime }c^{\prime }}^{\prime } &=&\left(
M_{1}\right) _{a^{\prime }}^{~a}\left( M_{2}\right) _{b^{\prime
}}^{~b}\left( M_{3}\right) _{c^{\prime }}^{~c}a_{abc};  \label{boss-1} \\
&&  \notag \\
M_{i} &\equiv &\left(
\begin{array}{cc}
\frak{a}_{i} & \frak{b}_{i} \\
\frak{c}_{i} & \frak{d}_{i}
\end{array}
\right) \in SL\left( 2,\mathbb{R}\right) ,~det\left( M_{i}\right)
=1,~\forall i=1,2,3,  \label{boss-2}
\end{eqnarray}
where each matrix pertains to the degrees of freedom of only one modulus (%
\textit{e.g.} $M_{1}$ to $s$, $M_{2}$ to $t$, $M_{3}$ to $u$). The
transformation (\ref{boss-1})-(\ref{boss-2}) of $\left( SL\left( 2,\mathbb{R}%
\right) \right) ^{3}\subset Sp\left( 8,\mathbb{R}\right) $ induces also a
\textit{linear fractional} (\textit{M\"{o}bius}) transformation on the
moduli $z^{i}$ as follows (no summation on repeated indices; once again, we
report Eq. (5.3) of \cite{GLS-1} for simplicity's sake; also recall Eq. (\ref
{boss-3-bis})):
\begin{equation}
z^{\prime i}=\frac{\frak{a}_{i}z^{i}+\frak{b}_{i}}{\frak{c}_{i}z^{i}+\frak{d}%
_{i}}.  \label{boss-3}
\end{equation}

As done in \cite{GLS-1} and \cite{Cai-Pang}, we use the configuration $D0-D6$
as \textit{``pivot''} in order to perform the transformation (\ref{boss-1})-(%
\ref{boss-3}). Indeed, such a BH charge configuration supports only non-BPS $%
Z\neq 0$ attractors, as it can be easily realized by computing the
corresponding \textit{quartic} $U$-invariant, given by Eq. (\ref{I4}): $%
\mathcal{I}_{4}\left( \Gamma _{D0-D6}\right) <0$ (see also the treatment of
\cite{Ceresole}). Thus, we want to transform from the configuration $D0-D6$
(corresponding to charges $\left( q_{0},p^{0}\right) $, which we denote here
$\left( q,p\right) $) to the most general configuration $D0-D2-D4-D6$,
corresponding to all BH charges switched on: $\left(
q_{0},q_{i},p^{i},p^{0}\right) $. By exploiting the transformation (\ref
{boss-1})-(\ref{boss-3}), the parameters $\frak{a}_{i},\frak{b}_{i},\frak{c}%
_{i},\frak{d}_{i}$ of the $M_{i}$s dualizing from $D0-D6$ to $D0-D2-D4-D6$
must satisfy the following set of constraints:
\begin{eqnarray}
-q_{0} &=&-\frak{a}_{1}\frak{a}_{2}\frak{a}_{3}q+\frak{b}_{1}\frak{b}_{2}%
\frak{b}_{3}p;  \notag \\
&&  \notag \\
q_{i} &=&-\frac{1}{2}s_{ijk}\frak{c}_{i}\frak{a}_{j}\frak{a}_{k}q+\frac{1}{2}%
s_{ijk}\frak{d}_{i}\frak{b}_{j}\frak{b}_{k}p;  \notag \\
&&  \notag \\
p^{i} &=&-\frac{1}{2}s_{ijk}\frak{a}_{i}\frak{c}_{j}\frak{c}_{k}q+\frac{1}{2}%
s_{ijk}\frak{b}_{i}\frak{d}_{j}\frak{d}_{k}p;  \notag \\
&&  \notag \\
p^{0} &=&-\frak{c}_{1}\frak{c}_{2}\frak{c}_{3}q+\frak{d}_{1}\frak{d}_{2}%
\frak{d}_{3}p,  \label{a-1}
\end{eqnarray}
where $s_{ijk}\equiv \left| \epsilon _{ijk}\right| $. Notice that the system
(\ref{a-1}) admits solutions iff\ the condition (\ref{a-2}) is met; this
implies the transformation (\ref{boss-1})-(\ref{boss-3}) to belong to the $U$%
-duality orbit $\mathcal{O}_{non-BPS,Z\neq 0}$ given by Eq. (\ref{a-3}). The
sign of the BH charges $q$ and $p$ is actually irrelevant for the condition (%
\ref{a-2}) to be satisfied; thus, without loss of any generality, one can
choose \textit{e.g.} $q>0$, $p>0$. Within such a choice, the explicit form
of the matrices $M_{i}$s under consideration (and of their inverse) reads as
follows:
\begin{eqnarray}
M_{i} &=&-\frac{sgn\left( \xi \right) }{\sqrt{\left( \varsigma _{i}+\varrho
_{i}\right) \xi }}\left(
\begin{array}{cc}
\varsigma _{i}\xi & -\varrho _{i} \\
\xi & 1
\end{array}
\right) \Leftrightarrow M_{i}^{-1}=-\frac{sgn\left( \xi \right) }{\sqrt{%
\left( \varsigma _{i}+\varrho _{i}\right) \xi }}\left(
\begin{array}{cc}
1 & \varrho _{i} \\
-\xi & \varsigma _{i}\xi
\end{array}
\right) ;  \label{b-1} \\
&&  \notag \\
&&  \notag \\
\xi &\equiv &\left( \frac{p}{q}\right) ^{1/3}\left[ \frac{%
2p^{1}p^{2}p^{3}+p^{0}\left( \sqrt{-\mathcal{I}_{4}}-p^{\Lambda }q_{\Lambda
}\right) }{2p^{1}p^{2}p^{3}-p^{0}\left( \sqrt{-\mathcal{I}_{4}}-p^{\Lambda
}q_{\Lambda }\right) }\right] ^{1/3}\in \mathbb{R};  \label{b-2} \\
&&  \notag \\
&&  \notag \\
\varsigma _{i} &\equiv &\frac{\sqrt{-\mathcal{I}_{4}}+p^{\Lambda }q_{\Lambda
}-2p^{i}q_{i}}{s_{ijk}p^{j}p^{k}-2p^{0}q_{i}}\in \mathbb{R}~\text{(no sum on
}i\text{)};  \label{b-3} \\
&&  \notag \\
&&  \notag \\
\varrho _{i} &\equiv &\frac{\sqrt{-\mathcal{I}_{4}}-p^{\Lambda }q_{\Lambda
}+2p^{i}q_{i}}{s_{ijk}p^{j}p^{k}-2p^{0}q_{i}}\in \mathbb{R}~\text{(no sum on
}i\text{)}.  \label{b-4}
\end{eqnarray}
The definitions (\ref{b-3}) and (\ref{b-4}), together with Eq. (\ref{I4}),
imply that (no sum on $i$)
\begin{equation}
\varsigma _{i}\varrho _{i}=-\frac{s_{ijk}q_{j}q_{k}+2q_{0}p^{i}}{%
s_{ijk}p^{j}p^{k}-2p^{0}q_{i}}.  \label{b-5}
\end{equation}
As expected since the transformation (\ref{boss-1})-(\ref{boss-3}) belongs
to the orbit $\mathcal{O}_{non-BPS,Z\neq 0}$ of the $U$-duality, it leaves $%
\mathcal{I}_{4}$ unchanged:
\begin{equation}
\mathcal{I}_{4}\left( \Gamma _{D0-D2-D4-D6}\right) =-\left( p^{\Lambda
}q_{\Lambda }\right)
^{2}+4%
\sum_{i<j}p^{i}q_{i}p^{j}q_{j}-4p^{0}q_{1}q_{2}q_{3}+4q_{0}p^{1}p^{2}p^{3}=-%
\left( pq\right) ^{2}=\mathcal{I}_{4}\left( \Gamma _{D0-D6}\right) .
\end{equation}

It should be stressed that the transformation (\ref{boss-1})-(\ref{boss-3})
(along with Eqs. (\ref{b-1})-(\ref{b-4})) is \textit{not} the most general
transformation of $\mathcal{O}_{non-BPS,Z\neq 0}$ mapping the $D0-D6$ into
the $D0-D2-D4-D6$ configuration (and \textit{vice versa}). Indeed, it may be
further generalized by replacing $\xi $ with a triplet $\xi _{i}$,
constrained by $\xi _{1}\xi _{2}\xi _{3}=\xi ^{3}$. Such a \textit{%
two-parameter generalization} of the above transformation indicates, as
mentioned above, the presence of a real, $2$-dim. \textit{moduli space}
(namely $\left( SO\left( 1,1\right) \right) ^{2}$ \cite{ferrara4,TT2})
\textit{all along the non-BPS }$\mathit{Z\neq 0}$\textit{\ attractor flow};
this will become evident when looking at the explicit form of such a flow,
presented further below.

\subsection{\label{D0-D6}$D0-D6$: the Most General Flow and \textit{Fake
Superpotential}}

The most general non-BPS $Z\neq 0$ attractor flow in the $D0-D6$
configuration reads as follows \cite{GLS-1}:
\begin{eqnarray}
exp\left[ -4U_{non-BPS,Z\neq 0}\left( \tau \right) \right] &=&\left[
a-\left( -\mathcal{I}_{4}\right) ^{1/4}\tau \right] \left[ k^{1}-\left( -%
\mathcal{I}_{4}\right) ^{1/4}\tau \right] \left[ k^{2}-\left( -\mathcal{I}%
_{4}\right) ^{1/4}\tau \right] \left[ k^{3}-\left( -\mathcal{I}_{4}\right)
^{1/4}\tau \right] -b^{2};  \notag \\
&&  \label{ugen} \\
\ x_{non-BPS,Z\neq 0}^{i}\left( \tau \right) &=&\xi _{0}^{-1}e^{\alpha
_{i}}\cdot  \notag \\
&&\cdot \frac{\left[ k^{j}-\left( -\mathcal{I}_{4}\right) ^{1/4}\tau \right] %
\left[ k^{l}-\left( -\mathcal{I}_{4}\right) ^{1/4}\tau \right] -\left[
a-\left( -\mathcal{I}_{4}\right) ^{1/4}\tau \right] \left[ k^{i}-\left( -%
\mathcal{I}_{4}\right) ^{1/4}\tau \right] }{\left[ k^{j}-\left( -\mathcal{I}%
_{4}\right) ^{1/4}\tau \right] \left[ k^{l}-\left( -\mathcal{I}_{4}\right)
^{1/4}\tau \right] +\left[ a-\left( -\mathcal{I}_{4}\right) ^{1/4}\tau %
\right] \left[ k^{i}-\left( -\mathcal{I}_{4}\right) ^{1/4}\tau \right] -2{b}}%
;  \notag \\
&&  \label{xgenB} \\
y_{non-BPS,Z\neq 0}^{i}\left( \tau \right) &=&2\xi _{0}^{-1}e^{\alpha
_{i}}\cdot  \notag \\
&&\cdot \frac{exp\left[ -2U_{non-BPS,Z\neq 0}\left( \tau \right) \right] }{%
\left[ k^{j}-\left( -\mathcal{I}_{4}\right) ^{1/4}\tau \right] \left[
k^{l}-\left( -\mathcal{I}_{4}\right) ^{1/4}\tau \right] +\left[ a-\left( -%
\mathcal{I}_{4}\right) ^{1/4}\tau \right] \left[ k^{i}-\left( -\mathcal{I}%
_{4}\right) ^{1/4}\tau \right] -2{b}},  \notag \\
&&  \label{ygenB}
\end{eqnarray}
where
\begin{equation}
\xi _{0}\equiv (p/q)^{1/3},  \label{b-2-bis}
\end{equation}
$a\in \mathbb{R}_{0}$, $b$, $k^{i}\in \mathbb{R}$ ($k^{i}$s cannot all
vanish), and the triplet of real constants $\alpha _{i}$ satisfies the
constraint
\begin{equation}
\sum_{i}\alpha _{i}=0.  \label{b-2-tris}
\end{equation}

It is worth pointing out that the $D0-D6$ configuration supports \textit{%
axion-free} non-BPS $Z\neq 0$ attractor flow(s); when considering the
\textit{near-horizon limit}, and thus the critical, charge-dependent values
of the moduli, this is consistent with the analysis performed in \cite
{TT,TT2,Ceresole}. An \textit{axion-free} attractor flow solution of Eqs. (%
\ref{ugen})-(\ref{ygenB}) can be obtained \textit{e.g.} by putting
\begin{equation}
k^{i}=a~\forall i=1,2,3,
\end{equation}
and it reads as follows:
\begin{eqnarray}
exp\left[ -4U_{non-BPS,Z\neq 0,axion-free}\left( \tau \right) \right] &=&%
\left[ a-\left( -\mathcal{I}_{4}\right) ^{1/4}\tau \right] ^{4}-b^{2};
\notag  \label{night!-1} \\
\ x_{non-BPS,Z\neq 0,axion-free}^{i}\left( \tau \right) &=&0;  \notag
\label{night!-2} \\
y_{non-BPS,Z\neq 0,axion-free}^{i}\left( \tau \right) &=&\xi
_{0}^{-1}e^{\alpha _{i}}\sqrt{\frac{\left[ a-\left( -\mathcal{I}_{4}\right)
^{1/4}\tau \right] ^{2}+b}{\left[ a-\left( -\mathcal{I}_{4}\right)
^{1/4}\tau \right] ^{2}-b}}.  \label{night!-3}
\end{eqnarray}

The non-BPS $Z\neq 0$ \textit{fake superpotential} of the \textit{first
order formalism} can be computed to have the following form in the $D0-D6$
configuration:
\begin{eqnarray}
&&\mathcal{W}_{non-BPS,Z\neq 0}(z,\overline{z},q,p)=\frac{1}{4}e^{K/2}\left[
\prod_{i}\left| q^{1/3}+p^{1/3}e^{-\alpha _{i}}z^{i}\right| \right] \cdot
\notag \\
&&  \notag \\
\cdot &&\left[ 1+\sum_{i<j}\frac{\left( q^{2/3}-p^{2/3}e^{-2\alpha
_{i}}\left| z^{i}\right| ^{2}\right) \left( q^{2/3}-p^{2/3}e^{-2\alpha
_{j}}\left| z^{j}\right| ^{2}\right) -e^{-\alpha _{i}-\alpha
_{j}}q^{2/3}p^{2/3}(z^{i}-\overline{z}^{\overline{i}})(z^{j}-\overline{z}^{%
\overline{j}})}{\left| q^{1/3}+p^{1/3}e^{-\alpha _{i}}z^{i}\right|
^{2}\,\left| q^{1/3}+p^{1/3}e^{-\alpha _{j}}z^{j}\right| ^{2}}\right] .
\notag \\
&&  \label{D0D6fake}
\end{eqnarray}
The \textit{axion-free} version of such a \textit{fake superpotential} (%
\textit{e.g.} pertaining to the solution (\ref{night!-1})-(\ref{night!-3}))
reads as follows:
\begin{eqnarray}
&&\mathcal{W}_{non-BPS,Z\neq 0,axion-free}(y,q,p)=\frac{1}{2^{3}\sqrt{2}}%
\frac{1}{\sqrt{y^{1}y^{2}y^{3}}}\left[ \prod_{l}\left|
q^{1/3}-ip^{1/3}e^{-\alpha _{l}}y^{l}\right| \right] \cdot  \notag \\
&&  \notag \\
\cdot &&\left[ 1+\sum_{i<j}\frac{\left[ q^{2/3}-p^{2/3}e^{-2\alpha
_{i}}\left( y^{i}\right) ^{2}\right] \left[ q^{2/3}-p^{2/3}e^{-2\alpha
_{j}}\left( y^{j}\right) ^{2}\right] +4e^{-\alpha _{i}-\alpha
_{j}}q^{2/3}p^{2/3}y^{i}y^{j}}{\left| q^{1/3}-ip^{1/3}e^{-\alpha
_{i}}y^{i}\right| ^{2}\,\left| q^{1/3}-ip^{1/3}e^{-\alpha _{j}}y^{j}\right|
^{2}}\right] .  \label{D0D6fake-axion-free}
\end{eqnarray}

Now, by exploiting the \textit{first order formalism} \cite{Fake-Refs} for $%
d=4$ extremal BHs \cite{Cer-Dal-1,ADOT-1} (see also \cite{FHM} and \cite
{Gnecchi-1}), one can compute the relevant BH parameters of the non-BPS $%
Z\neq 0$ attractor flow of $d=4$ $stu$ model in the $D0-D6$ configuration,
starting from the expression of the non-BPS $Z\neq 0$ \textit{fake
superpotential} $\mathcal{W}_{non-BPS,Z\neq 0}$ given by Eq. (\ref{D0D6fake}%
).

Eqs. (\ref{D0D6fake}) and (\ref{d-1}) yield, after some algebra, the
following expression for the \textit{ADM mass}:
\begin{gather}  \label{D0D6ADM}
M_{ADM,non-BPS,Z\neq 0}\left( z_{\infty },\overline{z}_{\infty },\Gamma
_{D0-D6}\right) =\frac{P}{2^{7/2}}\,\left[ \prod_{i}\sqrt{\left[ \left(
\Lambda ^{i}\right) ^{-1}+B^{i}\right] ^{2}+1}\right] \cdot  \notag \\
\cdot \left\{ 1+\sum_{i<j}\frac{\left[ \left( \Lambda ^{i}\right)
^{-2}-\left( B^{i}\right) ^{2}-1\right] \left[ \left( \Lambda ^{j}\right)
^{-2}-\left( B^{j}\right) ^{2}-1\right] +4\left( \Lambda ^{i}\right)
^{-1}\left( \Lambda ^{j}\right) ^{-1}}{\left[ \left[ \left( \Lambda
^{i}\right) ^{-1}+B^{i}\right] ^{2}+1\right] \left[ \left[ \left( \Lambda
^{j}\right) ^{-1}+B^{j}\right] ^{2}+1\right] }\right\} ,  \notag \\
\end{gather}
where the quantities
\begin{equation}
\Lambda ^{i}\equiv \xi _{0}y_{\infty }^{i};~~B^{i}\equiv \frac{x_{\infty
}^{i}}{y_{\infty }^{i}},~~P\equiv p\sqrt{y_{\infty }^{1}y_{\infty
}^{2}y_{\infty }^{3}},~~Q\equiv \frac{q}{\sqrt{y_{\infty }^{1}y_{\infty
}^{2}y_{\infty }^{3}}}  \label{e-1}
\end{equation}
were introduced, and, for simplicity's sake, the $\alpha _{i}$s were chosen
\textit{all} to vanish (i.e. $\alpha _{i}=0~\forall i=1,2,3$). $P$ and $Q$
are the \textit{dressed charges}, \textit{i.e.} a sort of \textit{%
asymptotical redefinition} of the charges pertaining to $D6$ and $D0$
branes, respectively. On the other hand, $\Lambda ^{i}$ and $B^{i}$ are
usually named \textit{(asymptotical brane) fluxes} and $B$\textit{-fields},
respectively.

Eq. (\ref{D0D6ADM}) (along with the definitions (\ref{e-1})) coincides with
Eq. (5.56) of \cite{GLS-1}, provided that the following condition is met
(see Eq. (5.44) of \cite{GLS-1}):
\begin{equation}
\Lambda ^{1}\left[ 1+\left( B^{1}\right) ^{2}\right] -\left( \Lambda
^{1}\right) ^{-1}=\Lambda ^{2}\left[ 1+\left( B^{2}\right) ^{2}\right]
-\left( \Lambda ^{2}\right) ^{-1}=\Lambda ^{3}\left[ 1+\left( B^{3}\right)
^{2}\right] -\left( \Lambda ^{3}\right) ^{-1}.
\end{equation}
As observed in \cite{GLS-1}, Eq. (\ref{D0D6ADM}) (along with the definitions
(\ref{e-1})) yields the \textit{marginal bound} \cite{Marginal-Refs} to be
\textit{saturated}, because $M_{ADM,non-BPS,Z\neq 0}$ is equal to the sum of
the \textit{ADM masses} of four $D6$-branes with appropriate fluxes (for
further detail, see the discussion in \cite{GLS-1}).

Concerning the \textit{(covariant) scalar charges}, they can be
straightforwardly computed by recalling Eqs. (\ref{D0D6fake}) and (\ref{d-2}%
), but their expressions are rather cumbersome. For simplicity's sake, here
we limit ourselves to give the \textit{scalar charges} for the \textit{``}$%
t^{3}$\textit{-degeneration''} of the $stu$ model (in which all charges and
moduli are equal, insensitive to $i$-index; see \textit{e.g.} Eq. (\ref
{night-1}) below, as well as the treatment in Sect. 5 of \cite{BMOS-1}). By
denoting $t\equiv x-iy$, the \textit{covariant scalar charges} of axion and
dilaton in the $D0-D6$ configuration respectively read
\begin{eqnarray}
\Sigma _{x,non-BPS,Z\neq 0}\left( x_{\infty },y_{\infty },\Gamma
_{D0-D6}\right) &=&\sqrt{2}\,P\,x_{\infty }\frac{\left( \Lambda
^{-1}B+B^{2}+1\right) }{\sqrt{\left( \Lambda ^{-1}+B\right) ^{2}+1}}; \\
&&  \notag \\
\Sigma _{y,non-BPS,Z\neq 0}\left( x_{\infty },y_{\infty },\Gamma
_{D0-D6}\right) &=&-\frac{P\,y_{\infty }}{\sqrt{2}}\,\frac{\left[ \Lambda
^{-4}+\Lambda ^{-3}B+\Lambda ^{-1}B\left( B^{2}-1\right) +B^{4}-1\right] }{%
\sqrt{\left( \Lambda ^{-1}+B\right) ^{2}+1}}.  \notag \\
&&
\end{eqnarray}

\subsection{\label{D0-D2-D4-D6}$D0-D2-D4-D6$: the Most General Flow and
\textit{Fake Superpotential}}

Now, by performing the $U$-duality transformation (\ref{boss-1}-(\ref{boss-3}%
) (along with Eqs. (\ref{b-1})-(\ref{b-5})) and using the most general
non-BPS $Z\neq 0$ attractor flow in the $D0-D6$ configuration given by Eqs. (%
\ref{ugen})-(\ref{ygenB}), it is a matter of long but straightforward
computations to determine the most general non-BPS $Z\neq 0$ attractor flow
in the most general configuration, namely in the $D0-D2-D4-D6$ one, in which
\textit{all} BH charges are switched on. It reads as follows\footnote{%
In the particular case in which $b=0$, the expression of $exp\left[
-4U_{non-BPS,Z\neq 0}\left( \tau \right) \right] $ can be recast in the form
\begin{equation*}
exp\left[ -4U_{non-BPS,Z\neq 0}\left( \tau \right) \right] =-\mathcal{I}%
_{4}\left( \mathcal{H}\left( \tau \right) \right) ,
\end{equation*}
consistently with the results of \cite{K2-bis} and \cite{GLS-1}, and on the
same ground of (the first of) Eqs. (\ref{BPS-sol}) and (\ref
{non-BPS-Z=0-flow}), respectively holding for the $\frac{1}{2}$-BPS and
non-BPS $Z=0$ attractor flows.} (the moduli are here denoted as $\mathcal{Z}%
^{i}\equiv \mathcal{X}^{i}-i\mathcal{Y}^{i}$; $i\neq j\neq l$ and no sum on
repeated $i-$indices throughout):
\begin{eqnarray}
exp\left[ -4U_{non-BPS,Z\neq 0}\left( \tau \right) \right] &=&h_{0}\left(
\tau \right) h_{1}\left( \tau \right) h_{2}\left( \tau \right) h_{3}\left(
\tau \right) -b^{2};  \label{sol} \\
&&  \notag \\
\ \mathcal{X}_{non-BPS,Z\neq 0}^{i}\left( \tau \right) &=&\frac{\left\{
\begin{array}{l}
\varsigma _{i}e^{2\alpha _{i}}\nu ^{2}\left[ h_{j}\left( \tau \right)
h_{l}\left( \tau \right) +h_{0}\left( \tau \right) h_{i}\left( \tau \right)
+2b\right] + \\
\\
+e^{\alpha _{i}}\nu (\varsigma _{i}-\varrho _{i})\left[ h_{j}\left( \tau
\right) h_{l}\left( \tau \right) -h_{0}\left( \tau \right) h_{i}\left( \tau
\right) \right] + \\
\\
-\varrho _{i}\left[ h_{j}\left( \tau \right) h_{l}\left( \tau \right)
+h_{0}\left( \tau \right) h_{i}\left( \tau \right) -2b\right]
\end{array}
\right\} }{\left\{
\begin{array}{l}
e^{2\alpha _{i}}\nu ^{2}\left[ h_{j}\left( \tau \right) h_{l}\left( \tau
\right) +h_{0}\left( \tau \right) h_{i}\left( \tau \right) +2b\right] + \\
\\
+2e^{\alpha _{i}}\nu \left[ h_{j}\left( \tau \right) h_{l}\left( \tau
\right) -h_{0}\left( \tau \right) h_{i}\left( \tau \right) \right] + \\
\\
+h_{j}\left( \tau \right) h_{l}\left( \tau \right) +h_{0}\left( \tau \right)
h_{i}\left( \tau \right) -2b
\end{array}
\right\} };  \notag \\
&&  \label{sol-1} \\
&&  \notag \\
\mathcal{Y}_{non-BPS,Z\neq 0}^{i}\left( \tau \right) &=&\frac{2\nu e^{\alpha
_{i}}(\varsigma _{i}+\varrho _{i})exp\left[ -2U_{non-BPS,Z\neq 0}\left( \tau
\right) \right] }{\left\{
\begin{array}{l}
e^{2\alpha _{i}}\nu ^{2}\left[ h_{j}\left( \tau \right) h_{l}\left( \tau
\right) +h_{0}\left( \tau \right) h_{i}\left( \tau \right) +2b\right] + \\
\\
+2e^{\alpha _{i}}\nu \left[ h_{j}\left( \tau \right) h_{l}\left( \tau
\right) -h_{0}\left( \tau \right) h_{i}\left( \tau \right) \right] + \\
\\
+h_{j}\left( \tau \right) h_{l}\left( \tau \right) +h_{0}\left( \tau \right)
h_{i}\left( \tau \right) -2b
\end{array}
\right\} },  \notag \\
&&  \label{sol-2}
\end{eqnarray}
where $\varsigma _{i}$ and $\varrho _{i}$ have been defined in Eqs. (\ref
{b-3}) and (\ref{b-4}), respectively. The constants $\alpha _{i}$ and $b$
have been introduced in Eqs. (\ref{ugen})-(\ref{ygenB}). Furthermore, the
new quantities (see Eqs. (\ref{b-2}) and (\ref{b-2-bis}), as well)
\begin{eqnarray}
\nu &\equiv &\frac{\xi }{\xi _{0}}=\left[ \frac{2p^{1}p^{2}p^{3}+p^{0}\left(
\sqrt{-\mathcal{I}_{4}}-p^{\Lambda }q_{\Lambda }\right) }{%
2p^{1}p^{2}p^{3}-p^{0}\left( \sqrt{-\mathcal{I}_{4}}-p^{\Lambda }q_{\Lambda
}\right) }\right] ^{1/3}\in \mathbb{R;}  \label{sol-3} \\
&&  \notag \\
h_{\Lambda }\left( \tau \right) &\equiv &b_{\Lambda }-\left( -\mathcal{I}%
_{4}\right) ^{1/4}\tau ,  \label{sol-4}
\end{eqnarray}
where $b_{\Sigma }$ are real constants, have been defined.

By performing the \textit{near-horizon} (\textit{i.e.} $\tau \rightarrow
-\infty $) \textit{limit}, Eqs. (\ref{sol-1}) and (\ref{sol-2}) respectively
yield the following \textit{critical} values of the moduli (the subscript
\textit{``}$H$\textit{''} stands for \textit{``}horizon\textit{'')}:
\begin{eqnarray}
&&\mathcal{X}_{non-BPS,Z\neq 0,H}^{i}\equiv lim_{\tau \rightarrow -\infty }\
\mathcal{X}_{non-BPS,Z\neq 0}^{i}\left( \tau \right) =\frac{\varsigma
_{i}e^{2\alpha _{i}}\nu ^{2}-\varrho _{i}}{e^{2\alpha _{i}}\nu ^{2}+1};
\label{attrx} \\
&&  \notag \\
&&\mathcal{Y}_{non-BPS,Z\neq 0,H}^{i}\equiv lim_{\tau \rightarrow -\infty }\
\mathcal{Y}_{non-BPS,Z\neq 0}^{i}\left( \tau \right) =\frac{1}{2}\frac{%
e^{\alpha _{i}}(\varsigma _{i}+\varrho _{i})\nu }{e^{2\alpha _{i}}\nu ^{2}+1}%
=\frac{\sqrt{-\mathcal{I}_{4}}e^{\alpha _{i}}\nu }{\left(
s_{ijk}p^{j}p^{k}-2p^{0}q_{i}\right) \left( e^{2\alpha _{i}}\nu
^{2}+1\right) }.  \notag \\
&&  \label{attry}
\end{eqnarray}
It is worth pointing out that the $D0-D2-D4-D6$ configuration does \textit{%
not} support \textit{axion-free} non-BPS $Z\neq 0$ attractor flow(s); when
considering the \textit{near-horizon limit}, and thus the critical,
charge-dependent values of the moduli, this is consistent with the analysis
performed in \cite{TT,TT2,Ceresole}.

The solution (\ref{sol})-(\ref{sol-2}) (along with the definitions (\ref
{sol-3}) and (\ref{sol-4})) generalizes the result of \cite{K2-bis}. As
mentioned in Sect. \ref{Introduction}, in \cite{K2-bis} it was shown that,
within the (non-BPS $Z\neq 0$-supporting branches of the) $D2-D6$ (\textit{%
electric}) and $D0-D2-D4-D6$ configurations, in absence of (some of the) $B$%
\textit{-fields} the attractor flow solution can be obtained by replacing
the $Sp\left( 8,\mathbb{R}\right) $-covariant vector $\Gamma $ of charges
(defined by Eq. (\ref{Gamma})) with the $Sp\left( 8,\mathbb{R}\right) $%
-covariant vector $\mathcal{H}\left( \tau \right) $ of harmonic functions
(defined by Eqs. (\ref{H})-(\ref{constraints})) in the corresponding
critical, horizon solution.

For the $\frac{1}{2}$-BPS and non-BPS $Z=0$ attractor flows, respectively
treated in Sects. \ref{BPS-Flow} and \ref{Non-BPS-Z=0-Flow}, such a
procedure allows one to determine the most general attractor flow solution
starting from the corresponding critical, horizon solution.

On the other hand, for the non-BPS $Z\neq 0$ attractor flow such a procedure
fails in presence of non-vanishing $B$\textit{-fields}. In other words, it
can be shown that the completely general non-BPS $Z\neq 0$ attractor flow
solution (\ref{sol})-(\ref{sol-2}) is \textit{not} a solution of the \textit{%
would-be non-BPS }$Z\neq 0$\textit{\ stabilization Eqs.} (see \textit{e.g.}
the treatments of \cite{AoB-book}, \cite{K2} and \cite{BFMY})
\begin{eqnarray}
\mathcal{H}^{T}\left( \tau \right) &=&2e^{K\left( z\left( \tau \right) ,%
\overline{z}\left( \tau \right) \right) }Im\left[ W\left( z\left( \tau
\right) \right) \left(
\begin{array}{c}
\overline{X}^{\Lambda }\left( \overline{z}\left( \tau \right) \right) \\
\\
\overline{F}_{\Lambda }\left( \overline{z}\left( \tau \right) \right)
\end{array}
\right) \right. +  \notag \\
&&+\frac{i}{2}\frac{\overline{W}\left( \overline{z}\left( \tau \right)
\right) }{\left| W\right| ^{2}\left( z\left( \tau \right) ,\overline{z}%
\left( \tau \right) \right) }\overline{C}_{\overline{i}\overline{j}\overline{%
k}}\left( z\left( \tau \right) ,\overline{z}\left( \tau \right) \right) g^{i%
\overline{i}}\left( z\left( \tau \right) ,\overline{z}\left( \tau \right)
\right) g^{j\overline{j}}\left( z\left( \tau \right) ,\overline{z}\left(
\tau \right) \right) g^{k\overline{k}}\left( z\left( \tau \right) ,\overline{%
z}\left( \tau \right) \right) \cdot  \notag \\
&&\left. \cdot \left( D_{j}W\right) \left( z\left( \tau \right) ,\overline{z}%
\left( \tau \right) ,\mathcal{H}\left( \tau \right) \right) \left(
D_{k}W\right) \left( z\left( \tau \right) ,\overline{z}\left( \tau \right) ,%
\mathcal{H}\left( \tau \right) \right) \left(
\begin{array}{c}
\left( D_{i}X^{\Lambda }\right) \left( z\left( \tau \right) ,\overline{z}%
\left( \tau \right) \right) \\
\\
\left( D_{i}F_{\Lambda }\right) \left( z\left( \tau \right) ,\overline{z}%
\left( \tau \right) \right)
\end{array}
\right) \right] ,  \notag \\
&&  \label{non-BPS-Z<>0-stabilization-Eqs.}
\end{eqnarray}
which can be obtained from the \textit{non-BPS }$\mathit{Z\neq 0}$\textit{\
Attractor Eqs.} (see \textit{e.g.} the treatments in \cite{AoB-book} and
\cite{BFMY})
\begin{eqnarray}
\Gamma ^{T} &=&2e^{K\left( z,\overline{z}\right) }Im\left[ W\left( z\right)
\left(
\begin{array}{c}
\overline{X}^{\Lambda }\left( \overline{z}\right) \\
\\
\overline{F}_{\Lambda }\left( \overline{z}\right)
\end{array}
\right) \right. +  \notag \\
&&+\frac{i}{2}\frac{\overline{W}\left( \overline{z}\right) }{\left| W\right|
^{2}\left( z,\overline{z}\right) }\overline{C}_{\overline{i}\overline{j}%
\overline{k}}\left( z,\overline{z}\right) g^{i\overline{i}}\left( z,%
\overline{z}\right) g^{j\overline{j}}\left( z,\overline{z}\right) g^{k%
\overline{k}}\left( z,\overline{z}\right) \cdot  \notag \\
&&\left. \cdot \left( D_{j}W\right) \left( z,\overline{z},\Gamma \right)
\left( D_{k}W\right) \left( z,\overline{z},\Gamma \right) \left(
\begin{array}{c}
\left( D_{i}X^{\Lambda }\right) \left( z,\overline{z}\right) \\
\\
\left( D_{i}F_{\Lambda }\right) \left( z,\overline{z}\right)
\end{array}
\right) \right]  \label{non-BPS-Z<>0-AEs}
\end{eqnarray}
by simply replacing $\Gamma $ with $\mathcal{H}\left( \tau \right) $. On the
other hand, the non-BPS $Z\neq 0$ attractor flow solutions obtained in \cite
{K2-bis} for the $D2-D6$ and $D0-D2-D4-D6$ configurations (both without $B$%
\textit{-fields}) consistently do satisfy Eq. (\ref
{non-BPS-Z<>0-stabilization-Eqs.}).

Furthermore, Eq. (\ref{non-BPS-Z<>0-AEs}) is the \textit{near-horizon} ($%
\tau \rightarrow -\infty $) \textit{limit }of Eq. (\ref
{non-BPS-Z<>0-stabilization-Eqs.}), as it has to be.

The issue concerning whether (in all non-BPS $Z\neq 0$-supporting
configurations) the \textit{actual} \textit{non-BPS }$Z\neq 0$\textit{\
stabilization Eqs.} (if any) admit a \textit{(}$\frac{1}{2}$\textit{%
-)BPS-like }reformulation in terms of a non-BPS $Z\neq 0$ \textit{fake
superpotential} (whose general form is given by Eq. (\ref{D0D2D4D6fake})
below) is open, and its investigation is left for future work.

Next, we can compute the non-BPS $Z\neq 0$ \textit{fake superpotential} of
the \textit{first order formalism} in the $D0-D2-D4-D6$ configuration. To do
this, we apply the $U$-duality transformation (\ref{boss-1}-(\ref{boss-3})
(along with Eqs. (\ref{b-1})-(\ref{b-5})) to the expression of the non-BPS $%
Z\neq 0$ \textit{fake superpotential} in the $D0-D6$ configuration given by
Eq. (\ref{D0D6fake}), and, by noticing that $\mathcal{W}$ does \textit{not}
have any further covariance property under such a transformation, after some
algebra one achieves the following result:
\begin{eqnarray}
&&\mathcal{W}_{non-BPS,Z\neq 0}(\mathcal{Z},\overline{\mathcal{Z}}%
,p^{0},p^{1},p^{2},p^{3},q_{0},q_{1},q_{2},q_{3})=  \notag \\
&&  \notag \\
&=&\frac{1}{4}\frac{\nu ^{3/2}\left( -\mathcal{I}_{4}\right) ^{1/4}}{\sqrt{%
\prod_{i}(\varsigma _{i}+\varrho _{i})}}e^{K/2}\left[ \prod_{i}\left|
\varsigma _{i}-\mathcal{Z}^{i}+(\varrho _{i}+\mathcal{Z}^{i})e^{-\alpha
_{i}}\nu ^{-1}\right| \right] \cdot  \notag \\
&&  \notag \\
&&\cdot \left( 1+\sum_{i<j}\frac{\left[ |\varsigma _{i}-\mathcal{Z}%
^{i}|^{2}-|\varrho _{i}+\mathcal{Z}^{i}|^{2}e^{-2\alpha _{i}}\nu ^{-2}\right]
\left[ |\varsigma _{j}-\mathcal{Z}^{j}|^{2}-|\varrho _{j}+\mathcal{Z}%
^{j}|^{2}e^{-2\alpha _{j}}\nu ^{-2}\right] }{\left| \varsigma _{i}-\mathcal{Z%
}^{i}+(\varrho _{i}+\mathcal{Z}^{i})e^{-\alpha _{i}}\nu ^{-1}\right|
^{2}\,\left| \varsigma _{j}-\mathcal{Z}^{j}+(\varrho _{j}+\mathcal{Z}%
^{j})e^{-\alpha _{j}}\nu ^{-1}\right| ^{2}}+\right.  \notag \\
&&  \notag \\
&&\qquad -\left. \sum_{i<j}\frac{e^{-\alpha _{i}-\alpha _{j}}\nu
^{-2}(\varsigma _{i}+\varrho _{i})(\varsigma _{j}+\varrho _{j})(\mathcal{Z}%
^{i}-\overline{\mathcal{Z}}^{\overline{i}})(\mathcal{Z}^{j}-\overline{%
\mathcal{Z}}^{\overline{j}})}{\left| \varsigma _{i}-\mathcal{Z}^{i}+(\varrho
_{i}+\mathcal{Z}^{i})e^{-\alpha _{i}}\nu ^{-1}\right| ^{2}\,\left| \varsigma
_{j}-\mathcal{Z}^{j}+(\varrho _{j}+\mathcal{Z}^{j})e^{-\alpha _{j}}\nu
^{-1}\right| ^{2}}\right) .  \label{D0D2D4D6fake}
\end{eqnarray}
Consistently with the \textit{first order formalism} \cite{Fake-Refs} for $%
d=4$ extremal BHs \cite{Cer-Dal-1,ADOT-1} (see also \cite{FHM} and \cite
{Gnecchi-1}), it is easy to check that the \textit{near-horizon} \textit{%
limit} of $\mathcal{W}_{non-BPS,Z\neq 0}^{2}$ yields the square root of $-%
\mathcal{I}_{4}$ (given by Eq. (\ref{I4})), or equivalently the square root
of the \textit{Cayley's hyperdeterminant }$Det\left( \Psi \right) $:
\begin{gather}
\mathcal{W}_{non-BPS,Z\neq 0,H}^{2}(\Gamma _{D0-D2-D4-D6})\equiv  \notag \\
\equiv lim_{\tau \rightarrow -\infty }\mathcal{W}_{non-BPS,Z\neq 0}^{2}(%
\mathcal{Z}\left( \tau \right) ,\overline{\mathcal{Z}}\left( \tau \right)
,\Gamma _{D0-D2-D4-D6})=  \notag \\
=\sqrt{-\mathcal{I}_{4}}=\sqrt{Det\left( \Psi \right) }=\frac{%
S_{BH,non-BPZ,Z\neq 0}(\Gamma _{D0-D2-D4-D6})}{\pi },
\end{gather}
where in the last step the Bekenstein-Hawking entropy-area formula \cite
{hawking2} was used.

Now, as done for the $D0-D6$ configuration in the previous Subsection, by
exploiting the \textit{first order formalism} \cite{Fake-Refs} for $d=4$
extremal BHs \cite{Cer-Dal-1,ADOT-1} (see also \cite{FHM} and \cite
{Gnecchi-1}), one can compute the relevant BH parameters, such as the
\textit{ADM mass} (Eq. (\ref{d-1})) and the \textit{covariant scalar charges}
(Eq. (\ref{d-2})), starting from the \textit{fake superpotential} $\mathcal{W%
}_{non-BPS,Z\neq 0}$ given by Eq. (\ref{D0D2D4D6fake}). The computations are
long but straightforward, and they yield cumbersome results (also \textit{%
e.g.} in the limit of \textit{``}$t^{3}$\textit{-degeneration''}, see Eq. (%
\ref{night-1}) below), which we thus decide to omit here. We will explicitly
analyze some particular configurations in Sect. \ref{Analysis-Particular}.

However, it is easy to realize that Eq. (\ref{D0D2D4D6fake}) implies the
\textit{marginal bound} \cite{Marginal-Refs} to be \textit{saturated},
because (see Eq. (\ref{d-1}))
\begin{gather}
M_{ADM,non-BPS,Z\neq 0,}(\mathcal{Z}_{\infty },\overline{\mathcal{Z}}%
_{\infty },\Gamma _{D0-D2-D4-D6})=  \notag \\
=\mathcal{W}_{non-BPS,Z\neq 0}(\mathcal{Z}_{\infty },\overline{\mathcal{Z}}%
_{\infty },\Gamma _{D0-D2-D4-D6})\equiv  \notag \\
\equiv lim_{\tau \rightarrow 0^{-}}\mathcal{W}_{non-BPS,Z\neq 0}(\mathcal{Z}%
\left( \tau \right) ,\overline{\mathcal{Z}}\left( \tau \right) ,\Gamma
_{D0-D2-D4-D6})
\end{gather}
is equal to the sum of the \textit{ADM masses} of four $D6$-branes with
appropriate fluxes (for further detail on definition of such brane fluxes,
see the related discussion in \cite{GLS-1}). Thus, generalizing the related
results of \cite{GLS-1} and \cite{Cai-Pang}, it can be stated that the
\textit{marginal stability} holds for the most general non-BPS $Z\neq 0$
attractor flow of the $\mathcal{N}=2$, $d=4$ $stu$ model.

\section{\label{Analysis-Particular}Analysis of Particular Configurations}

In this Section we analyze in depth some particularly simple configurations,
generalizing some results in literature \cite{K2-bis,Hotta, GLS-1,Cai-Pang}.

\subsection{\label{D0-D4}\textit{Magnetic} ($D0-D4$)}

The configuration $D0-D4$ (also named \textit{magnetic}) of the $stu$ model
has been previously treated in \cite{Hotta} and \cite{GLS-1}. In this case,
the quantities of the $U$-duality transformation (\ref{boss-1})-(\ref{boss-3}%
) along $\mathcal{O}_{non-BPS,Z\neq 0}$ defined by Eqs. (\ref{b-2})-(\ref
{b-4}) undergo a major simplification:
\begin{equation}
\xi =\xi _{0};~~\varsigma _{i}=\varrho _{i}=\sqrt{\frac{-q_{0}p^{i}}{\frac{1%
}{2}s_{ijk}p^{j}p^{k}}}.
\end{equation}
Correspondingly, the non-BPS $Z\neq 0$ attractor flow (\ref{sol})-(\ref
{sol-2}) acquires the following form (as above, the moduli are here denoted
as $\mathcal{Z}^{i}\equiv \mathcal{X}^{i}-i\mathcal{Y}^{i}$; $i\neq j\neq l$
and no sum on repeated $i-$indices throughout):
\begin{eqnarray}
&&exp\left[ -4U_{non-BPS,Z\neq 0}\left( \tau \right) \right] =h_{0}\left(
\tau \right) h_{1}\left( \tau \right) h_{2}\left( \tau \right) h_{3}\left(
\tau \right) -b^{2};  \label{electric-1} \\
&&  \notag \\
\ &&\mathcal{X}_{non-BPS,Z\neq 0}^{i}\left( \tau \right) =\varsigma _{i}\cdot
\notag \\
&&  \notag \\
&&\cdot \frac{e^{2\alpha _{i}}\left[ h_{j}\left( \tau \right) h_{l}\left(
\tau \right) +h_{0}\left( \tau \right) h_{i}\left( \tau \right) +2b\right] -%
\left[ h_{j}\left( \tau \right) h_{l}\left( \tau \right) +h_{0}\left( \tau
\right) h_{i}\left( \tau \right) -2b\right] }{\left\{
\begin{array}{l}
e^{2\alpha _{i}}\left[ h_{j}\left( \tau \right) h_{l}\left( \tau \right)
+h_{0}\left( \tau \right) h_{i}\left( \tau \right) +2b\right] +2e^{\alpha
_{i}}\left[ h_{j}\left( \tau \right) h_{l}\left( \tau \right) -h_{0}\left(
\tau \right) h_{i}\left( \tau \right) \right] + \\
\\
+h_{j}\left( \tau \right) h_{l}\left( \tau \right) +h_{0}\left( \tau \right)
h_{i}\left( \tau \right) -2b
\end{array}
\right\} };  \notag \\
&&  \notag \\
&&  \label{electric-2} \\
&&\mathcal{Y}_{non-BPS,Z\neq 0}^{i}\left( \tau \right) =4\varsigma _{i}\cdot
\notag \\
&&  \notag \\
&&\cdot \frac{e^{\alpha _{i}}exp\left[ -2U_{non-BPS,Z\neq 0}\left( \tau
\right) \right] }{\left\{
\begin{array}{l}
e^{2\alpha _{i}}\left[ h_{j}\left( \tau \right) h_{l}\left( \tau \right)
+h_{0}\left( \tau \right) h_{i}\left( \tau \right) +2b\right] +2e^{\alpha
_{i}}\left[ h_{j}\left( \tau \right) h_{l}\left( \tau \right) -h_{0}\left(
\tau \right) h_{i}\left( \tau \right) \right] + \\
\\
+h_{j}\left( \tau \right) h_{l}\left( \tau \right) +h_{0}\left( \tau \right)
h_{i}\left( \tau \right) -2b
\end{array}
\right\} }.  \notag \\
&&  \label{electric-3}
\end{eqnarray}
It is worth pointing out that the $D0-D4$ configuration supports \textit{%
axion-free} non-BPS $Z\neq 0$ attractor flow(s); when considering the
\textit{near-horizon limit}, and thus the critical, charge-dependent values
of the moduli, this is consistent with the analysis performed in \cite
{TT,TT2,Ceresole}. An \textit{axion-free} attractor flow solution of Eqs. (%
\ref{electric-1})-(\ref{electric-3}) can be obtained \textit{e.g.} by
putting
\begin{eqnarray}
\alpha _{i} &=&0~\forall i=1,2,3;  \label{e-1-tris} \\
b &=&0,  \label{e-1-tris-2}
\end{eqnarray}
and it reads as follows:
\begin{eqnarray}
&&exp\left[ -4U_{non-BPS,Z\neq 0,axion-free}\left( \tau \right) \right]
=h_{0}\left( \tau \right) h_{1}\left( \tau \right) h_{2}\left( \tau \right)
h_{3}\left( \tau \right) ;  \notag  \label{night!-4} \\
\ &&\mathcal{X}_{non-BPS,Z\neq 0,axion-free}^{i}\left( \tau \right) =0;
\notag  \label{night!-5} \\
&&\mathcal{Y}_{non-BPS,Z\neq 0,axion-free}^{i}\left( \tau \right)
=4\varsigma _{i}\frac{\sqrt{h_{0}\left( \tau \right) h_{1}\left( \tau
\right) h_{2}\left( \tau \right) h_{3}\left( \tau \right) }}{4h_{j}\left(
\tau \right) h_{l}\left( \tau \right) }.  \label{night!-6}
\end{eqnarray}
Furthermore, within the additional assumption (\ref{e-1-tris}), Eqs. (\ref
{electric-1})-(\ref{electric-3}) yield the solution obtained in \cite{GLS-1}.

Always considering a framework in which the assumption (\ref{e-1-tris})
holds true, Eqs. (\ref{D0D2D4D6fake}) yields that the non-BPS $Z\neq 0$
\textit{fake superpotential} in the $D0-D4$ configuration has the following
expression:
\begin{gather}
\left. \mathcal{W}_{non-BPS,Z\neq 0}\right| _{\alpha _{i}=0~\forall i}\left(
\mathcal{Z},\overline{\mathcal{Z}},\Gamma _{D0-D4}\right) =e^{K/2}\cdot
\notag \\
\cdot \left[ -q_{0}+\frac{p^{1}}{2}\left( \mathcal{Z}^{2}\overline{\mathcal{Z%
}}^{3}+\mathcal{Z}^{3}\overline{\mathcal{Z}}^{2}\right) +\frac{p^{2}}{2}%
\left( \mathcal{Z}^{1}\overline{\mathcal{Z}}^{3}+\mathcal{Z}^{3}\overline{%
\mathcal{Z}}^{1}\right) +\frac{p^{3}}{2}\left( \mathcal{Z}^{2}\overline{%
\mathcal{Z}}^{1}+\mathcal{Z}^{1}\overline{\mathcal{Z}}^{2}\right) \right] .
\label{D0D4fake}
\end{gather}
The \textit{axion-free} version of such a \textit{fake superpotential} (%
\textit{e.g.} pertaining to the solution (\ref{night!-4})-(\ref{night!-6}))
reads as follows:
\begin{equation}
\mathcal{W}_{non-BPS,Z\neq 0,axion-free}\left( \mathcal{Y},\Gamma
_{D0-D4}\right) =\frac{1}{2\sqrt{2}}\frac{\left[ -q_{0}+p^{1}\mathcal{Y}^{2}%
\mathcal{Y}^{3}+p^{2}\mathcal{Y}^{1}\mathcal{Y}^{3}+p^{3}\mathcal{Y}^{1}%
\mathcal{Y}^{2}\right] }{\sqrt{\mathcal{Y}^{1}\mathcal{Y}^{2}\mathcal{Y}^{3}}%
}.  \label{D0D4fake-axion-free}
\end{equation}

The existence of a \textit{first order formalism} in the non-BPS $Z\neq 0$%
-supporting (branch of the) $D0-D4$ configuration of the $stu$ model, based
on the \textit{fake superpotential} given by Eq. (\ref{D0D4fake}), gives a
simple explanation of the \textit{integrability} of the equations of motion
of scalars, answering to the question raised in Appendix A of \cite{GLS-1}.

Now, as done above for the $D0-D6$ and $D0-D2-D4-D6$ configurations, by
exploiting the \textit{first order formalism} for $d=4$ extremal BHs, one
can compute the relevant BH parameters, such as the \textit{ADM mass} and
the \textit{covariant scalar charges}, starting from the \textit{fake
superpotential} $\left. \mathcal{W}_{non-BPS,Z\neq 0}\right| _{\alpha
_{i}=0~\forall i}$ given by Eq. (\ref{D0D4fake}).

Concerning the \textit{ADM mass}, by recalling Eq. (\ref{d-1}) and using Eq.
(\ref{D0D4fake}) one obtains an explicit expression which, after introducing
suitable \textit{dressed charges} (see Eq. (\ref{f-1})) and putting (see Eq.
(\ref{e-1}))
\begin{equation}
B^{1}=B^{2}=B^{3}=B,  \label{e-1-bis}
\end{equation}
is given by Eq. (4.6) of \cite{GLS-1}, which we report here for
completeness' sake:
\begin{gather}
\left. M_{ADM,non-BPS,\,Z\neq 0}\right| _{\alpha _{i}=0~\forall i}\left(
\mathcal{Z}_{\infty },\overline{\mathcal{Z}}_{\infty },\Gamma
_{D0-D4}\right) =  \notag \\
=lim_{\tau \rightarrow 0^{-}}\left. \mathcal{W}_{non-BPS,Z\neq 0}\right|
_{\alpha _{i}=0~\forall i}\left( \mathcal{Z}\left( \tau \right) ,\overline{%
\mathcal{Z}}\left( \tau \right) ,\Gamma _{D0-D4}\right) =  \notag \\
=\frac{1}{2\sqrt{2}}\left[ \left| Q_{0}\right| +\left( 1+B^{2}\right)
\sum_{i}P^{i}\right] ,  \label{i-1}
\end{gather}
where the \textit{dressed charges} are defined as follows (no summation on
repeated indices; notice the different definitions with respect to the $%
D0-D6 $ configuration, whose \textit{dressed charges} are given by Eq. (\ref
{e-1})):
\begin{equation}
Q_{0}\equiv \frac{q_{0}}{\sqrt{\mathcal{Y}_{\infty }^{1}\mathcal{Y}_{\infty
}^{2}\mathcal{Y}_{\infty }^{2}}},~~P^{i}\equiv \frac{\sqrt{\mathcal{Y}%
_{\infty }^{1}\mathcal{Y}_{\infty }^{2}\mathcal{Y}_{\infty }^{2}}}{\mathcal{Y%
}_{\infty }^{i}}\,p^{i}.  \label{f-1}
\end{equation}

By recalling Eq. (\ref{d-2}) and using Eq. (\ref{D0D4fake}), one can compute
the \textit{covariant scalar charges} of the non-BPS $Z\neq 0$ attractor
flow in the $D0-D4$ configuration. Within the simplifying assumptions (\ref
{e-1-tris}) and (\ref{e-1-bis}), one obtains the following explicit
expressions ($i\neq j\neq l$, no sum on repeated indices):
\begin{gather}
\Sigma _{\mathcal{X},i,non-BPS,Z\neq 0}\left( \mathcal{X}_{\infty },\mathcal{%
Y}_{\infty },\Gamma _{D0-D4}\right) \equiv  \notag \\
\equiv lim_{\tau \rightarrow 0^{-}}\left( \frac{\partial \left. \mathcal{W}%
_{non-BPS,Z\neq 0}\right| _{\alpha _{m}=0~\forall m}}{\partial \mathcal{X}%
^{i}}\right) \left( \mathcal{Z}\left( \tau \right) ,\overline{\mathcal{Z}}%
\left( \tau \right) ,\Gamma _{D0-D4}\right) =  \notag \\
=\sqrt{2}\,\mathcal{X}_{\infty }^{i}\,(P^{j}+P^{l});
\end{gather}
\begin{gather}
\Sigma _{\mathcal{Y},i,non-BPS,Z\neq 0}\left( \mathcal{X}_{\infty },\mathcal{%
Y}_{\infty },\Gamma _{D0-D4}\right) \equiv  \notag \\
\equiv lim_{\tau \rightarrow 0^{-}}\left( \frac{\partial \left. \mathcal{W}%
_{non-BPS,Z\neq 0}\right| _{\alpha _{m}=0~\forall m}}{\partial \mathcal{Y}%
^{i}}\right) \left( \mathcal{Z}\left( \tau \right) ,\overline{\mathcal{Z}}%
\left( \tau \right) ,\Gamma _{D0-D4}\right) =  \notag \\
=\frac{\mathcal{Y}_{\infty }^{i}}{\sqrt{2}}\left( -\left| Q_{0}\right|
-2P^{i}+(1-B^{2})\sum_{k}P^{k}\right) ,
\end{gather}
where the split in \textit{axionic scalar charges} $\Sigma _{\mathcal{X},i}$
and \textit{dilatonic scalar charges} $\Sigma _{\mathcal{Y},i}$ was
performed.\smallskip

It is here worth computing the difference between the squared non-BPS $Z\neq
0$ \textit{fake superpotential} and the squared absolute value of the $%
\mathcal{N}=2$, $d=4$ central charge along the non-BPS $Z\neq 0$ attractor
flow. This amounts to computing the difference generalizing the \textit{BPS
bound} \cite{gibbons2} to the whole attractor flow (in the non-BPS $Z\neq 0$%
-supporting branch of the \textit{magnetic} charge configuration):
\begin{equation}
\Delta \left( \mathcal{Y},\Gamma \right) \equiv \mathcal{W}_{non-BPS,Z\neq
0}^{2}-\left| Z\right| ^{2}=\frac{|q_{0}|}{2}\left( \frac{p^{1}}{\mathcal{Y}%
^{1}}+\frac{p^{2}}{\mathcal{Y}^{2}}+\frac{p^{3}}{\mathcal{Y}^{3}}\right) >0.
\label{gap}
\end{equation}
$\Delta $ is dilaton-dependent and strictly positive all along the non-BPS $%
Z\neq 0$ attractor flow. At the infinity, by using the \textit{dressed
charges} defined by Eq. (\ref{f-1}), the result given by Eq. (4.8) of \cite
{GLS-1} is recovered. Thus, the \textit{BPS bound} \cite{gibbons2} holds not
only at the BH event horizon ($r=r_{H}$), but actually (in a
dilaton-dependent way) all along the non-BPS $Z\neq 0$ attractor flow (%
\textit{i.e.} $\forall r\geqslant r_{H}$). \medskip

Of course, by relaxing the simplifying conditions (\ref{e-1-tris}) and/or (%
\ref{e-1-bis}), \textit{i.e.} by considering non-vanishing $\alpha _{i}$s
(constrained by Eq. (\ref{b-2-tris})) and/or different, $i$-indexed $B$%
\textit{-fields}, a much richer situation arises, but the main features of
the framework, outlined above, are left unchanged.

\subsection{\label{D2-D6}\textit{Electric} ($D2-D6$)}

The configuration $D2-D6$ (also named \textit{electric}) of the $stu$ model
has been previously treated in \cite{K2-bis} and \cite{Cai-Pang}.
Analogously to what happens in the $D0-D4$ (\textit{magnetic})
configuration, in this case the quantities of the $U$-duality transformation
(\ref{boss-1})-(\ref{boss-3}) along $\mathcal{O}_{non-BPS,Z\neq 0}$ defined
by Eqs. (\ref{b-2})-(\ref{b-4}) undergo a major simplification (the prime
denotes the charges in the considered configuration):
\begin{equation}
\xi =-\xi _{0};~~\varsigma _{i}=\varrho _{i}=-\sqrt{\frac{\frac{1}{2}%
s_{ijk}q_{j}^{\prime }q_{k}^{\prime }}{p^{\prime 0}q_{i}^{\prime }}}.
\end{equation}

Correspondingly, the non-BPS $Z\neq 0$ attractor flow (\ref{sol})-(\ref
{sol-2}) acquires the following form (as above, the moduli are here denoted
as $\mathcal{Z}^{i}\equiv \mathcal{X}^{i}-i\mathcal{Y}^{i}$; $i\neq j\neq l$
and no sum on repeated $i-$indices throughout):
\begin{eqnarray}
&&exp\left[ -4U_{non-BPS,Z\neq 0}\left( \tau \right) \right] =h_{0}\left(
\tau \right) h_{1}\left( \tau \right) h_{2}\left( \tau \right) h_{3}\left(
\tau \right) -b^{2};  \label{magnetic-1} \\
&&  \notag \\
\ &&\mathcal{X}_{non-BPS,Z\neq 0}^{i}\left( \tau \right) =\varsigma _{i}\cdot
\notag \\
&&  \notag \\
&&\cdot \frac{e^{2\alpha _{i}}\left[ h_{j}\left( \tau \right) h_{l}\left(
\tau \right) +h_{0}\left( \tau \right) h_{i}\left( \tau \right) +2b\right] -%
\left[ h_{j}\left( \tau \right) h_{l}\left( \tau \right) +h_{0}\left( \tau
\right) h_{i}\left( \tau \right) -2b\right] }{\left\{
\begin{array}{l}
e^{2\alpha _{i}}\left[ h_{j}\left( \tau \right) h_{l}\left( \tau \right)
+h_{0}\left( \tau \right) h_{i}\left( \tau \right) +2b\right] -2e^{\alpha
_{i}}\left[ h_{j}\left( \tau \right) h_{l}\left( \tau \right) -h_{0}\left(
\tau \right) h_{i}\left( \tau \right) \right] + \\
\\
+h_{j}\left( \tau \right) h_{l}\left( \tau \right) +h_{0}\left( \tau \right)
h_{i}\left( \tau \right) -2b
\end{array}
\right\} };  \notag \\
&&  \label{magnetic-2} \\
&&  \notag \\
&&\mathcal{Y}_{non-BPS,Z\neq 0}^{i}\left( \tau \right) =-4\varsigma _{i}\cdot
\notag \\
&&  \notag \\
&&\cdot \frac{e^{\alpha _{i}}exp\left[ -2U_{non-BPS,Z\neq 0}\left( \tau
\right) \right] }{\left\{
\begin{array}{l}
e^{2\alpha _{i}}\left[ h_{j}\left( \tau \right) h_{l}\left( \tau \right)
+h_{0}\left( \tau \right) h_{i}\left( \tau \right) +2b\right] -2e^{\alpha
_{i}}\left[ h_{j}\left( \tau \right) h_{l}\left( \tau \right) -h_{0}\left(
\tau \right) h_{i}\left( \tau \right) \right] + \\
\\
+h_{j}\left( \tau \right) h_{l}\left( \tau \right) +h_{0}\left( \tau \right)
h_{i}\left( \tau \right) -2b
\end{array}
\right\} }.  \notag \\
&&  \label{magnetic-3}
\end{eqnarray}
It is worth pointing out that the $D2-D6$ configuration supports \textit{%
axion-free} non-BPS $Z\neq 0$ attractor flow(s); when considering the
\textit{near-horizon limit}, and thus the critical, charge-dependent values
of the moduli, this is consistent with the analysis performed in \cite
{TT,TT2,Ceresole}. An \textit{axion-free} attractor flow solution of Eqs. (%
\ref{magnetic-1})-(\ref{magnetic-3}) can be obtained \textit{e.g.} by
assuming the conditions given by Eqs. (\ref{e-1-tris}) and (\ref{e-1-tris-2}%
), and it reads as follows:
\begin{eqnarray}
&&exp\left[ -4U_{non-BPS,Z\neq 0,axion-free}\left( \tau \right) \right]
=h_{0}\left( \tau \right) h_{1}\left( \tau \right) h_{2}\left( \tau \right)
h_{3}\left( \tau \right) ;  \label{night!!-1} \\
\ &&\mathcal{X}_{non-BPS,Z\neq 0,axion-free}^{i}\left( \tau \right) =0;
\notag  \label{night!!-2} \\
&&\mathcal{Y}_{non-BPS,Z\neq 0,axion-free}^{i}\left( \tau \right) =-\frac{%
\varsigma _{i}}{h_{i}\left( \tau \right) }\sqrt{\frac{h_{1}\left( \tau
\right) h_{2}\left( \tau \right) h_{3}\left( \tau \right) }{h_{0}\left( \tau
\right) }}.  \label{night!!-3}
\end{eqnarray}
As done for the \textit{magnetic} configuration, in order to further
simplify Eqs. (\ref{magnetic-1})-(\ref{magnetic-3}) and (\ref{night!!-1})-(%
\ref{night!!-3}), one can consider the particular case constrained by Eq. (%
\ref{e-1-tris}). Within such an additional assumption, the solution obtained
in \cite{Cai-Pang}, generalizing the one of \cite{K2-bis}, is recovered.

Furthermore, within the simplifying assumption (\ref{e-1-tris}), Eq. (\ref
{D0D2D4D6fake}) yields that the non-BPS $Z\neq 0$ \textit{fake superpotential%
} in the $D2-D6$ configuration has the following expression:
\begin{gather}
\left. \mathcal{W}_{non-BPS,Z\neq 0}\right| _{\alpha _{i}=0~\forall i}\left(
\mathcal{Z},\overline{\mathcal{Z}},\Gamma _{D2-D6}\right) =e^{K/2}\left|
\mathcal{Z}^{1}\right| \left| \mathcal{Z}^{2}\right| \left| \mathcal{Z}%
^{3}\right| \cdot  \notag \\
\cdot \left[ p^{\prime }{}^{0}+\frac{q_{1}^{\prime }}{2}\frac{\left(
\mathcal{Z}^{2}\overline{\mathcal{Z}}^{3}+\mathcal{Z}^{3}\overline{\mathcal{Z%
}}^{2}\right) }{\left| \mathcal{Z}^{2}\right| ^{2}\left| \mathcal{Z}%
^{3}\right| ^{2}}+\frac{q_{2}^{\prime }}{2}\frac{\left( \mathcal{Z}^{1}%
\overline{\mathcal{Z}}^{3}+\mathcal{Z}^{3}\overline{\mathcal{Z}}^{1}\right)
}{\left| \mathcal{Z}^{1}\right| ^{2}\left| \mathcal{Z}^{3}\right| ^{2}}+%
\frac{q_{3}^{\prime }}{2}\frac{\left( \mathcal{Z}^{2}\overline{\mathcal{Z}}%
^{1}+\mathcal{Z}^{1}\overline{\mathcal{Z}}^{2}\right) }{\left| \mathcal{Z}%
^{1}\right| ^{2}\left| \mathcal{Z}^{2}\right| ^{2}}\right] .
\label{D2D6fake}
\end{gather}
The \textit{axion-free} version of such a \textit{fake superpotential} (%
\textit{e.g.} pertaining to the solution (\ref{night!!-1})-(\ref{night!!-3}%
)) reads as follows:
\begin{equation}
\left. \mathcal{W}_{non-BPS,Z\neq 0}\right| _{\alpha _{i}=0~\forall
i,~axion-free}\left( \mathcal{Z},\overline{\mathcal{Z}},\Gamma
_{D2-D6}\right) =\frac{1}{2\sqrt{2}}\sqrt{\mathcal{Y}^{1}\mathcal{Y}^{2}%
\mathcal{Y}^{3}}\left[ p^{\prime }{}^{0}+\frac{q_{1}^{\prime }}{\mathcal{Y}%
^{2}\mathcal{Y}^{3}}+\frac{q_{2}^{\prime }}{\mathcal{Y}^{1}\mathcal{Y}^{3}}+%
\frac{q_{3}^{\prime }}{\mathcal{Y}^{1}\mathcal{Y}^{2}}\right] ,
\label{D2D6fake-axion-free}
\end{equation}
coinciding with the \textit{fake superpotential} given by Eq. (4-20) of \cite
{Cer-Dal-1}.

The existence of a \textit{first order formalism} in the non-BPS $Z\neq 0$%
-supporting (branch of the) $D2-D6$ configuration of the $stu$ model, based
on the \textit{fake superpotential} given by Eq. (\ref{D2D6fake}), gives a
explanation of the \textit{integrability} of the equations of motion of
scalars supported by the \textit{electric} configuration (see the treatment
of \cite{Cai-Pang}).

Let us now consider the \textit{``}$t^{3}$\textit{-degeneration''} of the $%
stu$ model, in which all charges and moduli are equal, insensitive to $i$%
-index; in the considered configuration this amounts to putting
\begin{equation}
\mathcal{Z}^{1}=\mathcal{Z}^{2}=\mathcal{Z}^{3}=\mathcal{Z},\qquad
q_{1}^{\prime }=q_{2}^{\prime }=q_{3}^{\prime }=q^{\prime }/3
\label{night-1}
\end{equation}
(see the treatment in Sect. 5 of \cite{BMOS-1}). By doing so, Eq. (\ref
{D2D6fake}) yields the non-BPS $Z\neq 0$ \textit{fake superpotential} given
by Eq. (5.5) of \cite{Cer-Dal-1}, which we report here for completeness'
sake:
\begin{equation}
\left. \mathcal{W}_{non-BPS,Z\neq 0}\right| _{\alpha _{i}=0~\forall
i,~t^{3}-deg.}=\left| \frac{p^{\prime }{}^{0}\left( \mathcal{Z}\right) ^{2}%
\overline{\mathcal{Z}}+q^{\prime }\mathcal{Z}}{\sqrt{-i(\mathcal{Z}-%
\overline{\mathcal{Z}})^{3}}}\right| .  \label{D2D6fake-t^3-deg}
\end{equation}
The \textit{axion-free} version of such a \textit{fake superpotential} (%
\textit{e.g.} pertaining to \textit{``}$t^{3}$\textit{-degeneration''} of
the solution (\ref{night!!-1})-(\ref{night!!-3})) reads as follows:
\begin{equation}
\mathcal{W}_{non-BPS,Z\neq 0,axion-free,~t^{3}-deg.}=\frac{1}{2\sqrt{2}}%
\left| \frac{p^{\prime }{}^{0}\mathcal{Y}^{3}+q^{\prime }\mathcal{Y}}{\sqrt{%
\mathcal{Y}^{3}}}\right| .  \label{D2D6fake-t^3-deg-axion-free}
\end{equation}

Now, as done above for the $D0-D6$, $D0-D2-D4-D6$ and $D0-D4$
configurations, by exploiting the \textit{first order formalism} for $d=4$
extremal BHs, one can compute the relevant BH parameters, such as the
\textit{ADM mass} and the \textit{covariant scalar charges}, starting from
the \textit{fake superpotential} $\left. \mathcal{W}_{non-BPS,Z\neq
0}\right| _{\alpha _{i}=0~\forall i}$ given by Eq. (\ref{D2D6fake}).

Concerning the \textit{ADM mass}, by recalling Eq. (\ref{d-1}) and using
Eqs. (\ref{D2D6fake}), (\ref{e-1}) and (\ref{e-1-bis}), one obtains an
explicit expression which, after introducing suitable \textit{dressed charges%
} (see Eq. (\ref{l-1})), reads as follows:
\begin{gather}
\left. M_{ADM,non-BPS,\,Z\neq 0}\right| _{\alpha _{i}=0~\forall i}\left(
\mathcal{Z}_{\infty },\overline{\mathcal{Z}}_{\infty },\Gamma
_{D2-D6}\right) =  \notag \\
=lim_{\tau \rightarrow 0^{-}}\left. \mathcal{W}_{non-BPS,Z\neq 0}\right|
_{\alpha _{i}=0~\forall i}\left( \mathcal{Z}\left( \tau \right) ,\overline{%
\mathcal{Z}}\left( \tau \right) ,\Gamma _{D2-D6}\right) =  \notag \\
=\frac{\sqrt{1+B^{2}}}{2\sqrt{2}}\left[ (1+B^{2})P^{\prime
}{}^{0}+\sum_{i}Q_{i}^{\prime }\right] ,  \label{m-1}
\end{gather}
where the \textit{dressed charges} are defined as follows (no summation on
repeated indices; notice the different definitions with respect to the $%
D0-D6 $ and $D0-D4$ configurations, whose \textit{dressed charges} are given
by Eqs. (\ref{e-1}) and (\ref{f-1}), respectively):
\begin{equation}
P^{\prime }{}^{0}\equiv p^{\prime }{}^{0}\sqrt{\mathcal{Y}_{\infty }^{1}%
\mathcal{Y}_{\infty }^{2}\mathcal{Y}_{\infty }^{2}},\quad Q_{i}^{\prime
}\equiv \frac{\mathcal{Y}_{\infty }^{i}}{\sqrt{\mathcal{Y}_{\infty }^{1}%
\mathcal{Y}_{\infty }^{2}\mathcal{Y}_{\infty }^{2}}}\,q_{i}^{\prime }.
\label{l-1}
\end{equation}
Up to redefinition of the \textit{dressed charges}, Eq. (\ref{m-1}) is
nothing but Eq. (5.2) of \cite{Cai-Pang}.

By recalling Eq. (\ref{d-2}) and using Eq. (\ref{D2D6fake}), one can compute
the \textit{covariant scalar charges} of the non-BPS $Z\neq 0$ attractor
flow in the $D2-D6$ configuration. Within the simplifying assumptions (\ref
{e-1-tris}) and (\ref{e-1-bis}), one obtains the following explicit
expressions (no sum on repeated indices):
\begin{gather}
\Sigma _{\mathcal{X},i,non-BPS,Z\neq 0}\left( \mathcal{X}_{\infty },\mathcal{%
Y}_{\infty },\Gamma _{D2-D6}\right) \equiv  \notag \\
\equiv lim_{\tau \rightarrow 0^{-}}\left( \frac{\partial \left. \mathcal{W}%
_{non-BPS,Z\neq 0}\right| _{\alpha _{m}=0~\forall m}}{\partial \mathcal{X}%
^{i}}\right) \left( \mathcal{Z}\left( \tau \right) ,\overline{\mathcal{Z}}%
\left( \tau \right) ,\Gamma _{D2-D6}\right) =  \notag \\
=\sqrt{2}\,\frac{\mathcal{X}_{\infty }^{i}}{\sqrt{1+B^{2}}}\,\left[
(1+B^{2})P^{\prime }{}^{0}+Q_{i}^{\prime }\right] ;
\end{gather}
\begin{gather}
\Sigma _{\mathcal{Y},i,non-BPS,Z\neq 0}\left( \mathcal{X}_{\infty },\mathcal{%
Y}_{\infty },\Gamma _{D2-D6}\right) \equiv  \notag \\
\equiv lim_{\tau \rightarrow 0^{-}}\left( \frac{\partial \left. \mathcal{W}%
_{non-BPS,Z\neq 0}\right| _{\alpha _{m}=0~\forall m}}{\partial \mathcal{Y}%
^{i}}\right) \left( \mathcal{Z}\left( \tau \right) ,\overline{\mathcal{Z}}%
\left( \tau \right) ,\Gamma _{D2-D6}\right) =  \notag \\
=\frac{\mathcal{Y}_{\infty }^{i}}{\sqrt{2}\,\sqrt{1+B^{2}}}\left[ \left(
B^{4}-1\right) P^{\prime }{}^{0}-2Q_{i}^{\prime }+\left( 1+B^{2}\right)
^{2}\sum_{j}Q_{j}^{\prime }\right] ,
\end{gather}
where, as for the $D0-D4$ configuration, the split in \textit{axionic scalar
charges} $\Sigma _{\mathcal{X},i}$ and \textit{dilatonic scalar charges} $%
\Sigma _{\mathcal{Y},i}$ was performed.\smallskip

As done for the \textit{magnetic} configuration in Subsect. \ref{D0-D4},
also for \textit{electric} configuration it is worth computing the
difference between the squared non-BPS $Z\neq 0$ \textit{fake superpotential}
and the squared absolute value of the $\mathcal{N}=2$, $d=4$ central charge
along the non-BPS $Z\neq 0$ attractor flow:
\begin{equation}
\Delta \left( \mathcal{X},\mathcal{Y},\Gamma \right) \equiv \mathcal{W}%
_{non-BPS,Z\neq 0}^{2}-\left| Z\right| ^{2}=\frac{p^{\prime 0}}{2}\left(
q_{1}^{\prime }\frac{(\mathcal{X}^{1})^{2}+(\mathcal{Y}^{1})^{2}}{\mathcal{Y}%
^{1}}+q_{2}^{\prime }\frac{(\mathcal{X}^{2})^{2}+(\mathcal{Y}^{2})^{2}}{%
\mathcal{Y}^{2}}+q_{3}^{\prime }\frac{(\mathcal{X}^{3})^{2}+(\mathcal{Y}%
^{3})^{2}}{\mathcal{Y}^{3}}\right) >0.  \label{gap1}
\end{equation}
Differently from what happens for the \textit{magnetic} configuration, for
\textit{electric} configuration $\Delta $ does depend also on axions, but
nevertheless it is still strictly positive all along the non-BPS $Z\neq 0$
attractor flow. At infinity, by using the \textit{dressed charges} defined
by Eq. (\ref{l-1}), the following result is achieved:
\begin{equation}
\Delta \left( \mathcal{X}_{\infty },\mathcal{Y}_{\infty },\Gamma \right) =%
\frac{P^{\prime 0}}{2}(1+B^{2})\sum_{i}Q_{i}^{\prime }.
\end{equation}
Thus, the \textit{BPS bound} \cite{gibbons2} holds not only at the BH event
horizon ($r=r_{H}$), but actually (in a scalar-dependent way) all along the
non-BPS $Z\neq 0$ attractor flow (\textit{i.e.} $\forall r\geqslant r_{H}$).
\medskip

Of course, by relaxing the simplifying conditions (\ref{e-1-tris}) and/or (%
\ref{e-1-bis}), \textit{i.e.} by considering non-vanishing $\alpha _{i}$s
(constrained by Eq. (\ref{b-2-tris})) and/or different, $i$-indexed $B$%
\textit{-fields}, a much richer situation arises, but the main features of
the framework, outlined above, are left unchanged.\medskip

By noticing that the $D0-D4$ (\textit{magnetic}) and $D2-D6$ (\textit{%
electric}) configurations are reciprocally \textit{dual} in $d=4$ and
recalling the treatment of Subsect. \ref{U-Duality-Transf}, it is worth
computing the matrices $M_{i,D0-D4\longrightarrow D2-D6}$ representing the $%
U $-duality transformation along the charge orbit $\mathcal{O}%
_{non-BPS,Z\neq 0}$ which connects (the non-BPS $Z\neq 0$-supporting \textit{%
branches} of) such two charge configurations. In order to do this, we
exploit the treatment given in Subsect. \ref{U-Duality-Transf}, by
performing the following steps:
\begin{equation}
\overset{\left( q_{0},p^{i}\right) }{D0-D4}\longrightarrow \overset{\left(
q,p\right) }{D0-D6}\longrightarrow \overset{\left( q_{i}^{\prime },p^{\prime
0}\right) }{D2-D6}.
\end{equation}
For the step $D0-D4\longrightarrow D0-D6$, we consider $M_{i}^{-1}$ given by
Eq. (\ref{b-1}), along with the definitions (\ref{b-2})-(\ref{b-4})
specified for the configuration $D0-D4$, obtaining $M_{i,D0-D4%
\longrightarrow D0-D6}^{-1}$. Thence, for the the step $D0-D6\longrightarrow
D2-D6$, we take $M_{i}$ given by Eq. (\ref{b-1}), along with the definitions
(\ref{b-2})-(\ref{b-4}) specified for the configuration $D2-D6$, obtaining $%
M_{i,D0-D6\longrightarrow D2-D6}$. Thus (no sum on repeated index $i=1,2,3$
throughout; also recall Eq. (\ref{boss-1})):
\begin{eqnarray}
\left( M_{i,D0-D4\longrightarrow D2-D6}\right) _{a^{\prime }}^{~b^{\prime }}
&=&\left( M_{i,D0-D6\longrightarrow D2-D6}\right) _{a^{\prime }}^{~a}\left(
M_{i,D0-D4\longrightarrow D0-D6}^{-1}\right) _{a}^{~b^{\prime }}=  \notag \\
&&  \notag \\
&=&
\begin{pmatrix}
0 & -\sqrt[4]{{-\frac{q_{0}p^{i}\,s_{ijk}q_{j}^{\prime }q_{k}^{\prime }}{%
p^{\prime }{}^{0}q_{i}^{\prime }\,s_{ijk}p^{j}p^{k}}}} \\
\sqrt[4]{{-\frac{p^{\prime }{}^{0}q_{i}^{\prime }\,s_{ijk}p^{j}p^{k}}{%
q_{0}p^{i}\,s_{ijk}q_{j}^{\prime }q_{k}^{\prime }}}} & 0
\end{pmatrix}
.
\end{eqnarray}
Consequently, by recalling Eqs. (\ref{boss-3}) and (\ref{b-5}) one can
directly relate the non-BPS $Z\neq 0$ attractor flows $\mathcal{Z}%
_{non-BPS,Z\neq 0,D2-D6}^{i}\left( \tau \right) $ and $\mathcal{Z}%
_{non-BPS,Z\neq 0,D0-D4}^{i}\left( \tau \right) $ (respectively given by
Eqs. (\ref{magnetic-1})-(\ref{magnetic-3}) and (\ref{electric-1})-(\ref
{electric-3}); recall that $\mathcal{Z}^{i}\left( \tau \right) =\mathcal{X}%
^{i}\left( \tau \right) -i\mathcal{Y}^{i}\left( \tau \right) $) by the
following expression, explicitly showing the \textit{duality} between the $%
D0-D4$ (\textit{magnetic}) and $D2-D6$ (\textit{electric}) configurations in
$d=4$:
\begin{equation}
\mathcal{Z}_{non-BPS,Z\neq 0,D2-D6}^{i}\left( \tau \right) =-\sqrt{-\frac{%
q_{0}\,p^{i}\,s_{ijk}q\prime _{j}q\prime _{k}}{p\prime ^{0}\,q\prime
_{i}\,s_{ijk}p^{j}p^{k}}}\,\frac{1}{\mathcal{Z}_{non-BPS,Z\neq
0,D0-D4}^{i}\left( \tau \right) }.
\end{equation}

\subsection{\label{D0-D2-D4}$D0-D2-D4$}

The configuration $D0-D2-D4$ of the $stu$ model has been previously treated
in \cite{Cai-Pang}. In this case, the quantities of the $U$-duality
transformation (\ref{boss-1})-(\ref{boss-3}) along $\mathcal{O}%
_{non-BPS,Z\neq 0}$ defined by Eqs. (\ref{b-2})-(\ref{b-4}) have the
following form (no summation on repeated index $i=1,2,3$ throughout):
\begin{equation}
\xi =\xi _{0};~~\varsigma _{i}=\frac{\sqrt{-\mathcal{I}_{4}}%
+p^{l}q_{l}-2p^{i}q_{i}}{s_{ijk}p^{j}p^{k}};~~\varrho _{i}=\frac{\sqrt{-%
\mathcal{I}_{4}}-p^{l}q_{l}+2p^{i}q_{i}}{s_{ijk}p^{j}p^{k}}.
\end{equation}
Within the additional assumption (\ref{e-1-tris}) (considered for
simplicity' sake), the non-BPS $Z\neq 0$ attractor flow (\ref{sol})-(\ref
{sol-2}) correspondingly acquires the following form (as above, the moduli
are here denoted as $\mathcal{Z}^{i}\equiv \mathcal{X}^{i}-i\mathcal{Y}^{i}$%
; $i\neq j\neq l$, and no sum on repeated $i-$indices throughout):
\begin{eqnarray}
&&exp\left[ -4U_{non-BPS,Z\neq 0}\left( \tau \right) \right] =h_{0}\left(
\tau \right) h_{1}\left( \tau \right) h_{2}\left( \tau \right) h_{3}\left(
\tau \right) -b^{2};  \label{D0D2D4-1} \\
&&  \notag \\
\ &&\mathcal{X}_{non-BPS,Z\neq 0}^{i}\left( \tau \right) =\frac{\sqrt{-%
\mathcal{I}_{4}}}{s_{ikm}p^{k}p^{m}}\frac{b}{h_{j}\left( \tau \right)
h_{l}\left( \tau \right) }+\frac{p^{n}q_{n}-2p^{i}q_{i}}{s_{ikm}p^{k}p^{m}};
\label{D0D2D4-2} \\
&&  \notag \\
&&\mathcal{Y}_{non-BPS,Z\neq 0}^{i}\left( \tau \right) =\frac{\sqrt{-%
\mathcal{I}_{4}}}{s_{ikm}p^{k}p^{m}}\frac{exp\left[ -2U_{non-BPS,Z\neq
0}\left( \tau \right) \right] }{h_{j}\left( \tau \right) h_{l}\left( \tau
\right) }.  \label{D0D2D4-3}
\end{eqnarray}
This is nothing but the solution obtained in \cite{Cai-Pang}.

It is worth pointing out that, as the general case $D0-D2-D4-D6$ (see
Subsect. \ref{D0-D2-D4-D6}), the $D0-D2-D4$ configuration does \textit{not}
support \textit{axion-free} non-BPS $Z\neq 0$ attractor flow(s); when
considering the \textit{near-horizon limit}, and thus the critical,
charge-dependent values of the moduli, this is consistent with the analysis
performed in \cite{TT,TT2,Ceresole}.

Furthermore, always within the simplifying assumption (\ref{e-1-tris}), Eq. (%
\ref{D0D2D4D6fake}) yields that the non-BPS $Z\neq 0$ \textit{fake
superpotential} in the $D0-D2-D4$ configuration has the following
expression:
\begin{gather}
\left. \mathcal{W}_{non-BPS,Z\neq 0}\right| _{\alpha _{i}=0~\forall i}\left(
\mathcal{Z},\overline{\mathcal{Z}},\Gamma _{D0-D2-D4}\right) =  \notag \\
=e^{K/2}\left[ -q_{0}-\frac{q_{1}}{2}\left( \mathcal{Z}^{1}+\overline{%
\mathcal{Z}}^{1}\right) -\frac{q_{2}}{2}\left( \mathcal{Z}^{2}+\overline{%
\mathcal{Z}}^{2}\right) -\frac{q_{3}}{2}\left( \mathcal{Z}^{3}+\overline{%
\mathcal{Z}}^{3}\right) +\right.  \notag \\
\left. +\frac{p^{1}}{2}\left( \mathcal{Z}^{2}\overline{\mathcal{Z}}^{3}+%
\mathcal{Z}^{3}\overline{\mathcal{Z}}^{2}\right) +\frac{p^{2}}{2}\left(
\mathcal{Z}^{1}\overline{\mathcal{Z}}^{3}+\mathcal{Z}^{3}\overline{\mathcal{Z%
}}^{1}\right) +\frac{p^{3}}{2}\left( \mathcal{Z}^{1}\overline{\mathcal{Z}}%
^{2}+\mathcal{Z}^{2}\overline{\mathcal{Z}}^{1}\right) \right] .
\label{D0D2D4fake}
\end{gather}
The existence of a \textit{first order formalism} in the non-BPS $Z\neq 0$%
-supporting (branch of the) $D0-D2-D4$ configuration of the $stu$ model,
based on the \textit{fake superpotential} given by Eq. (\ref{D0D2D4fake}),
gives a explanation of the \textit{integrability} of the equations of motion
of scalars supported by such a configuration (see the treatment of \cite
{Cai-Pang}).

Now, by exploiting the \textit{first order formalism} for $d=4$ extremal
BHs, one can compute the relevant BH parameters, such as the \textit{ADM mass%
} and the \textit{covariant scalar charges}, starting from the \textit{fake
superpotential} $\left. \mathcal{W}_{non-BPS,Z\neq 0}\right| _{\alpha
_{i}=0~\forall i}$ given by Eq. (\ref{D0D2D4fake}).

Concerning the \textit{ADM mass}, by recalling Eq. (\ref{d-1}) and using Eq.
(\ref{D0D2D4fake}) one obtains the following result:
\begin{gather}
\left. M_{ADM,non-BPS,\,Z\neq 0}\right| _{\alpha _{i}=0~\forall i}\left(
\mathcal{Z}_{\infty },\overline{\mathcal{Z}}_{\infty },\Gamma
_{D0-D2-D4}\right) =  \notag \\
=lim_{\tau \rightarrow 0^{-}}\left. \mathcal{W}_{non-BPS,Z\neq 0}\right|
_{\alpha _{i}=0~\forall i}\left( \mathcal{Z}\left( \tau \right) ,\overline{%
\mathcal{Z}}\left( \tau \right) ,\Gamma _{D0-D2-D4}\right) =  \notag \\
=\frac{1}{2\sqrt{2}}\left[ \left| Q_{0}\right|
-\sum_{i}Q_{i}\,B_{i}+\sum_{i}P^{i}+\sum_{i\neq j\neq k}P^{i}B_{j}B_{k}%
\right] ,  \label{r-1}
\end{gather}
where the \textit{dressed charges} are defined by Eqs. (\ref{f-1}) and (\ref
{l-1}).

By recalling Eq. (\ref{d-2}) and using Eq. (\ref{D0D2D4fake}), one can
compute the \textit{covariant scalar charges} of the non-BPS $Z\neq 0$
attractor flow in the $D0-D2-D4$ configuration. Always within the assumption
(\ref{e-1-tris}), one obtains the following explicit expressions (no sum on
repeated index $i=1,2,3$):
\begin{gather}
\Sigma _{\mathcal{X},i,non-BPS,Z\neq 0}\left( \mathcal{X}_{\infty },\mathcal{%
Y}_{\infty },\Gamma _{D0-D2-D4}\right) \equiv  \notag \\
\equiv lim_{\tau \rightarrow 0^{-}}\left( \frac{\partial \left. \mathcal{W}%
_{non-BPS,Z\neq 0}\right| _{\alpha _{m}=0~\forall m}}{\partial \mathcal{X}%
^{i}}\right) \left( \mathcal{Z}\left( \tau \right) ,\overline{\mathcal{Z}}%
\left( \tau \right) ,\Gamma _{D0-D2-D4}\right) =  \notag \\
=\sqrt{2}\mathcal{Y}_{\infty }^{i}\,(s_{ijk}P^{j}B_{k}-Q_{i});  \label{r-2}
\end{gather}
\begin{gather}
\Sigma _{\mathcal{Y},i,non-BPS,Z\neq 0}\left( \mathcal{X}_{\infty },\mathcal{%
Y}_{\infty },\Gamma _{D0-D2-D4}\right) \equiv  \notag \\
\equiv lim_{\tau \rightarrow 0^{-}}\left( \frac{\partial \left. \mathcal{W}%
_{non-BPS,Z\neq 0}\right| _{\alpha _{m}=0~\forall m}}{\partial \mathcal{Y}%
^{i}}\right) \left( \mathcal{Z}\left( \tau \right) ,\overline{\mathcal{Z}}%
\left( \tau \right) ,\Gamma _{D0-D2-D4}\right) =  \notag \\
=\frac{\mathcal{Y}_{\infty }^{i}}{\sqrt{2}}\left[ -\left| Q_{0}\right|
-2P^{i}+\sum_{j}Q_{j}\,B_{j}+\sum_{j}P^{j}-\sum_{i\neq j\neq
k}P^{i}B_{j}B_{k}\right] ,  \label{r-3}
\end{gather}
where, as above, the split in \textit{axionic scalar charges} $\Sigma _{%
\mathcal{X},i}$ and \textit{dilatonic scalar charges} $\Sigma _{\mathcal{Y}%
,i}$ was performed, and the definition (\ref{e-1}) of $B$\textit{-fields}
was used.\smallskip

As done for the \textit{magnetic} and\textit{\ electric} configurations
(respectively in Subsects. \ref{D0-D4} and \ref{D2-D6}), also for $D0-D2-D4$
configuration it is worth computing the difference between the squared
non-BPS $Z\neq 0$ \textit{fake superpotential} and the squared absolute
value of the $\mathcal{N}=2$, $d=4$ central charge along the non-BPS $Z\neq
0 $ attractor flow. For simplicity's sake, we decide to perform computations
in the \textit{``}$t^{3}$\textit{-degeneration''} of the $stu$ model, in
which all charges and moduli are equal, insensitive to $i$-index; in the
considered configuration this amounts to putting (see the treatment in Sect.
5 of \cite{BMOS-1})
\begin{equation}
\mathcal{Z}^{1}=\mathcal{Z}^{2}=\mathcal{Z}^{3}=\mathcal{Z}\equiv \mathcal{X}%
-i\mathcal{Y},~~q_{1}=q_{2}=q_{3}=q/3,~~p^{1}=p^{2}=p^{3}=p,
\label{t^3-degeneration}
\end{equation}
which generalizes Eq. (\ref{night-1}), also implying the assumption (\ref
{e-1-bis}). Then, one obtains the following result:
\begin{equation}
\Delta \left( \mathcal{Y},\Gamma \right) \equiv \mathcal{W}_{non-BPS,Z\neq
0}^{2}-\left| Z\right| ^{2}=\frac{1}{8\mathcal{Y}}\left(
12|q_{0}|p-q^{2}\right) >0.  \label{gap2}
\end{equation}
In this case, $\Delta $ is strictly positive all along the non-BPS $Z\neq 0$
attractor flow, due to the fact that $\mathcal{I}_{4}$ is strictly negative.
At infinity, by using the \textit{dressed charges} defined by Eqs. (\ref{f-1}%
) and (\ref{l-1}), the following result is achieved:
\begin{equation}
\Delta \left( \mathcal{Y}_{\infty },\Gamma \right) =\frac{1}{8}\left(
12|Q_{0}|P-Q^{2}\right) .
\end{equation}
Thus, the \textit{BPS bound} \cite{gibbons2} is found to hold not only at
the BH event horizon ($r=r_{H}$), but actually (in a scalar-dependent way)
all along the non-BPS $Z\neq 0$ attractor flow (\textit{i.e.} $\forall
r\geqslant r_{H}$). \medskip

It is here worth pointing out that, by exploiting the procedure outlined in
Subsect. \ref{U-Duality-Transf}, the results (\ref{gap}), (\ref{gap1}) and (%
\ref{gap2}) can be related one to the others by performing suitable $U$%
-duality transformations. In such a way (or equivalently by recalling the
results of Sect. \ref{stu} and Subsect. \ref{D0-D2-D4-D6}), one can also
compute $\Delta $ for the non-BPS $Z\neq 0$-supporting branch of the most
general (\textit{i.e.} $D0-D2-D4-D6$) BH charge configuration.

Of course, by relaxing the simplifying condition (\ref{e-1-tris}) and/or the
\textit{``}$t^{3}$\textit{-degeneration''} described by Eq. (\ref
{t^3-degeneration}) (in turn implying the condition(\ref{e-1-bis})), a much
richer situation arises, but the main features of the framework, outlined
above, are left unchanged.

\section{\label{Conclusion}Conclusion}

In the present paper the analysis and solution of the equations of motion of
the scalar fields of the so-called $stu$ model \cite
{Magnific-7,Duff-stu,BKRSW,Shmakova,TT,Saraikin-Vafa-1,TT2,BMOS-1},
consisting of $\mathcal{N}=2$, $d=4$ \textit{ungauged} supergravity coupled
to $3$ Abelian vector multiplets whose complex scalars span the special
K\"{a}hler manifold $\frac{G}{H}=\left( \frac{SU\left( 1,1\right) }{U\left(
1\right) }\right) ^{3}$, has been performed in full detail. The obtained
results complete and generalize the ones already present in literature \cite
{K2-bis, Hotta, GLS-1,Cai-Pang}.

The $3$ classes of \textit{non-degenerate} attractor flows of the $stu$
model have been presented and/or determined in full generality, and their
features have been studied and compared. We sketchily list the essential
facts below:

\begin{itemize}
\item  The $\frac{1}{2}$-BPS attractor flow, known since \cite{Cvetic-Youm}-
\nocite{Tseytlin,Gauntlett,Bala,BPS-flow-1,BPS-flow-2,BPS-flow-3,Bates-Denef}
\cite{bala-foam} (as well as the third of Refs. \cite{Marginal-Refs}), has
been reviewed. It corresponds to $\mathcal{I}_{4}>0$, and it does \textit{not%
} yield any associated \textit{moduli space}, at the BH event horizon nor
along the flow. The most general flow solution (\ref{BPS-sol}) can be
obtained starting from the known most general horizon, critical solution
\cite{BKRSW,Shmakova,TT,Saraikin-Vafa-1,TT2,BMOS-1,K2-bis}, and replacing
the BH charges with suitable harmonic functions. Correspondingly, the $\frac{%
1}{2}$\textit{-BPS Attractor Eqs. }(\ref{BPS-AEs}), determining the
attractor solution at BH event horizon, can be extended to the whole flow
into the so-called $\frac{1}{2}$\textit{-BPS stabilization Eqs. }(\ref
{BPS-stabilization-Eqs.}) by simply substituting the BH charges with the
corresponding harmonic functions in the radial parameter $\tau $. The
\textit{(first order) }$\frac{1}{2}$\textit{-BPS} \textit{fake superpotential%
} (which is nothing but the absolute value of the $\mathcal{N}=2$, $d=4$
\textit{central charge function} $Z$ given by Eq. (\ref{g-1}), and thus it
is manifesly $H$-invariant) has been explicitly determined, and the relevant
BH parameters, namely the \textit{gravitational ADM mass} (\ref{M-BPS-gen})
and \textit{(covariant) scalar charges }, have been computed, as functions
of BH charges and (spatially) asymptotical scalars. The \textit{marginal
bound} has been shown to be never saturated, and thus \textit{marginal
stability} \cite{Marginal-Refs} does not hold for $\frac{1}{2}$-BPS states
in the considered framework.

\item  The non-BPS $Z=0$ attractor flow, hitherto unknown (up to a short
comment in \cite{GLS-1}), has been studied and derived in full generality.
As the $\frac{1}{2}$-BPS attractor flow, it corresponds to $\mathcal{I}%
_{4}>0 $, and it does \textit{not} yield any associated \textit{moduli space}%
, at the BH event horizon nor along the flow. Due to the underlying triality
symmetry of the $stu$ model, $3$ different \textit{``polarizations'' }of the
results are possible, and choosing one of them does not imply any loss of
generality. The most general flow solution (\ref{non-BPS-Z=0-flow}) can be
obtained starting from the known most general horizon, critical solution
\cite{BMOS-1}, and replacing the BH charges with suitable harmonic
functions. Correspondingly, the \textit{non-BPS }$\mathit{Z=0}$ \textit{%
Attractor Eqs. }(\ref{non-BPS-Z=0-AEs}), determining the attractor solution
at BH event horizon, can be extended to the whole flow into the so-called
\textit{non-BPS }$\mathit{Z=0}$\textit{\ stabilization Eqs. }(\ref
{non-BPS-Z=0-stabilization-Eqs.}) by simply substituting the BH charges with
the corresponding harmonic functions in the radial parameter $\tau $. The
\textit{(first order) non-BPS }$\mathit{Z=0}$ \textit{fake superpotential}
has been explicitly determined, and it is given by the manifestly $H$%
-invariant Eqs. (\ref{Ws}) and (\ref{Ws-geom}) (or equivalently, by Eqs. (%
\ref{Wt}) and (\ref{Wt-geom}), or (\ref{Wu}) and (\ref{Wu-geom})). The
relevant BH parameters, namely the \textit{gravitational ADM mass} (\ref
{M-non-BPS-Z=0-gen}) and \textit{(covariant) scalar charges }, have been
computed, as functions of BH charges and (spatially) asymptotical scalars.
Furthermore, the \textit{BPS bound} \cite{gibbons2} has been shown to hold
all along the non-BPS $Z=0$ attractor flow (\textit{i.e.} $\forall
r\geqslant r_{H}$), and not only at the BH event horizon ($r=r_{H}$). The
\textit{marginal bound} has been shown to be never saturated, and thus
\textit{marginal stability} \cite{Marginal-Refs} does not hold for non-BPS $%
Z=0$ states in the considered framework. The strict similarity between $%
\frac{1}{2}$BPS and non-BPS $Z=0$ attractor flows can be explained by
noticing that both such flows can be uplifted to the \textit{same} $\frac{1}{%
8}$-BPS \textit{non-degenerate} attractor flow of $\mathcal{N}=8$, $d=4$
supergravity (see \textit{e.g.} the discussion in \cite{BMOS-1}).

\item  The non-BPS $Z\neq 0$ attractor flow, studied in \cite{K2-bis, Hotta,
GLS-1,Cai-Pang} in various configurations, has been here studied and
analyzed in full generality. It corresponds to $\mathcal{I}_{4}<0$, and it
yields an associated \textit{moduli space} $\left( SO\left( 1,1\right)
\right) ^{2}$ (which is nothing but the scalar manifold of the $d=5$ uplift
of the $stu$ model), both at the BH event horizon \cite{ferrara4,TT2} and
along the flow \cite{GLS-1}. Consistently with the analysis performed in
\cite{Hotta} for the $D0-D4$ (\textit{magnetic}) configuration, the most
general flow solution (\ref{sol})-(\ref{sol-2}), supported by the (relevant
branch of the) $D0-D2-D4-D6$ configuration (with all charges switched on),
cannot be obtained starting from the known most general horizon, critical
solution, and replacing the BH charges with suitable harmonic functions. The
opposite claim of \cite{K2-bis} for the $D2-D6$ (\textit{electric}) and $%
D0-D2-D4-D6$ configurations is actually due to the fact that (some) $B$%
\textit{-fields} were chosen to vanish therein. Correspondingly, the \textit{%
non-BPS }$\mathit{Z\neq 0}$ \textit{Attractor Eqs. }(\ref{non-BPS-Z<>0-AEs}%
), determining the attractor solution at BH event horizon, cannot be
extended to the whole flow into the so-called \textit{non-BPS }$\mathit{%
Z\neq 0}$\textit{\ stabilization Eqs. }(whose \textit{would-be} version is
given by Eq. (\ref{non-BPS-Z<>0-stabilization-Eqs.})) by simply substituting
the BH charges with the corresponding harmonic functions in the radial
parameter $\tau $. The \textit{(first order) non-BPS }$\mathit{Z\neq 0}$
\textit{fake superpotential} has been explicitly determined, and it is given
by the manifestly non-$H$-invariant Eq. (\ref{D0D2D4D6fake}), which
reproduces the few known results \cite{Cer-Dal-1} in the corresponding
particular cases. The relevant BH parameters, namely the \textit{%
gravitational ADM mass} and \textit{(covariant) scalar charges}, have been
computed in Subsect. \ref{D0-D6} and Sect. \ref{Analysis-Particular} for
various configurations, as functions of BH charges and (spatially)
asymptotical scalars. Furthermore, the \textit{BPS bound} \cite{gibbons2}
has been shown to hold all along the non-BPS $Z\neq 0$ attractor flow (%
\textit{i.e.} $\forall r\geqslant r_{H}$), and not only at the BH event
horizon ($r=r_{H}$). This has been explicitly computed for non-BPS $Z\neq 0$%
-supporting branches of \textit{magnetic }(Subsect. \ref{D0-D4}), \textit{%
electric} (Subsect. \ref{D2-D6}) and $D0-D2-D4$ (Subsect. \ref{D0-D2-D4})
configurations. On the other hand, by exploiting the procedure outlined in
Subsect. \ref{U-Duality-Transf}, or equivalently by recalling the results of
Sect. \ref{stu} and Subsect. \ref{D0-D2-D4-D6}, one can also prove the
\textit{BPS bound} to hold for the non-BPS $Z\neq 0$-supporting branch of
the most general (\textit{i.e.} $D0-D2-D4-D6$) BH charge configuration. From
the very Eq. (\ref{D0D2D4D6fake}), the \textit{marginal bound} turns out to
be saturated, and thus \textit{marginal stability} \cite{Marginal-Refs}
holds for non-BPS $Z\neq 0$ states in the considered framework. The manifest
non-$H$-invariance of the general \textit{(first order) non-BPS }$\mathit{%
Z\neq 0}$ \textit{fake superpotential} (\ref{D0D2D4D6fake}) seems clash with
the $H$-invariance imposed by Eq. (2.21) of \cite{Cer-Dal-1} (or
equivalently by Eq. (13) of \cite{ADOT-1}), reported as eq. (\ref{cc-1})
above, relating the \textit{fake superpotential }and the \textit{warp factor}
$U\left( \tau \right) $ appearing in the \textit{Ansatz} (\ref{c-1})for the
static, spherically symmetric, asymptotically flat, \textit{extremal} dyonic
BH metric. A way out to such an apparent contradiction consists in recalling
the treatment of Subsect. 2.2. of \cite{Cer-Dal-1}, and thus observing that
the fake superpotential is \textit{not} unique within the same attractor
flow, the various equivalent superpotentials being related through a matrix $%
R$ satisfying the conditions (2.28) and (2.29) of \cite{Cer-Dal-1} (see in
general Subsect. 2.2 of \cite{Cer-Dal-1}). If $R$ is scalar-dependent, it
may relate (manifestly) $H$-invariant \textit{fake superpotentials} to
(manifestly) non-$H$-invariant ones, and \textit{vice versa}. Thus, one may
state that a suitable scalar-dependent matrix $R$ exists, satisfying Eqs.
(2.28) and (2.29) of \cite{Cer-Dal-1}, and mapping the \textit{non-BPS }$%
\mathit{Z\neq 0}$ \textit{fake superpotential} (\ref{D0D2D4D6fake}) into an
equivalent, but (manifestly) $H$-invariant one. It would be interesting to
determine explicitly such a matrix; we leave such an issue for future
work.\bigskip
\end{itemize}

Various comments, remarks, ideas for further developments along the lines of
research considered in the present paper are listed below.

\begin{itemize}
\item  By exploiting the approach considered in Sect. 5 of \cite{BMOS-1},
the $stu$ can be consistently related to the so-called $st^{2}$ and $t^{3}$
models, respectively with $2$ and $1$ \textit{complex} scalars. Through such
a procedure, all the results obtained for the $stu$ model can be considered
to hold for such models. Furthermore, by performing the \textit{near-horizon}
(\textit{i.e.} $\tau \rightarrow -\infty $) \textit{limit} on the attractor
flow solutions, one obtains the corresponding attractor solution at the
event horizon of the extremal BH. This is particularly relevant for the
non-BPS $Z\neq 0$ horizon attractor solutions, hitherto analytically known
(in a rather intricate form) only for the $t^{3}$ model, so far the only $%
\mathcal{N}=2$, $d=4$ supergravity model based on \textit{cubic} special
K\"{a}hler geometry whose Attractor Eqs. had been completely solved. In the
\textit{near-horizon} \textit{limit}, the results of the present paper yield
the non-BPS $Z\neq 0$ horizon attractor solutions for both $st^{2}$ and $stu$
models.

\item  It should be recalled once again that the $stu$ model is a sector of
\textit{all} $\mathcal{N}>2$, $d=4$ supergravities, as well as of \textit{all%
} $\mathcal{N}=2$, $d=4$ supergravities based on homogeneous (both \textit{%
symmetric} \cite{GST1,Magnific-7,CFG,CVP} and \textit{non-symmetric} - see
\textit{e.g.} \cite{dWVVP,dWVP} -) scalar manifolds based on cubic
geometries. Thus, $stu$ model captures the essential features of extremal
BHs in all such theories (see \textit{e.g.} the $stu$ interpretation of $%
\mathcal{N}=8$, $d=4$ attractors , and the observations in \cite{GLS-1}).
Consequently, (the core of) the results holding for $\mathit{stu}$ model can
be thought to hold \textit{at least} for all such theories. For instance, it
would be interesting to try to extend them to some of the theories
considered in \cite{Gnecchi-1}, also in relation to the issue of the \textit{%
effective horizon radius} treated therein.

\item  The $stu$ model has been recently shown to be relevant for the
analogy between pure states of multipartite entanglement of \textit{qubits}
in \textit{quantum information theory} and extremal stringy BHs \cite{duff1}
(see also \cite{Levay-Saniga-Vrana} for further recent developments). In the
seventh of Refs. \cite{duff1} the relation between quantum information
theory and the theory of extremal stringy BHs was studied within the $stu$
model, showing that the \textit{three-qubit interpretation} of
supersymmetric, $\frac{1}{2}$-BPS attractors can be extended also to include
non-supersymmetric, non-BPS (both $Z\neq 0$ and $Z=0$) ones, performing a
classification of the attractor solutions based on the charge codes of
\textit{quantum error correction}. However, only \textit{double-extremal }%
solutions, with \textit{constant}, \textit{non-dynamical} scalars \textit{%
all along} the attractor flow, were discussed therein. Thus, as also
observed in \cite{Cai-Pang}, it would be interesting to extend the analysis
of the seventh of Refs. \cite{duff1} using the full general non-BPS (both $%
Z\neq 0$ and $Z=0$) attractor flow solutions obtained in the present paper.

\item  The existence of a \textit{first order formalism} for the equations
of motion of the scalar fields (also named \textit{attractor flow Eqs.}) in
the background of an extremal BH \cite{Cer-Dal-1,ADOT-1} \textit{in principle%
} implies the \textit{integrability} of such equations, regardless their
eventual intricate form. This answers to the question raised in Appendix A
of \cite{GLS-1}, and it is particularly relevant for the non-BPS $Z\neq 0$
attractor flow, as pointed out in Subsects. \ref{D0-D4}-\ref{D0-D2-D4}. It
would be interesting to study the integrability of the equations of motion
of the scalars in presence of \textit{quantum} (\textit{perturbative} and/or
\textit{non-perturbative}) corrections to the considered $stu$ model. For
instance, it would be interesting to study the attractor flow Eqs. for a
quantum corrected prepotential $f=stu+i\lambda $, with $\lambda \in \mathbb{R%
}$, which is the only correction which preserves the \textit{axion shift
symmetry} and modifies the geometry of the scalar manifold (see \cite{BFMS1}
and Refs. therein). A tempting ideas, inspired by the intriguing connection
between quantum information theory and extremal BHs mentioned at the
previous point, is to consider the quantum, axion-shift-consistent parameter
$\lambda $ as related to the \textit{quantum noise} of the system (see
\textit{e.g.} \cite{Quantum-Noise} and Refs. therein).

\item  As found in \cite{Ferrara-Gimon}, observed also in \cite{GLS-1} and
noticed in Sect. \ref{BPS-Flow}, an immediate consequence of the general
form of $\frac{1}{2}$-BPS attractor flow given by Eq. (\ref{BPS-sol}) is
that $\Gamma _{\infty }$ satisfies the $\frac{1}{2}$\textit{-BPS Attractor
Eqs. }\cite{BPS-flow-1}. This determines a sort of \textit{``Attractor
Mechanism at spatial infinity'',} mapping the $6$ \textit{real} moduli $%
\left( x^{1},x^{2},x^{3},y^{1},y^{2},y^{3}\right) $ into the $8$ \textit{real%
} constants $\left( p_{\infty }^{1},p_{\infty }^{2},p_{\infty
}^{3},q_{1,\infty },q_{2,\infty },q_{3,\infty }\right) $, arranged as $%
\Gamma _{\infty }$ and constrained by the $2$ \textit{real} conditions (\ref
{constraints}). As noticed in \cite{GLS-1}, the absence of \textit{flat}
directions in the $\frac{1}{2}$-BPS attractor flow (which is a general
feature of $\mathcal{N}=2$, $d=4$ \textit{ungauged} supergravity coupled to
Abelian vector multiplets, \textit{at least} as far as the metric of the
scalar manifold is strictly positive definite $\forall \tau \in \mathbb{R}%
^{-}$ \cite{FGK}) is crucial for the validity of the expression (\ref
{BPS-sol}). As pointed out in Sect. \ref{Non-BPS-Z=0-Flow}, the same holds
for the non-BPS $Z=0$ case. Indeed, a consequence of the general form of
non-BPS $Z=0$ attractor flow given by Eq. (\ref{non-BPS-Z=0-flow}) is that $%
\Gamma _{\infty }$ satisfies the \textit{non-BPS }$\mathit{Z=0}$ \textit{%
Attractor Eqs. }(see \textit{e.g.} \cite{AoB-book} and \cite{BFMY}),
determining a sort of \textit{``Attractor Mechanism at spatial infinity''}.
Analogously to what happens in the $\frac{1}{2}$-BPS case, the absence of
\textit{flat} directions in the non-BPS $Z=0$ attractor flow (which is
\textit{not} a general feature of $\mathcal{N}=2$, $d=4$ \textit{ungauged}
supergravity coupled to Abelian vector multiplets, but however holds for the
$stu$ model \cite{Ferrara-Marrani-1,ferrara4}) is crucial for the validity
of the expression (\ref{non-BPS-Z=0-flow}). In view of the crucial
differences among the non-BPS $Z\neq 0$ attractor flow and the $\frac{1}{2}$%
-BPS and non-BPS $Z=0$ attractor flows (such as the presence of a $2$-dim.
real \textit{moduli space} $\left( SO\left( 1,1\right) \right) ^{2}$ \textit{%
all along} the non-BPS $Z\neq 0$ attractor flow), it would be interesting to
investigate the non-BPS $Z\neq 0$ \textit{``Attractor Mechanism at spatial
infinity''}, if any.
\end{itemize}

\section*{Acknowledgements}

We would like to warmly acknowledge L. Andrianopoli, A. Ceresole, G.
Dall'Agata, R. D'Auria, E. Orazi, M. Trigiante, and especially E. G.
Gimon, for fruitful discussions and interest in this work. We also
thank A. Banijamali for careful reading of our manuscript and
correcting some typos.

A. M. would also like to thank the Department of Physics, Theory Unit Group
at CERN, where part of this work was done, for kind hospitality and
stimulating environment.

The work of S.B. has been supported in part by the European Community Human
Potential Program under contract MRTN-CT-2004-005104 \textit{``Constituents,
fundamental forces and symmetries of the Universe''}.

The work of S.F.~has been supported in part by the European Community Human
Potential Program under contract MRTN-CT-2004-005104 \textit{``Constituents,
fundamental forces and symmetries of the Universe''}, in association with
INFN Frascati National Laboratories and by D.O.E.~grant DE-FG03-91ER40662,
Task C.

The work of A.M. has been supported by Museo Storico della Fisica e
Centro Studi e Ricerche ``\textit{Enrico Fermi''}, Rome, Italy, in
association with INFN-LNF, and by an INFN Visiting Theoretical
Fellowship at Stanford Institute for Theoretical Physics.

The work of A.Y. was supported in part by the grant INTAS-05-7928, in
association with INFN Frascati National Laboratories. 

\end{document}